\documentclass[a4paper,11pt]{article}

\usepackage{subcaption}
\usepackage{jheppub} %
\usepackage{physics}

\usepackage{bookmark}                    
\usepackage{amssymb}
\usepackage{amsmath}
\usepackage{mathstyle}
\usepackage{iftex}
\ifLuaTeX
\usepackage{fontspec} %
\else
\usepackage[utf8]{inputenc} %
\fi
\usepackage{float}
\usepackage{xcolor}
\usepackage{cancel}
\usepackage{breqn}
\usepackage{bm}
\usepackage{enumitem} %
\usepackage{mathtools}
\usepackage{braket}

\usepackage{comment}
\usepackage{amstext} %
\usepackage{lineno}
\usepackage{xspace} %
\usepackage{accsupp} %
\def\det{\mathop{\rm det}\nolimits}

\def\Tr{\mathop{\rm Tr}\nolimits}
\def\bbra{{\langle\kern-2.5pt\langle}}
\def\kket{{\rangle\kern-2.5pt\rangle}}
\def\Bbra{{\Big\langle\kern-3.5pt\Big\langle}}
\def\Kket{{\Big\rangle\kern-3.5pt\Big\rangle}}

\newcommand   \half{\frac 1 2}
\newcommand   \lptl{\raise .8ex\hbox{$^\leftarrow$} \hspace{-9pt} \partial}
\newcommand   \lrptl{\raise .8ex\hbox{$^\leftrightarrow$} \hspace{-9pt} \partial}

\newcommand   \SO    {\mathrm{SO}}

\newcommand   \Sp    {\mathrm{Sp}}

\newcommand   \cL {\mathcal{L}}

\newcommand   \cN {\mathcal{N}}
\newcommand   \cO {\mathcal{O}}

\newcommand{\be}{
  \begin{equation}
  \begin{aligned}
}
\newcommand{\ee}{
  \end{aligned}
  \end{equation}
}

\makeatletter
\DeclareRobustCommand\bea{\@ifnextchar[{\@@bea}{\@bea}}
\def\@@bea[#1]#2\eea{\begin{subequations}\begin{align}#2\end{align}\label{#1}\end{subequations}}
\def\@bea#1\eea{\begin{subequations}\begin{align}#1\end{align}\end{subequations}}
\makeatother

\usepackage{array}   %
\newcolumntype{L}{>{$}l<{$}} %

\usepackage[compat=1.1.0]{tikz-feynman}
\usepackage{cleveref}

\title{\boldmath Melonic limits of the quartic Yukawa model and general features of melonic CFTs}
\author[a,1]{Ludo Fraser-Taliente,\note{Corresponding author.}}
\author[a]{and John Wheater}

\affiliation[a]{Rudolf Peierls Centre for Theoretical Physics, University of Oxford,\\Parks Road, Oxford OX1 3PU, UK}

\emailAdd{ludovic.fraser-taliente@physics.ox.ac.uk}

\makeatletter
\edef\savedcodes{\catcode`\noexpand\_=\the\catcode`\_}
\edef\@tempa{\csname opt@newtxmath.sty\endcsname}
\def\@tempb{{subscriptcorrection}}
\expandafter\expandafter\expandafter
  \in@\expandafter\@tempb\expandafter{\@tempa}
\ifin@
  \catcode`\_=12 %
\fi
\makeatother
\usepackage{tensor}
\newcommand{\od}[2]{\frac{\mathrm{d}#1}{\mathrm{d}#1}}
\savedcodes

\newcommand\hp{h_p}
\newcommand\hw{h_w}
\newcommand\hmelonic{\ensuremath{h_{\text{melonic}}}\xspace}
\newcommand\hprismatic{\ensuremath{h_{\text{prismatic}}}\xspace}
\newcommand\lammelonic{\ensuremath{\lambda_{\text{melonic}}}\xspace}
\newcommand{\hlammelonic}{\ensuremath{h\lambda_{\text{melonic}}}\xspace}
\newcommand{\hlamprismatic}{\ensuremath{h\lambda_{\text{prismatic}\xspace}}}

\abstract{%
We study a set of large-$N$ tensor field theories with a rich structure of fixed points, encompassing both the melonic and prismatic CFTs observed previously in the conformal limits of other tensor theories and in the generalised Sachdev-Ye-Kitaev (SYK) model. The tensor fields interact via an $\mathrm{O}(N)^3$-invariant generalisation of the quartic Yukawa model, $\phi^2\bar{\psi}\psi+\phi^6$. To understand the structure of IR/UV fixed points, we perform a partial four-loop perturbative analysis in $D=3-\epsilon$. We identify the flows between the melonic and prismatic fixed points in the bosonic and fermionic sectors, finding an apparent line of fixed points in both. We reproduce these fixed points non-perturbatively using the Schwinger-Dyson equations, and in addition identify the supersymmetric fixed points in general dimension. Selecting a particular fermionic fixed point, we study its conformal spectrum non-perturbatively, comparing it to the sextic prismatic model. In particular, we establish the dimensional windows in which this theory remains stable. We comment on the structure of large-$N$ melonic CFTs across various dimensions, noting a number of features which we expect to be common to any such theory.
}


\begin{document}
\maketitle
\flushbottom

\newpage

\section{Introduction}

The large-$N$ limit is essentially just mean field theory. In a quantum field theory containing large numbers of degrees of freedom and a large symmetry group $G$, $G$-invariant quantities self-average; they then simplify, or even in some cases become analytically solvable.

The large-$N$ vector $\phi_i$ and matrix $\phi_{ij}$ scalar field theories, invariant under $\mathrm{O}(N)$ and $\mathrm{O}(N)^2$ respectively, are well known. The natural continuation is to consider higher rank fundamental fields; that is, standard field theories in $\mathbb{R}^d$, but with fields transforming in a representation of the symmetry group $\mathrm{O}(N)^{r>2}$.  
These have, at their renormalisation group fixed points, a new class of non-trivial CFTs; the \textit{melonic CFTs} \cite{Gurau:2024nzv}, which have rich behaviour, and yet are analytically tractable. 

These theories lie, perhaps surprisingly, between the vector models and the matrix models. Recall that the diagrammatic expansion of the vector models is dominated by contributions from the cactus/snail diagrams in the large-$N$ limit, which are completely summable via a geometric series; they are in a sense too simple to be interesting, leading to \textit{ultralocal} dynamics. The diagrammatic expansion of the matrix models is dominated by the subset of Feynman diagrams that can be drawn on the plane; they are \textit{not} so directly summable, and analytic progress is more difficult. %

The graphs that dominate in the melonic limit of tensor models, and indeed after the disorder average of the SYK model, are the \textit{melonic} graphs, which are a simpler subset of the planar graphs: they are the graphs of leading \textit{(Gurau) degree} \cite{Gurau:2010ba}. They are simple enough to be summable, but complex enough to be interesting, and in that sense lie between the vector and matrix models -- in richness, but not in rank. In particular, unlike the vector model, we have non-trivial dynamics at order $N^0$, such as a non-zero anomalous dimension for operators; and, unlike the matrix model, it is straightforward to make exact statements. 
\begin{figure}
    \centering
    \includegraphics[width=0.5\textwidth]{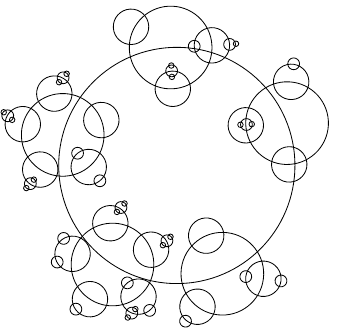}
    \caption{All graphs in the melonic limit are constructed from the iterated melon; here we show a high-order contribution to the self-energy in an arbitrary quartic melonic theory.}
    \label{fig:quarticMelon}
\end{figure}
The standard single field models \cite{Gross:2016kjj,Giombi:2017dtl,Benedetti:2019rja} have a relatively simple fixed point structure, that so far has been grouped into the categories of prismatic and melonic. In this paper we will study non-perturbatively the complete set of (symmetry-unbroken) large-$N$ melonic fixed points of the quartic Yukawa model; namely, the tensorial generalisation of a Dirac fermion and real scalar with interaction $\phi^2 \bar{\psi}\psi+\phi^6$.

This theory has both prismatic and melonic-type fixed points, with and without fermions. We will show that these fixed points match to those of Wilson-Fisher type found by a perturbative analysis around the upper critical dimension, $D=3-\epsilon$. This gives access to the flows between fixed points, which cannot be seen non-perturbatively here. Models of this kind have much richer structure than the melonic theory of a single field. We will observe a complex network of fixed points as the dimension is varied, and study their stability and unitarity, both perturbatively and non-perturbatively.

\subsection{Tensor models}

To establish our conventions, we begin by considering the standard single-field real bosonic tensor model \cite{Giombi:2017dtl,Benedetti:2019rja}. This model has some large number of real scalar fields $\phi_I$, $I=1,\ldots,\cN$, and without interaction terms has an $O(\cN)$ symmetry group. Particular choices of interaction can then break this down to a smaller subgroup $\mathrm{O}(N)^r$, with $N^r = \cN$, where we call $r$ the rank. The well-studied vector and matrix models correspond to $r=1$ and $r=2$ respectively \cite{Moshe:2003xn}. We will focus on the case $r=3$, but for a recent summary of rank $r>3$ tensors, see \cite{Jepsen:2023pzm}. 

Taking $r=3$, we are naturally led towards organising our $\cN=N^3$ fields into a tensorial field $\phi_{abc}$, $a,b,c=1,\ldots,N$, as then terms in the Lagrangian allowed by the symmetry are tensor invariants of these fields, such as $\sum_{abc} \phi_{abc}\phi_{abc}$. Note that each index\footnote{In the $\mathrm{O}(N)$ model, we do not distinguish raised and lowered indices.} here transforms in the fundamental of a different $\mathrm{O}(N)$ global symmetry, such that the action of the symmetry group $\mathrm{O}(N)^3$ is
\begin{equation}\begin{aligned}
\phi_{abc} \mapsto O_{a a'} P_{b b'} Q_{c c'} \phi_{a'b'c'},%
\end{aligned}\end{equation}
and we require that the Lagrangian is invariant under this global transformation of all the fields. In order to obtain a conformal field theory in $D\le 3$, we restrict to renormalisable interactions. The melonic limit is then taken by choosing an optimal scaling in $N$ for the coupling constants, and then taking $N \to \infty$ in the graphical expansion.

With only one field in the Lagrangian (and therefore only one field in the melonic-dominant Feynman diagrams) the scaling dimension of $\phi$ in the conformal large-$N$ limit is completely determined by dimensional analysis. With the full propagator $G(p)=\expval{\phi(p)\phi(-p)}$, in the melonic limit the IR Schwinger-Dyson equation is simply $G(p)^{-1} \propto \int_{k,l}G(p+k+l)G(k)G(l)$, which immediately gives that $D-2\Delta_\phi = 2D-3(D-2\Delta_\phi)$, and so $\Delta_\phi = D/4$ exactly \cite{Giombi:2017dtl}. However, when a second field is introduced, as in the prismatic model \cite{Giombi:2018qgp}, we have the possibility of more large-$N$ fixed points and scaling dimensions that are non-linear in $D$. 

\subsection{Fermionic models}

Previously the tensorial Gross-Neveu model has been studied in $D=3$ \cite{Delporte:2020rce}. This model is non-renormalisable at finite $N$, but becomes renormalisable in the large $N$ limit. The theories at finite and large $N$ are fundamentally different, which makes it less likely that interesting structure persists to finite $N$. This motivates looking for a theory of fermions in the large-$N$ limit where the precursor theory is renormalisable at finite $N$. However, the obvious step of studying a Yukawa-like interaction $\lambda \phi \bar{\psi}\psi+ g \phi^4$ is not possible for a rank-3 tensor model\footnote{We note that there is a Sachdev-Ye-Kitaev model-like disorder-averaged approach to this $D\le 4$ theory \cite{Prakash:2022gvb}, since the disorder average approach gives identical diagrams to the melonic theories at leading, but not subleading, order in $N$. However, the RG flow information and connection to the bosonic model is lost in doing so. Analogous models have also been studied in a supersymmetric quantum mechanical context \cite{Marcus:2018tsr}.}, and so we move on to what we call the \textit{quartic Yukawa model}. %

We consider a theory with two tensor fields -- one real scalar $\phi_{abc}$ and one Dirac fermion $\psi_{abc}$ -- with an interaction term given by the tensorial generalisation of
\begin{equation}\begin{aligned}
V(\phi, \psi) = \frac{\lambda}{2} \phi^2 \bar\psi \psi + \frac{h}{6!} \phi^6.
\end{aligned}\end{equation}
This is the unique extension of the tensorial sextic bosonic scalar field theory \cite{Giombi:2018qgp,Benedetti:2019rja,Harribey:2021xgh} to a theory containing fermions, which is still renormalisable in $D\le 3$. This Lagrangian without the $\phi^6$ potential was first studied -- in the vector large-$N$ limit -- by Popovi\'c in 1977 \cite{Popovic:1977cq} (commented on in \cref{sec:Popovic}); with the additional potential, it was studied in exactly three dimensions in \cite{Dilkes:1997vc,McKeon:1999vx}, and then with the addition of a Chern-Simons coupling in \cite{Jack:2016utw}; a non-tensor version with the scalar potential was briefly considered in \cite{Herzog:2022jlx}; with Majorana spinors instead, it gives the supersymmetric Wess-Zumino model $(\Phi^4)$ in 2+1 dimensions \cite{DeWolfe:2019etx}. We will analyse the melonic fixed points both perturbatively in $D=3-\epsilon$ and via the non-perturbative Schwinger-Dyson equations. Additionally, as we will see, identifying non-perturbatively all perturbative symmetry-unbroken fixed points requires the introduction of a non-dynamical auxiliary scalar field $X_{abc}$. %

Our analysis will find the known fixed points of the bosonic sector (the melonic \cite{Benedetti:2019rja,Harribey:2021xgh} and prismatic \cite{Giombi:2018qgp}), but will also uncover three new fermionic generalisations of these melonic fixed points. We will see in perturbation theory an apparent collision of these fixed points for a particular dimension of the gamma matrices, but it will be resolved by a non-perturbative analysis. An apparent line of fixed points in both the fermionic and bosonic sectors (in the same direction for both), will also be found, but we will be unable to establish if it is just an artifact of the order of the perturbative calculation.   Focusing on the simplest of these new CFTs, which we refer to by \lammelonic{}, we will investigate its spectrum as a function of $D$. We will frequently draw comparison to the simpler sextic prismatic fixed point (which we here call \hprismatic), and so various results will also be presented for it.

Specifically, we will consider the reality of the scaling dimensions of the spectrum, as a probe of stability and, separately, unitarity. Note that for non-integer-$D$, at high scaling dimension, we expect the so-called evanescent operators/negative-norm states to appear, making the CFT non-unitary \cite{Hogervorst:2015akt,DiPietro:2017vsp,Ji:2018yaf}; these operators disappear in integer $D$. Nonetheless, it is still interesting to consider to what extent we can consider these theories to be unitary, just as in the case of the bosonic melonic field theories \cite{Giombi:2017dtl,Benedetti:2019ikb}, precisely because they are the dimensional continuation of theories that become unitary in integer dimension \cite{Fei:2015oha}. %
Despite its fixed point occurring at a negative value of the coupling constant, it displays hallmarks of unitarity in the strict large-$N$ limit; this appears to be because the coupling only appears squared, thanks to the melonic dominance.

\subsection{Guide to the paper}

In \cref{sec:modelanalysis} we comment on the general features of the non-tensor $\phi^2 \bar\psi \psi + \phi^6$ model.  We present the results of a perturbative analysis in $D=3-\epsilon$ of this theory to third order in the coupling constants, in \cref{sec:betas}, both in full generality and specialising to the large-$N$ melonic limit; here we will identify the apparent line of fixed points in both the bosonic and fermionic sectors. In \cref{sec:SDEanalysis} we then use the large-$N$ melonic limit to analyse non-perturbatively the conformal field theories arising at the fixed points of the model, including a matching to the supersymmetric theory; we highlight a number of features which arise in the simpler setting of \hprismatic, before proceeding to the fermionic theories. In \cref{sec:bilinears}, this is then further developed in the specific case of the \lammelonic fixed point, where we use diagonalisation of the four-point kernel to obtain the exact spectrum $\{\Delta\}$ of bilinear operators in the OPE of the fundamental fields via a Bethe-Salpeter-like equation; this diagonalisation is exact in the melonic limit. First we explain the computation, and then in \cref{sec:bilinearsCalculationResults} we study the properties of the spectrum so obtained. We conclude in \cref{sec:outlook}, and provide a number of useful and technical results in the appendices, particularly some concerning the $\phi^2 \bar{\psi}\psi  +\phi^6$ model.

\section{General comments on the quartic Yukawa model} \label{sec:modelanalysis}

\subsection{Our model, \texorpdfstring{$\phi^2 \bar{\psi}\psi+\phi^6$}{phi^2 psibar psi plus phi^6}}

The most general renormalisable theory of a Dirac fermion and a real scalar field in $D\le 3$ is described by the following Lagrangian
\begin{equation} \label{eq:Neq1Lagrangian}
\mathcal{L}=-\bar{\psi}\left(\not \partial+M\right) \psi-\frac{1}{2}\phi (-\partial^2 + m^2) \phi -V_{\mathrm{int}}(\phi, \psi), \quad V_{\mathrm{int}}(\phi,\psi) \equiv \frac{\lambda}{2} \phi^{2} \overline{\psi} \psi + \frac{g}{4 !} \phi^{4} + \frac{h}{6!}\phi^6.
\end{equation}
We have: suppressed counterterms for convenience; assumed a $\mathbb{Z}_2$ to remove $\phi$-odd terms; taken the mostly positive $(-++\cdots)$ signature. The complex Dirac fermions are used because they exist in any dimension. The real scalar fields are partly used for simplicity, and partly used to ensure that we can straightforwardly access the prismatic fixed points. The result for a complex scalar field will be straightforward to obtain from the results below; we need only add an index of $\SO(2)\cong \mathrm{U}(1)$.

It will prove convenient, for the purposes of accessing the full range of IR CFTs (specifically, those of prismatic type), to also consider the addition of a non-dynamical real auxiliary field $X$; this is just as in the standard $\phi^4$ vector model \cite{Moshe:2003xn}.
\begin{equation}
\cL_{\mathrm{aux}} = -\half X^2 - V_{\mathrm{int,aux}}(X,\phi),  \quad V_{\mathrm{int,aux}}(X,\phi) = \frac{\rho}{3!} X \phi^3
\end{equation}
Since $X$ enters only quadratically, it can of course integrated out exactly, leading only to a shift in the value of $h$.

\subsection{Comments on the model}
We make the following observations about this model as a non-tensorial quantum field theory.
\begin{itemize}
    \item As in the case of the Wilson-Fisher fixed point, which exists in $D=4-\epsilon$ for $\phi^4$, we expect to be able to find a fixed point of the renormalisation group in $D=3-\epsilon$, where the renormalised coupling constants have a perturbative expansion around zero for $\epsilon \ll 1$; this is done in \cref{sec:vectorModelAnalysis}. In standard perturbative QFT, this analysis is trustworthy for small $\epsilon$, with the fixed point colliding with the trivial fixed point in $D=3$ exactly. However, in the case of the melonic theories, it is possible to exactly solve for the scaling dimensions of these theories for all values of $D$, while also matching on to the perturbative $\epsilon \ll 1$ expansion.
    \item It is possible to consistently set each of the $D=3$ marginal couplings, $\lambda$, and $h$ to zero by setting their bare values to zero. That this is possible for $\lambda$ is obvious, since then the fermions are non-interacting; that this is possible for $h$ is more surprising, and because the one-loop $\lambda^3$ contribution to $\expval{\phi^6}$ is not divergent (even though in 3D $\Tr(\gamma_\mu \gamma_\nu \gamma_\rho) \propto \epsilon_{\mu\nu\rho} \neq 0$). 
    \item Any value of $M\neq 0$ explicitly breaks parity symmetry in 3D. %
    \item To sidestep the confusion of fermions in non-integer dimensions (reviewed in \cite{Pannell:2023tzc,Jack:2023zjt}), we will only deal with Dirac fermions, which are well-defined in any integer dimension. We will follow the standard approach of leaving the dimension of the gamma matrices as a free parameter, $T \equiv \Tr[\mathbb{I}_s]$. The ratio of the number of degrees of fermionic degrees of freedom to the number of bosonic degrees of freedom will also prove a useful parametrisation, $r\equiv 2T$. We discuss this point further in \cref{sec:rparameter}.
\end{itemize}

\subsection{Indexology}

Now, let us consider the $\cal{N}$-vector version of the model. Sprinkling $O(\cN)$ indices:
\begin{align}\label{eq:NvectorLagrangian}
\begin{split}
\mathcal{L}&=-\bar{\psi}_I\left(\not \partial \delta_{IJ}+M_{IJ}\right) \psi_J 
- \frac{1}{2} \phi_I (-\partial^2 \delta_{IJ} + m^{2}_{IJ}) \phi_I \phi_J -V_{\mathrm{int}}(\phi,\psi)\\
 V_{\mathrm{int}}(\phi,\psi) &= \frac{\lambda_{(IJ)KL}}{2} \phi_I \phi_J \overline{\psi}_K \psi_L+\frac{g_{(IJKL)}}{4 !} \phi_I\phi_J\phi_K\phi_L + \frac{h_{(IJKLMN)}}{6!}\phi_I\phi_J\phi_K\phi_L\phi_M\phi_N
\end{split}\\
 \cL_{\mathrm{aux}} &=-\frac{1}{2} X_I X_I - \frac{\rho_{I(JKL)}}{3!} X_I \phi_J \phi_K \phi_{L}\label{eq:auxLagrangian}
\end{align}
Once again, note that integrating out the non-dynamical $X$ simply leads to a redefinition of $h_{(IJKLMN)}$. Thus, for the perturbative analysis, we do not need to deal with additional complication of the coupling $\rho$.

The beta functions and anomalous dimensions of the theory without auxiliary field (with non-marginal couplings set to zero) are given in general by  \eqref{eq:vectorBetas3mepsLam} and \eqref{eq:vectorBetas3mepsh}. By taking specific $\mathrm{O}(N)^3$-symmetric forms for $h$ and $\lambda$, the beta functions of the tensorial theory can be obtained by index contraction. Said forms are depicted visually in \cref{fig:potentialWithH}, and in symbolic form in \cref{app:potential}.

Depending on the precise values chosen for the coupling constants, the symmetry group in these cases may be $\mathrm{U}(N)^3$ instead: however, we need only ensure that we pick the faithfully acting subgroup of a product of three general linear groups -- if we account for the symmetry factors, the melonic limit is unchanged.

We will be using the term \textit{superindex} to refer to a grouped set of three indices of $\mathrm{O}(N)^3$. That is, $\phi_I = \phi_{i_r i_b i_g}$, with $i_r,i_g,i_b=1,\ldots,N$.

\subsection{Conventions for CFTs and RG flow}
\subsubsection{Conformal field theories}

We will be interested in the fixed points of the RG flow of a given quantum field theory, which typically are conformal field theories. Conformal field theories in $D$ dimensions are defined by their data, which consists of a set of operators and $\SO(D)$ representations $\{(\Delta_i, \rho_i)\}$, and three-point coefficients $C_{ijk}$. These are observable, in the sense that they could be measured (at least for gauge-invariant quantities). Free fermions and bosons are trivially conformal field theories, and have the following scaling dimensions $D$:
\begin{equation}
    \Delta_\phi^0 = \frac{D-2}{2}, \quad \Delta_\psi^0 = \frac{D-1}{2}
\end{equation}
We will be finding interacting fixed points in the large-$N$ limit. In that limit, we will find that the operator which in the free theory had scaling dimension $\Delta_\phi^0$ now has a scaling dimension $\Delta_\phi$. Therefore, in general, we define the anomalous dimension of an operator $\cO$ to be $\gamma_\cO$, defined by $\gamma_\cO = \Delta_\cO - \Delta_\cO^0$.

In \cref{sec:SDEanalysis,sec:bilinearCalculationOverview}, we will be calculating these dimensions using a non-perturbative approach; however, these dimensions are usually found via a perturbative renormalisation group analysis. We shall do this shortly in \cref{sec:betas}.

\subsubsection{Beta functions in RG}

To calculate the scaling dimension, we follow the usual process of taking the bare Lagrangian:
\begin{equation}\begin{aligned}
\cL = \cL_{\mathrm{kinetic}}  + \sum_i (g_i)_0 \mu^{D-d_i} \cO_i^{\mathrm{bare}},
\end{aligned}\end{equation}
where $(g_i)_0$ is dimensionless, with the dimensions made up by the scale parameter $\mu$, where $d_i$ is the classical scale parameter. For example, for the operators $\cO_i^{\mathrm{bare}} = (\phi_0)^n$ in scalar field theory, $d_i = n \frac{D-2}{2}$. 

Then we replace all bare coupling constants $(g_i)_0 = Z_{g_i} g_i$, and all bare fields $(\phi_i)_0 = Z_{\phi_i} \phi_i$ with their renormalised versions. Following dimensional regularisation and the imposition of a renormalisation scheme, these renormalisation constants $Z$ develop a dependence on the renormalisation scale $\mu$, which we describe via the beta functions $\beta[g_i]$ of each of the couplings
\begin{equation}\begin{aligned}
\beta[g_i] &= \dv{g_i}{\log \mu},
\end{aligned}\end{equation}
and the anomalous ($\gamma_{\phi_i}$) and scaling ($\Delta_{\phi_i}$) dimensions of each of the fields
\begin{equation}\begin{aligned}
\gamma_{\phi_i} &= \dv{\log Z_{\phi_i}}{\log\mu}, \quad \Delta_{\phi_i} = \Delta_{\phi_i}^{\mathrm{free}} + \gamma_{\phi_i}.
\end{aligned}\end{equation}

The fixed points are found by solving $\beta[g^*_i]=0$. The anomalous dimensions of the fields are then simply $\gamma_{\phi_i}|_{\{g_*\}}$. Around a particular fixed point, $\beta[g^*_i]=0$, we can then define the stability matrix 
\begin{equation}\begin{aligned}\label{eq:stabMat}
S_{ij} = \dv{\beta[g_i]}{g_j}|_{\{g_*\}},
\end{aligned}\end{equation}
the eigenvalues of which are $\Delta_i - D$, where $\Delta_i$ is the scaling dimension of the renormalised operator $\cO_i$ in the renormalised Lagrangian. Positive eigenvalues of $S_{ij}$ signify an irrelevant operator in the IR, while negative eigenvalues signify a relevant operator. Complex eigenvalues indicate the fixed point is non-unitary.%

\section{\texorpdfstring{$3-\epsilon$}{3-epsilon} beta functions at large N in the melonic limit} \label{sec:betas}

We first perform a standard perturbative analysis of the multi-field theory, for finite $N$, and completely arbitrary couplings. For a similar analysis of the marginal theory of scalars and fermions in 4D ($\phi \bar\psi \psi + \phi^4$), see \cite{Osborn:2020cnf,Pannell:2023tzc}. The Feynman loop integrals are standard (see \cref{app:loopIntegrals}), except for being in $D=3$; note, of course, that diagrams with different tensor structures may have identical momentum structure. %

\subsection{Vector beta functions in \texorpdfstring{$D=3-\epsilon$}{D=3-eps}} \label{sec:vectorBetaFunctions}

We calculate the beta functions and field anomalous dimensions for the Lagrangian \eqref{eq:NvectorLagrangian} in $\overline{MS}$ scheme. In the following: Greek indices are dummy indices which are summed over; $F$ and $G$ are the indices of an anti-fermion and fermion respectively; the Latin $B_i$ indices indicate the index of a boson, which must be symmetrised over with weight one; we set $s=1/(8\pi)$. Then, to the indicated order in the marginal coupling constants\footnote{Non-marginal coupling constants have been set to zero, as they could be obtained from these calculations via the \textit{dummy field method} of \cite{Martin:1993zk} (pedagogically reviewed in \cite{Schienbein:2018fsw}). We do not, for example, use $g_{IJKL}$ here, as to find $\epsilon$-perturbative fixed points, it, like the field masses, must be tuned to zero.} $\lambda_{(IJ)KL}$ and $h_{(IJKLMN)}$, we find

\begin{subequations}
\begin{align}
    \begin{split}
    \label{eq:vectorBetas3mepsLam}
    &\beta[\lambda]_{B_1 B_2 FG} =-\epsilon  \lambda _{B_1 B_2 FG}+ \frac{1}{3} s^4 h_{B_1 \beta \gamma \delta \varepsilon \zeta } h_{B_2 \beta \gamma \delta \varepsilon \mu } \lambda _{\zeta \mu FG}\\
    &\quad +2\lambda _{B_1 \beta FG}\left[\frac{s^2}{3} T\lambda _{B_2 \zeta \eta \theta } \lambda _{\beta \zeta \theta \eta } + \frac{s^4}{90}  h_{B_2 \kappa \gamma \delta \varepsilon \zeta } h_{\beta \kappa \gamma \delta \varepsilon \zeta}\right]\\ 
    &\quad+ \frac{1}{3} s^2 \left[\lambda _{B_1 B_2 F\delta } \lambda _{\varepsilon \zeta \delta \theta } \lambda _{\varepsilon \zeta \theta G}+\lambda _{B_1 B_2 \gamma G} \lambda _{\varepsilon \zeta \gamma \theta } \lambda _{\varepsilon \zeta \theta F}\right]\\
    &+\frac{1}{3} s^2\left(6 T \lambda _{B_1 \beta \gamma \delta } \lambda _{B_2 \zeta \delta \gamma } \lambda _{\beta \zeta FG}+6 \lambda _{B_1 B_2 \gamma \delta } \lambda _{\varepsilon \zeta \delta G} \lambda _{\varepsilon \zeta F\gamma }\right.\\
    &\quad \quad\left.+12 \lambda _{B_1 \beta \gamma G} \lambda _{\beta \zeta F\eta } \lambda _{B_2 \zeta \eta \gamma }+12 \lambda _{B_1 \beta F\delta } \lambda _{B_2 \zeta \delta \theta } \lambda _{\beta \zeta \theta G}\right) +O(\lambda^4,\ldots)|_{\mathrm{sym } B_i},
    \end{split}\\
    \begin{split}
    \label{eq:vectorBetas3mepsh}
    &\beta[h]_{B_1 B_2B_3B_4B_5B_6} = -2 \epsilon  h_{B_1 B_2B_3B_4B_5B_6}+\frac{20}{3} s^2 h_{B_1 B_2 B_3\eta \nu \xi }h_{B_4 B_5 B_6\eta \nu \xi } \\
    &\quad +6 h_{B_1 B_2B_3B_4B_5\nu } \left[\frac{s^2}{3} T \lambda _{B_6\rho \tau \sigma } \lambda _{\nu \rho \sigma \tau } +\frac{s^4}{90}  h_{B_6\rho \sigma \tau \upsilon \varphi } h_{\nu \rho \sigma \tau \upsilon \varphi }\right]\\
    &\quad +5 h_{B_1 B_2B_3B_4\nu \xi } (6 s^2 T \lambda _{B_5\nu \sigma \tau } \lambda _{B_6\xi \tau \sigma }+s^4 h_{B_5\nu \sigma \tau \upsilon \varphi } h_{B_6\xi \sigma \tau \upsilon \varphi })\\
    &\quad -\tfrac{15}{2} \pi ^2 s^4 h_{B_1B_2\beta \eta \nu \xi } h_{B_3B_4\beta \eta \upsilon \varphi } h_{B_5B_6\nu \xi \upsilon \varphi }-80 s^4 h_{B_1B_2B_3\eta \nu \xi } h_{B_4B_5\eta o\chi \psi } h_{B_6\nu \xi o\chi \psi }\\
    & \quad - 360 s^2 T \left(\lambda _{B_1B_2\gamma \delta } \lambda _{B_3\zeta \delta \theta }\lambda _{B_4B_5\theta \kappa } \lambda _{B_6\zeta \kappa \gamma } +\lambda _{B_1B_2\gamma \delta } \lambda _{B_3B_4\delta \kappa }\lambda _{B_5\zeta \kappa \eta } \lambda _{B_6\zeta \eta \gamma }  \right)\\
    &\quad +O(h^4,\lambda^5,\ldots) |_{\mathrm{sym } B_i},
    \end{split}\\
    &\gamma^\phi_{B_1B_2} = \frac{s^2}{3} T \lambda _{B_1\beta \gamma \delta }\lambda _{B_2\beta \delta \gamma } + \frac{s^4}{90} h_{B_1\beta \gamma \delta \varepsilon \zeta } h_{B_2\beta \gamma \delta \varepsilon \zeta } +O(\lambda^3,\ldots) |_{\mathrm{sym } B_i}, \label{eq:generalVecGammaPhi}\\
    &\gamma^\psi_{FG} = \frac{s^2}{3} \lambda _{\alpha \beta F\delta } \lambda_{\alpha \beta \delta G}+O(\lambda^3,\ldots) \label{eq:generalVecGammaPsi},
\end{align} \label{eq:generalVectorResults}

\end{subequations}
where we can isolate the contributions of the anomalous dimensions to the beta functions in square brackets. Note that in this formulation, to obtain the $N=1$ beta functions, we need only drop all the index structure -- for example, $\beta(\lambda) = -\epsilon \lambda + \frac{1}{45} s^2 (s^2 h ( 15 h \lambda + h \lambda) + \ldots) + \ldots$, etc. This is not a complete four-loop calculation, as we have not calculated the $O(\lambda^4)$ contributions to the anomalous dimension, or the $O(\lambda^5)$ contributions to $h$ which appear at four loops.

We have three checks on our results here.
\begin{enumerate}
    \item The anomalous dimensions precisely match the leading order conformal calculation of \eqref{eq:conformalAnalysisDims}
    \item We were able to reproduce some finite-$N$ two-loop beta functions calculations: those of the bosonic sector matched  appendix A of \cite{Giombi:2018qgp}; and, other than one discrepancy (see \cref{app:JackPooleDiscrepancy}), the full theory matched the $D=3$ results of \cite{Jack:2016utw} to the order calculated there. %
    Likewise, the bosonic sector matches the vectorial calculation of \cite{Pisarski:1982vz}, up to their definition of $\gamma_\varphi \equiv 2 \gamma_\phi$.
    \item The anomalous dimensions at the fixed points agree (to four loops) with the non-perturbative melonic analysis of the Schwinger-Dyson equations of \cref{sec:SDEanalysis}, which is a completely orthogonal calculation. As a reminder, the scaling/anomalous dimensions of physical operators $\gamma_{\cO}$ evaluated at fixed points are physical, and so scheme-independent: $\Delta_{\cO} = \Delta_{\cO,\mathrm{free}} + \gamma_{\cO}$. Away from fixed points, this is no longer true.
\end{enumerate}

\subsection{Simple example}

As described above, we can calculate these quantities for the tensor model by breaking each superindex $I=1,\ldots, \cN=N^3$ into three separate indices $(ijk)$, each transforming under a separate $\mathrm{O}(N)^3$. Taking the melonic limit then requires substituting suitable combinations of delta functions for each of the coupling constant. It is then an exercise in (automated) index contractions to evaluate each of the tensor beta functions above, and to decompose them to find the beta functions of each $\mathrm{O}(N)^3$-invariant coupling constant. Taking the large-$N$ limit, assuming no symmetry breaking of $\mathrm{O}(N)^3$, we obtain the required results for the large-$N$ melonic theory.

We will now illustrate this, taking a rank two (matrix) model of scalars for simplicity, because the procedure is identical but more amenable to compact presentation. Consider a matrix model with fields $\phi_{ab},\psi_{ab}$, each transforming in the $\Box\times \Box$ of $\mathrm{O}(N)\times \mathrm{O}(N)$. This has $\kappa_{IJKL}=\kappa_{(i_r i_b)(j_r j_b)(k_r k_b) (l_r l_b)}$\footnote{These braces do not mean symmetrisation.}. Note that capital letters indicate a superindex of $\mathrm{O}(N)^2$, $I=(i_r i_b)$, and so both the red and blue indices. We also impose the symmetry under \textit{colour-averaging}, which is switching $r \leftrightarrow b$. Then, to evaluate an example index contraction $\beta[\kappa]_{B_1 B_2 FG} \supset \alpha \kappa_{B_1 \alpha F\delta}\kappa_{\alpha B_2 \delta G}$, we first must find the symmetric tensorial form of $\kappa_{IJKL}$:
\begin{equation}\begin{aligned}
&V(\phi,\psi)=\frac{\kappa_{IJKL}}{2} \phi_I \phi_J \textcolor{red}{\bar{\psi}_K} \textcolor{gray}{\psi_L} \equiv \frac{\kappa_{dt}}{2} \times \frac{1}{2} \left(\phi_{ab} \phi_{cb} \bar\psi_{ad}\psi_{cd} +\phi_{ab} \phi_{ac} \bar\psi_{db}\psi_{dc}\right)\\ %
&\implies \kappa_{IJKL}=\kappa_{(i_r i_b)(j_r j_b)(k_r k_b) (l_r l_b)} = \frac{\kappa_{b}}{2}(\delta_{i_r k_r} \delta_{i_b j_b}\delta_{j_r l_r} \delta_{k_b l_b}+ \delta_{i_r j_r} \delta_{i_b k_b} \delta_{j_b l_b} \delta_{k_r l_r})|_{\mathrm{sym }I\leftrightarrow J}\\
& \equiv \frac{\kappa_{dt}}{2} O^{dt}_{(i_r i_b)(j_r j_b)(k_r k_b) (l_r l_b)}  \label{eq:exampleKappa}
\end{aligned}\end{equation}
Given an example expression for the beta function,
\begin{equation}\begin{aligned} \label{eq:exampleBeta}
&\beta[\kappa]_{B_1 B_2 F G} =\alpha \kappa_{B_1 \alpha F\delta}\kappa_{\alpha B_2 \delta G} |_{\mathrm{sym} B_1 \leftrightarrow B_2},
\end{aligned}\end{equation}
we can then substitute in \eqref{eq:exampleKappa} expression for $\kappa$, and perform the index contractions:
\begin{equation}\begin{aligned}
&\beta[\kappa]_{(i_r i_b)(j_r j_b)(k_r k_b) (l_r l_b)} = \kappa_{(i_r i_b)(j_r j_b)(\gamma_r \gamma_b) (\delta_r \delta_b)}  \kappa_{(k_r k_b)(l_r l_b)(\delta_r \delta_b) (\gamma_r \gamma_b)}|_{\mathrm{sym}}\\
&=\alpha \frac{N \kappa_{dt}^2}{8} O^{dt}+ \mathrm{other structures}\\
&=\beta[\kappa_{dt}] O^{dt} + \ldots,
\end{aligned}\end{equation}
where $|_{\mathrm{sym}}$ now indicates that we perform $\frac{1}{4!} \sum_{\mathrm{perms}}$ over the $2!$ permutations of the 2 multi-indices $I=(i_r,i_b),J=(j_r,j_b)$. %

Thus, if \eqref{eq:exampleBeta} were the general beta function, the coupling constant defined by $\kappa_{dt}$ would flow according to the beta function
\begin{equation}\begin{aligned}
\beta[\kappa_{dt}]= \alpha \frac{N \kappa_{dt}^2}{8}
\end{aligned}\end{equation}
These delta function combinations $O^i$ are much more conveniently represented visually:
\begin{equation}\begin{aligned}
V(\phi,\psi)=\frac{\kappa_{IJKL}}{2} \phi_I \phi_J \textcolor{red}{\bar{\psi}_K} \textcolor{gray}{\psi_L}= \frac{\kappa_b}{2} \times \frac{1}{2} \left(\begin{tikzpicture}[baseline={([yshift=-.5ex]current bounding box.center)}]
\draw[color=red] (1,0) -- (1,1) (0,1) -- (0,0);
\draw[color=blue] (1,1) -- (0,1) (0,0) -- (1,0);
\fill[color=black] (0,0) circle (0.05) (1,0) circle (0.05); 
\fill[color=gray] (1,1) circle (0.05);
\fill[color=red] (0,1) circle (0.05);
\end{tikzpicture}+\begin{tikzpicture}[baseline={([yshift=-.5ex]current bounding box.center)}]
\draw[color=blue] (1,0) -- (1,1) (0,1) -- (0,0);
\draw[color=red] (1,1) -- (0,1) (0,0) -- (1,0);
\fill[color=black] (0,0) circle (0.05) (1,0) circle (0.05); 
\fill[color=gray] (1,1) circle (0.05);
\fill[color=red] (0,1) circle (0.05);
\end{tikzpicture}\right).\\
\end{aligned}\end{equation}
Here, red lines indicate contraction of the first index, and blue lines indicate contraction of the second index between two fields. $\phi_I$ is indicated by a black dot; $\bar{\psi}_K$ a red dot; $\psi_J$ a grey dot. A red line between $\phi_I$ and $\phi_J$ corresponds to a delta function $\delta_{i_r j_r}$, etc. Thus, these graphical depictions map to an analytical expression of tensors. When we introduce the third index on these fields below, we shall use the natural generalisation of this notation, with additional green lines.

\newpage
\subsection{The \texorpdfstring{$\mathrm{O}(N)^3$}{O(N)\^3} model and its beta functions} \label{sec:findingPerturbativeBetas}

\begin{figure}[H]
\includegraphics[width=\textwidth]{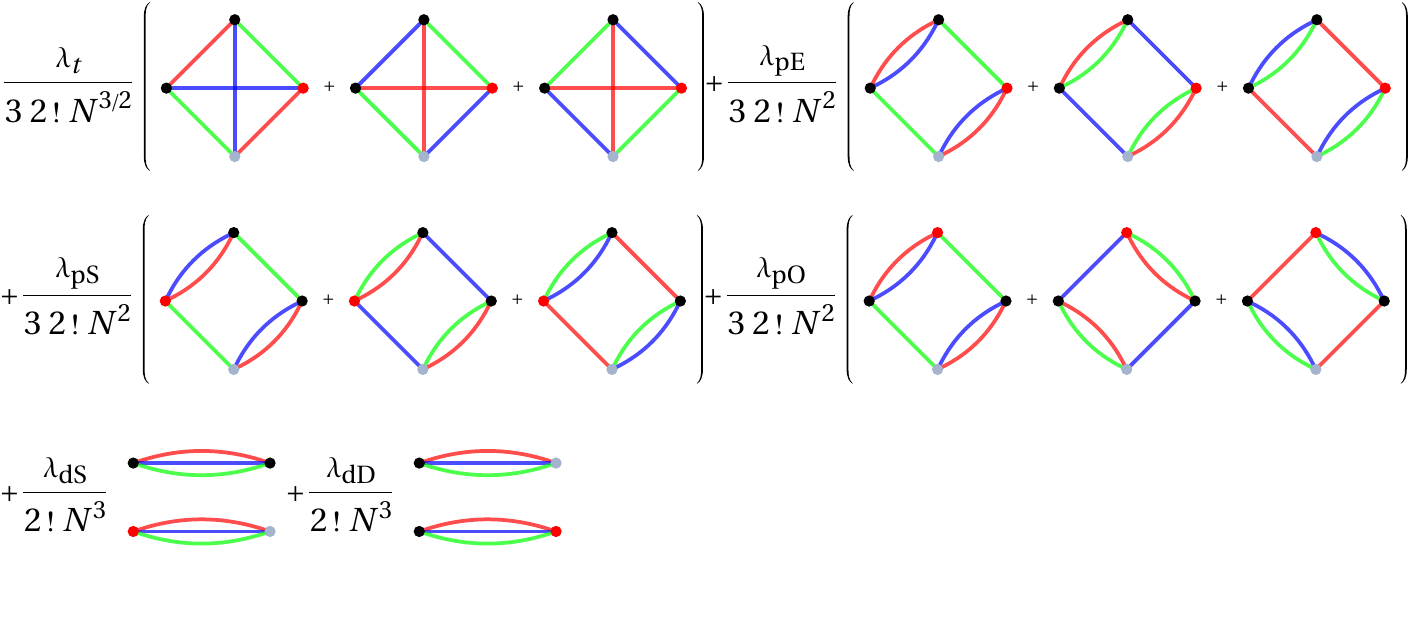}
\includegraphics[width=\textwidth]{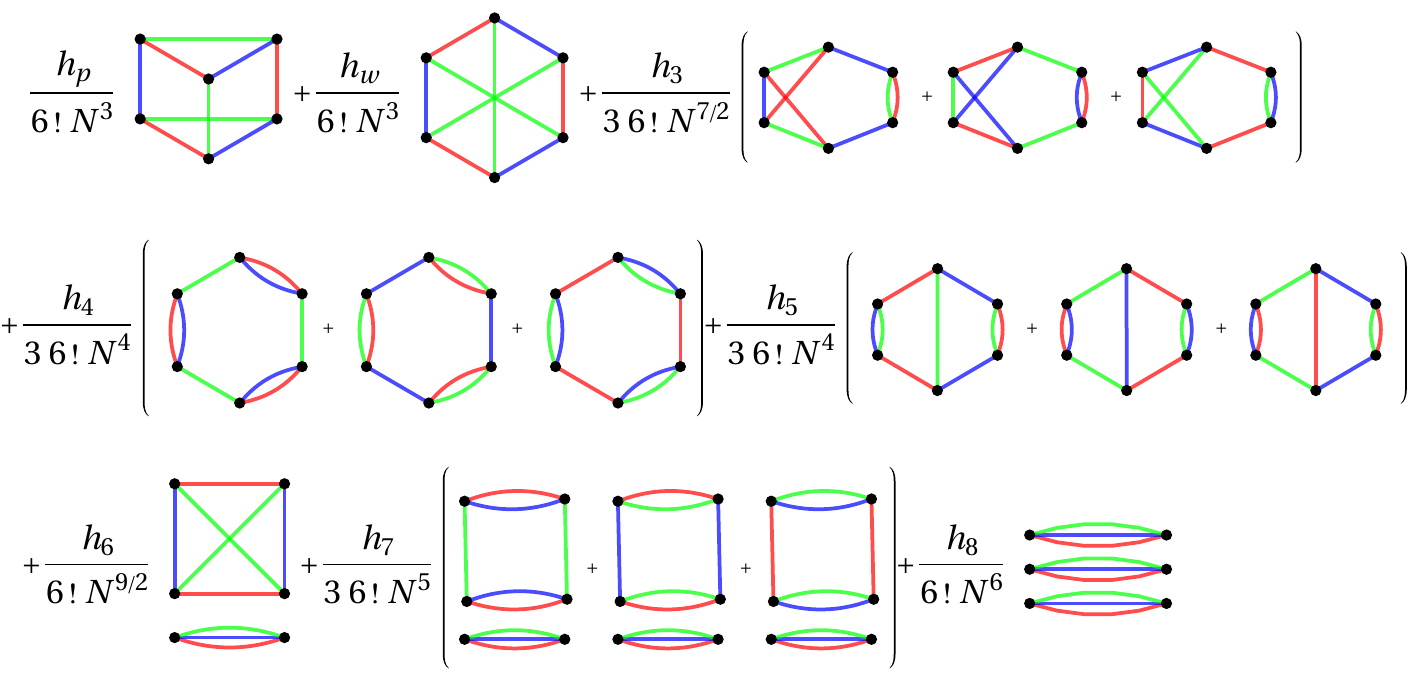}
\caption{The interaction terms $V_{\mathrm{int}}$ for the $\mathrm{O}(N)^3$ $\phi^2 \bar{\psi}\psi + \phi^6$ model. Black dots indicate $\phi_{abc}$s; red dots $\bar{\psi}_{abc}$s; grey dots $\psi_{abc}$s. Red, green, and blue lines indicate contractions between the first, second, and third index. Note the symmetry of this potential under the interchange of the three colours.} \label{fig:potentialWithH}
\end{figure}

The marginal sector of the potential for the tensorial version of the $\phi^2 \bar{\psi}\psi$ model that we will use is depicted graphically in \cref{fig:potentialWithH} (by marginal, we mean that the masses and $g_{IJKL}$ have been pre-emptively set to zero). This is the most general marginal potential in $D=3$ that satisfies an $\mathrm{O}(N)^3 \times S_3$ symmetry, where the $S_3$ corresponds to the permutation symmetry of each of the $\mathrm{O}(N)$ groups (i.e. the three colours). Without the $S_3$, for example, we would need a different coupling constant for each of the three 3-colourings of the $\lambda_t$ invariant \begin{tikzpicture}[baseline={([yshift=-.5ex]current bounding box.center)}]
\draw[color=black] (0.4,0) -- (0.4,0.4) (0,0.4) -- (0,0) (0.4,0.4) -- (0,0.4) (0,0) -- (0.4,0) (0,0) -- (0.4,0.4) (0,0.4) -- (0.4,0);
\fill[color=black] (0,0) circle (0.05) (0.4,0) circle (0.05); 
\fill[color=gray] (0.4,0.4) circle (0.05);
\fill[color=red] (0,0.4) circle (0.05);
\end{tikzpicture} visible in the figure.
\noindent A full analytic expression is given in \cref{app:potential}. In addition, these results can be used to reproduce the results of the complex sextic \cite{Benedetti:2019rja} and real prismatic \cite{Giombi:2018qgp} tensor models; see \cref{app:potentialcomparison} for details.

Applying the general formulae of \eqref{eq:generalVectorResults}, the field anomalous dimensions are 
\begin{subequations}
\begin{align}
    \gamma_\phi &=\frac{\lambda_t^2 s^2}{18}  T + \frac{s^4}{97200} \left(3 \hp^2+9 \hw^2-50 \lambda_t^4 T (22 T+20)\right) + O(\lambda_t^5,\ldots), \label{eq:anomDimPhiGeneralPert}\\
    \gamma_\psi &= \frac{\lambda_t^2 s^2}{18} - \frac{\lambda_t^4 s^4}{486} (1+8 T)+O(\lambda^5,\ldots), \label{eq:anomDimPsiGeneralPert}
\end{align}
\end{subequations}
with the field scaling dimensions at any fixed points being $\Delta_\phi= \frac{D-2}{2} + \gamma_\phi$, $\Delta_\psi= \frac{D-1}{2} + \gamma_\psi$. The only couplings that contribute to the two-point function of the fields, and therefore this anomalous dimension, in the large-$N$ limit, are $\{\lambda_t, \hp, \hw\}$; we term such couplings the \textit{dominant couplings}:
\begin{itemize}
    \item $\lambda_t$ is the tetrahedral interaction, dominant just as in the standard bosonic $\phi^4$ \cite{Giombi:2017dtl}.
    \item $\hp$ is the prismatic interaction, dominant just as in the prismatic model \cite{Giombi:2018qgp}.
    \item $\hw$ is the wheel graph (which is also the complete bipartite graph $K_{3,3}$, dominant just as in the bosonic $\phi^6$ model \cite{Benedetti:2019eyl}).
\end{itemize}
These three have beta functions that depend only on each other; if these three beta functions are zero, then the fixed point values of the remaining coupling constants are determined. Up to $O(\lambda_t^5,\dots)$, we have the dominant beta functions
\begin{equation}\begin{aligned} %
\beta \left(\lambda_t\right)&=-\epsilon  \lambda_t+\frac{1}{9}  (T+1) \lambda_t^3 s^2+ \frac{\lambda_t}{3}\left(\frac{\hp^2+3\hw^2}{5400} -\frac{\lambda_t^4}{162} (T (11 T+10) + 16 T +2) \right)s^4,\\
\beta \left(\hp\right)&=-2 \hp \epsilon + \hp  \left(\frac{1}{3} T \lambda_t^2 +\frac{\hp}{90} \right)s^2+ \hp\left(\frac{\hp^2+3 \hw^2}{5400}-\frac{\lambda_t^4}{162} T (11 T+10)\right)s^4,\\
\beta \left(\hw\right)&=-2 
\hw \epsilon +\hw\left(\frac{1}{3} T \lambda_t^2\right) s^2+ \hw\left(\frac{\hp^2+3 \hw^2}{5400}-\frac{ \lambda_t^4}{162} T (11 T+10)\right)s^4.%
\end{aligned}\end{equation}
The full set of 14 beta functions for the marginal coupling constants is for compactness given only in \cref{app:betas}\footnote{For the large-$N$ melonic theory, the momentum structure of the surviving diagrams means that it is simple enough to calculate all leading-$N$ corrections to four loops. We therefore can calculate to order $\sim h^3, \lambda^5$, rather than the $\sim h^3,\lambda^3$ general results given in \cref{sec:vectorBetaFunctions}. Thus, we indicate with $O(s^5)$ the order of 5-loop contributions, since when working in $D=3-\epsilon$, each loop order brings a factor of $s=1/(8\pi)$.}. 
We solve for the Wilson-Fisher-like fixed points of the flow ($\beta_i =0$) perturbatively in $\epsilon$ for $D =3-\epsilon$; we can trust the results for $\epsilon \ll 1$. %

The trivial and pure bosonic fixed points, where the fermions decouple (i.e. $\lambda_i=0$ exactly), are well known and correspond to the free theory, the sextic bosonic \cite{Benedetti:2019rja}, and the prismatic models \cite{Giombi:2018qgp}:
\begin{equation}\begin{aligned}\label{eq:bosonicFixedPoints}
\mathrm{trivial}:& \quad \mathrm{ all zero}\\
\hmelonic:&\quad  s^2 \hw = \pm 60s^2   \sqrt{\epsilon} \implies \gamma_\phi = \epsilon/3\\
\hprismatic:&\quad s^2  \hp = (180 \epsilon - 540 \epsilon^2) \implies \gamma_\phi = \epsilon^2 %
\end{aligned}\end{equation}
These formulae for the fixed points are perturbative solutions in $\epsilon$, truncated to the order shown. The fixed points with interacting fermions ($\lambda_i \neq 0$) are the following:
\begin{equation}\begin{aligned} \label{eq:fermionicFixedPoints}
\hlammelonic:&\, s \lambda_t  = \pm_1 \sqrt{3}\sqrt{\epsilon}, \,  s^2  \hw = \pm_2 30 i  \sqrt{2T-4} \sqrt{\epsilon}, \mathrm{ (independent signs)}\\
& \implies \gamma_\phi = \epsilon/3, \, \gamma_\psi = \epsilon/6\\
\lammelonic:& \, s \lambda_t = \pm \left(\frac{3 \sqrt{\epsilon}}{\sqrt{T+1}} + \frac{(11T^2 +26T +2) \epsilon^{3/2}}{4(T+1)^{5/2}}\right) \\
& \implies \gamma_\phi = \frac{\epsilon}{2} \frac{T}{T+1} + \frac{T(5T-8)}{12(T+1)^3} \epsilon^2, \quad \gamma_\psi = \frac{\epsilon}{2} \frac{1}{T+1}- \frac{T(5T-8)}{12(T+1)^3} \epsilon^2 \\
\hlamprismatic:& \, s \lambda_t = \pm \left(\frac{3 \sqrt{\epsilon}}{\sqrt{T+1}} + \frac{(4T^2 +19T -5) \epsilon^{3/2}}{2(T+1)^{5/2}}\right),\, s^2 \hp = -\frac{90(T-2)s^2}{T+1}\epsilon \\
& \implies \gamma_\phi = \frac{\epsilon}{2} \frac{T}{T+1} + \frac{(T-1)(2T-3)}{3(T+1)^3} \epsilon^2, \quad \gamma_\psi = \frac{\epsilon}{2} \frac{1}{T+1} - \frac{(T-1)(2T-3)}{12(T+1)^3} \epsilon^2 
\end{aligned}\end{equation}  %
We note that:
\begin{enumerate}
\item The particular names that we have given these fixed points will be justified during the SDE analysis in \cref{sec:SDEanalysis}, where we shall also re-derive these results to all orders in $\epsilon$. This will confirm that the anomalous dimensions in \hmelonic and \hlammelonic are exactly linear in $\epsilon$, and so have no higher order corrections. We re-iterate that the anomalous dimensions at fixed points are physical and thus must be scheme-independent. The values of the coupling constants, however, are not.
\item The precise order in $\epsilon$ that we can calculate to depends on the particular fixed point; for example, in $h\lambda_{\mathrm{melonic}}$, we are only able to calculate to $\sqrt{\epsilon}$ order, because the four-loop result for the $h$s only reaches order $h^3$. 
\item At the fermionic fixed points, $\gamma_\phi + \gamma_\psi=\epsilon/2$ to the order known. From \cref{sec:SDEanalysis}, we expect this to be true to all orders. This is a non-trivial check, as the implied cancellation of terms between \eqref{eq:anomDimPsiGeneralPert} and \eqref{eq:anomDimPhiGeneralPert} occurs only at the fixed point.

\end{enumerate}

\subsection{Values of the other coupling constants} \label{sec:othercouplingconstants}

Solving the remaining 11 beta functions requires non-zero values for some coupling constants. We summarise in the following table, where we give each coupling constant value at the fixed point to all known orders in $\epsilon=3-D$. All coupling constants not shown or blank are zero to the order calculated. First, the bosonic fixed points:
\begin{center}
$\begin{array}{l|l|l}
 \mathrm{} & \hmelonic & \hprismatic \\
 \hline
 \hp & \mathrm{} & 180 \epsilon -540 \epsilon ^2 \\
 \hw & \pm 60 \sqrt{\epsilon } & \mathrm{} \\
 h_4 & \mathrm{} & -90 \pi ^2 \epsilon ^2 \\
 h_5 & -270\pm (-540) \sqrt{\epsilon } & 2160 \epsilon ^2+1080 \epsilon  \\
 h_7 & \frac{2700}{7}  \pm 540 \sqrt{\epsilon }& 540 \epsilon -540 \left(\pi ^2-83\right) \epsilon ^2 \\
 h_8 &-\frac{1090}{7}  \pm (-180) \sqrt{\epsilon } & -180 \left(\pi ^2-126\right) \epsilon ^2 \\
\end{array}$
\end{center}
The fermionic fixed points: %
\begin{center}
$\begin{array}{l|l|l|l}
 \mathrm{} & \lammelonic & \hlammelonic & \hlamprismatic\\
 \hline
 \lambda_t & \pm 3 \sqrt{\frac{\epsilon }{1+T}} & \pm_1 \sqrt{3} \sqrt{\epsilon } & \pm 3 \sqrt{\frac{\epsilon }{1+T}} \\
 \hp & \mathrm{} & \mathrm{} & -\frac{90 (T-2) \epsilon }{T+1} \\
 \hw & \mathrm{} & \pm_2 30 i \sqrt{2} \sqrt{(T-2) \epsilon } & \mathrm{} \\
 h_5 & \mathrm{} & -\frac{405 (T-2)}{T-3} - 9 h_w & \frac{1080 (T-2) \epsilon }{(T+1) (3 T-2)} \\
 h_7 & \mathrm{} & \frac{2430 (T-2) (2 T-5)}{(T-3) (8 T-21)} + 9 h_w & -\frac{270 (T-2) (9 T+2) \epsilon }{(T+1) (3 T-2) (3 T-1)} \\
 h_8 & \mathrm{} & -\frac{5 (T-2) \left(808 T^2-3978 T+4905\right)}{(T-3) (2 T-5) (8 T-21)} - 3h_w & \frac{1620 (T-2) T (2 T+1) \epsilon }{(T+1) (3 T-2) (3 T-1) (5 T-1)} \\
\end{array}$
\end{center}
We now can make the following observations, recalling first that $T=2$ is a distinguished value, being the minimal dimension of the gamma matrices in $D=3$:
\begin{enumerate}
\item In \hlammelonic, $\pm_1$ can be chosen independently of $\pm_2$.
\item In \hmelonic, $h_{5,7,8}$ are not small, just as was found in \cite{Benedetti:2019rja}. The same is true for \hlammelonic, $h_{5,7,8}$, unless $T$ is taken near $2$. This might make us doubt the validity of the perturbative approach. However, there is still a possibility of the perturbative series being trustworthy if these large coupling constants always appear together with the dominant $\lambda_t, h_{1,2}$. In \cref{sec:SDEanalysis} we will identify these same fixed-points, but in a way that ignores the non-dominant coupling constants.
\item All of the fermionic fixed points seem (to this order) to reduce to the same fixed point for $T=2$. However, our non-perturbative analysis in \cref{sec:SDEanalysis} will show these fixed points to be distinct even at $T=2$. Therefore, this collision is an artefact of our inability to calculate to higher loop order. In particular, it is unrelated to supersymmetry.
\item If we were interested in calculating the $O(1/N)$ corrections here, we would have to take care with the way we take the large $N$ expansion. This is because beyond optimal scaling the large $N$ and small $\epsilon$ expansions do not commute \cite{Harribey:2021xgh, Jepsen:2023pzm}: we would need to first take $\epsilon N\to\infty$, and then $\epsilon\to 0$. %
\end{enumerate}

\subsection{Scaling dimensions of interaction terms} \label{sec:stabmats}

We can calculate the conformal dimensions of the interaction terms by calculating the eigenvalues of the stability matrix \eqref{eq:stabMat}. The stability matrices themselves are too large to reproduce, so we give only the eigenvalues here. Each of these eigenvalues equals $\Delta_\cO-D$, where $\Delta_\cO$ is the scaling dimension of one of the marginal $\mathrm{O}(N)^3$ singlet operators in the theory; these are linear combinations of the 14 operators appearing in the potential of \cref{fig:potentialWithH}. Where it is easy to do so, we also indicate the direction in the space of coupling constants that a given eigenvalue corresponds to (to leading order). 

First, we give the bosonic theories, where the $\lambda$s do not mix with the $h$s:
{\scriptsize
\begin{align*} %
\hmelonic &: h\mathrm{s}: \left\{h_8: 30 \epsilon ,14 \epsilon ,10 \epsilon , 6 \epsilon ,2 \epsilon ,0,\left(
\begin{array}{cc}
 4 \epsilon  & 1 \\
 0 & 4 \epsilon  \\
\end{array}
\right)\right\} + \lambda \mathrm{s}\mathrm{: } \left\{\frac{29 \epsilon }{3},\frac{5 \epsilon }{3},-\frac{\epsilon }{3},-\frac{\epsilon }{3},-\frac{\epsilon }{3},-\frac{\epsilon }{3}\right\}\\
\hprismatic &: h\mathrm{s}: \{6 \epsilon ,2 \epsilon ,2 \epsilon,-2 \epsilon ,-2 \epsilon ,-2 \epsilon ,-2 \epsilon ,-2 \epsilon \} + \lambda \mathrm{s}\mathrm{: } \left\{-\epsilon ,-\epsilon ,-\epsilon , -\epsilon , -\epsilon ,-\epsilon\right\}.
\end{align*}
}
Next, we give the theories with coupled fermions, where the two types do mix:
{\scriptsize
\begin{align*}
\lammelonic &: \Bigg\{h_8: \frac{2 (5 T-1) \epsilon }{T+1},\frac{2 (3 T-1) \epsilon }{T+1}, \lambda_{d_S}:  3 \epsilon ,\frac{2 (2 T-1) \epsilon }{T+1},\frac{2 (2 T-1) \epsilon }{T+1},2 \epsilon ,\frac{(3 T-2) \epsilon }{T+1},\frac{6 \epsilon }{T+1},\epsilon ,\\
& \quad \quad \quad \quad \quad \frac{2 (T-1) \epsilon }{T+1},\frac{2 \epsilon }{T+1},\frac{(T-2) \epsilon }{T+1},\frac{(T-2) \epsilon }{T+1},-\frac{2 \epsilon }{T+1}\Bigg\}\\
\hlammelonic &: \left\{2 \epsilon ,\frac{1}{3} \left(7-2 T+\sqrt{4 (T-2)^2 + 9}\right) \epsilon ,\frac{2 \epsilon }{3},\frac{1}{3} (7-2 T) \epsilon ,0,-2 (T-3) \epsilon ,-\frac{4}{3} (T-3) \epsilon ,-\frac{2}{3} (T-3) \epsilon ,-\frac{2 \epsilon }{3},\right. \\
& \quad \quad \quad \quad \quad \left. \frac{1}{3} \left(7-2 T-\sqrt{4 (T-2)^2 + 9}\right) \epsilon ,(11-4 T) \epsilon ,\frac{2}{3} (21-8 T) \epsilon ,2 (5-2 T) \epsilon ,6 (5-2 T) \epsilon \right\}\\
\hlamprismatic &: \bigl\{\frac{2 (5 T-1) \epsilon }{T+1},\frac{2 (3 T-1) \epsilon }{T+1},3 \epsilon ,\frac{2 (2 T-1) \epsilon }{T+1},2 \epsilon ,\frac{(3 T-2) \epsilon }{T+1},\frac{6 \epsilon }{T+1},\frac{6 \epsilon }{T+1},\epsilon ,\frac{2 \epsilon }{T+1},\\
& \quad \quad \quad \quad \quad \frac{2 \epsilon }{T+1},\frac{(T-2) \epsilon }{T+1},-\frac{(T-2) \epsilon }{T+1},-\frac{2 \epsilon }{T+1}\bigr\}.
\end{align*}
}

\noindent All the fixed points that differ only by $\pm_i$ have the same eigenvalues: this suggests that the seemingly distinct fixed points related by switching those signs describe the same CFT. We might have expected this from the diagrammatic expansion, where all diagrams are constructed out of melons, and hence the dominant couplings always appear squared.

Some of these operators are marginally irrelevant ($\Delta_{\cO}>D$), and some are marginally relevant ($\Delta_{\cO} < D$) -- in the case of the fermionic fixed points, which are which depends on $T$: therefore these CFTs are saddle points of RG flow.

\begin{enumerate}
\item Despite \hmelonic being the theory of a real scalar field, in the strict large-$N$ limit it is identical to the complex sextic bosonic model of \cite{Benedetti:2019rja}, as mentioned above. Indeed, the $\mathrm{O}(N)$ invariants that do not descend from $\mathrm{U}(N)$ invariants (see \cref{app:potentialcomparison}) are zero in this theory. The stability matrix is not diagonalisable in this case, but has Jordan normal form, showing it to be a logarithmic CFT\footnote{See, for example \cite{Cardy:2013rqg,Hogervorst:2016itc}. This also occurs in the strict large-$N$ limit of the Fishnet Conformal Field Theories (FCFTs) of \cite{Kazakov:2022dbd}.} at the fixed point, and so automatically non-unitary, to leading order in $N$. Working to higher orders in the coupling constants will not modify this. However, at subleading orders in $N$, this is lifted, due to new $h^2,h^3$ terms in the beta functions \cite{Harribey:2021xgh}. 
\item \hlammelonic has a diagonalisable stability matrix, unlike its bosonic cousin \hmelonic.
\item The zero eigenvalues in the case of both \hmelonic and \hlammelonic are in the same direction $-\hp + 3 h_5 - 3 h_7 + h_8$, indicating that these CFTs appear to sit on a line of fixed points. However, this free parameter would appear at an order higher than we were able to calculate, so this degeneracy could be lifted. This deformation would have broken the $\mathrm{U}(N)$ symmetry of the complex sextic model (see \cref{app:potentialcomparison}), and therefore was invisible to \cite{Benedetti:2019rja}.
\item None of the other theories are obviously non-unitary; indeed, for all values of $T$, even $T<0$, all of these anomalous dimensions for all fermionic fixed points are real; likewise for $\epsilon<0$. This is particularly notable in the case of \hlammelonic, which has complex coupling constants. This observation agrees with what is seen in the bilinear analysis of \lammelonic in \cref{sec:bilinears}, where at least perturbatively close to $D=3-\epsilon$, the bilinear spectrum is found to be fully real.
\end{enumerate}

\subsection{The flow system for the three dominant coupling constants}

\begin{figure}
    \centering
    \includegraphics[width=1.0\textwidth]{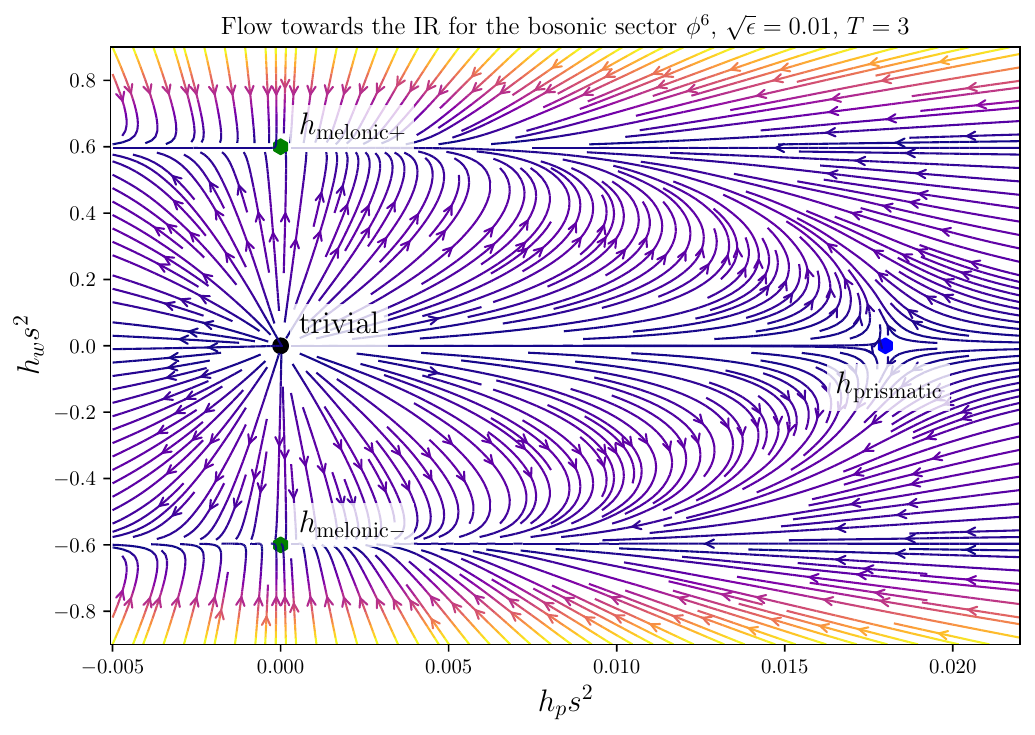}
    \caption{Flow towards the IR for the bosonic sector of the theory -- when the fermions are decoupled. We expect the two \hmelonic fixed points to represent the same CFT, as they only differ by the sign of the coupling constant. All non-trivial fixed points shown here are saddle points.}
    \label{fig:flowSystemForBosonicSector}
\end{figure}

\begin{figure}
    \includegraphics[width=1.0\textwidth]{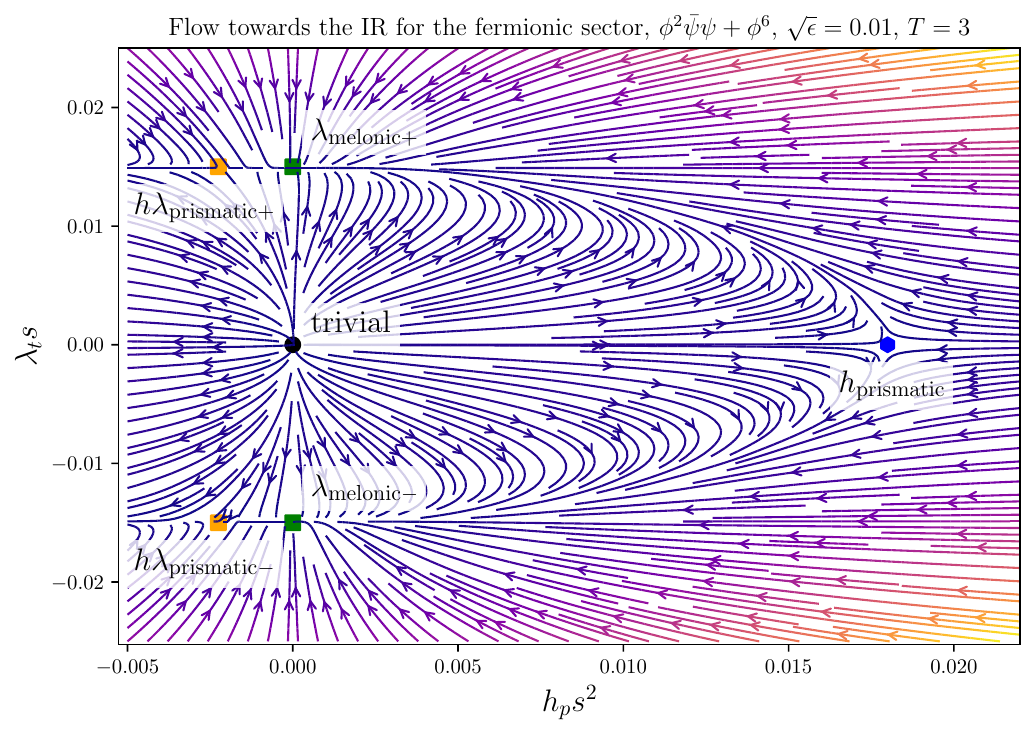}
    \caption{Flow towards the IR for the fermionic sector of the theory; this plot shares its $x$-axis with \cref{fig:flowSystemForBosonicSector}, and also displays both the trivial fixed point and \lammelonic. We once again expect the two \lammelonic fixed points to refer to the same CFT; likewise for \hlamprismatic. All non-trivial fixed points shown here are saddle points.}
    \label{fig:flowSystemForFermionicSector}
\end{figure}

To understand the relationship between these CFTs in the space of coupling constants, it is convenient to reduce to the dominant coupling constants $(\lambda_t, \hp, \hw)$, and consider their flow under the renormalisation group towards the IR. For concreteness, in the following plots of the flow we take $\sqrt{\epsilon} =1/100$. In \cref{fig:flowSystemForBosonicSector} we show the bosonic sector: the flow between the existing sextic bosonic models in the literature, \hmelonic \cite{Benedetti:2019rja,Harribey:2021xgh} and \hprismatic \cite{Giombi:2018qgp}. In \cref{fig:flowSystemForFermionicSector}, we show the sector with coupled fermions with $T=2$, which avoids the collision of the perturbative coupling constants visible in \cref{eq:fermionicFixedPoints} at $T=2$. 
As shown by the signs of the eigenvalues, all the non-trivial fixed points here are saddle points of RG flow; IR-stable to some deformations, but IR-unstable to others. The fate of the fixed points perturbed in the unstable directions is unknown.

We are unable to illustrate the location of \hlammelonic on this plot, due to the imaginary nature of $\hw$ at that fixed point. However, if we restrict to the subspace $\hp=0$, we can write down the beta functions for the squared coupling constants $H_w=\hw^2$ and $\Lambda_t=\lambda_t^2$. 
\begin{equation}\begin{aligned}
\beta(\Lambda_t)|_{\hp=0}&= -2\epsilon  \Lambda _t +\frac{2}{9}  (T+1) \Lambda_t^2 s^2 + \frac{\Lambda_t}{3} \left(\frac{H_w}{900}-\frac{\Lambda _t^2}{81} (T (11 T+10) + 16 T+2) \right)s^4+O\left(\Lambda_t^4,\ldots\right)\\
\beta(H_w)|_{\hp=0}&=-4 \epsilon H_w + \frac{2}{3} T H_w   \Lambda_t s^2+ \left(\frac{H_w^2}{900}-\frac{\Lambda _t^2}{81} H_w T (11 T+10) \right)s^4+O\left(\Lambda_t^4,\ldots\right)
\end{aligned}\end{equation}
The fixed points here occur for real values of these squared coupling constants -- albeit negative in the case of $\hlammelonic$:
\begin{equation}\begin{aligned}
\hmelonic &: \, s^4 H_w= 3600\epsilon + O(\epsilon^2),\\
\lammelonic &: \, s^2 \Lambda_t =  \frac{9 \epsilon }{T+1}+\frac{3 (T (11 T+26)+2) \epsilon ^2}{2 (T+1)^3}+O\left(\epsilon ^3\right)\\
\hlammelonic &: \,  s^4 H_w = -1800(T-2)\epsilon + O(\epsilon^2), s^2 \Lambda_t = 3\epsilon + O(\epsilon^2)
\end{aligned}\end{equation}
The reason that we could not do this for $\hp$ is due to the quadratic term appearing in the beta function; this is because $\hp$ is only non-zero at the fixed points of prismatic-type theories, where it is not true that the coupling constants only appears squared. In such theories, the actually melonic coupling is $\rho X \phi^3$, so only $\rho^2$ appears. However, since $\rho^2 \sim \hp$, an $\hp^2$ term appears in the beta function.

\begin{figure}[H]
    \centering
    \includegraphics[width=1.0\textwidth]{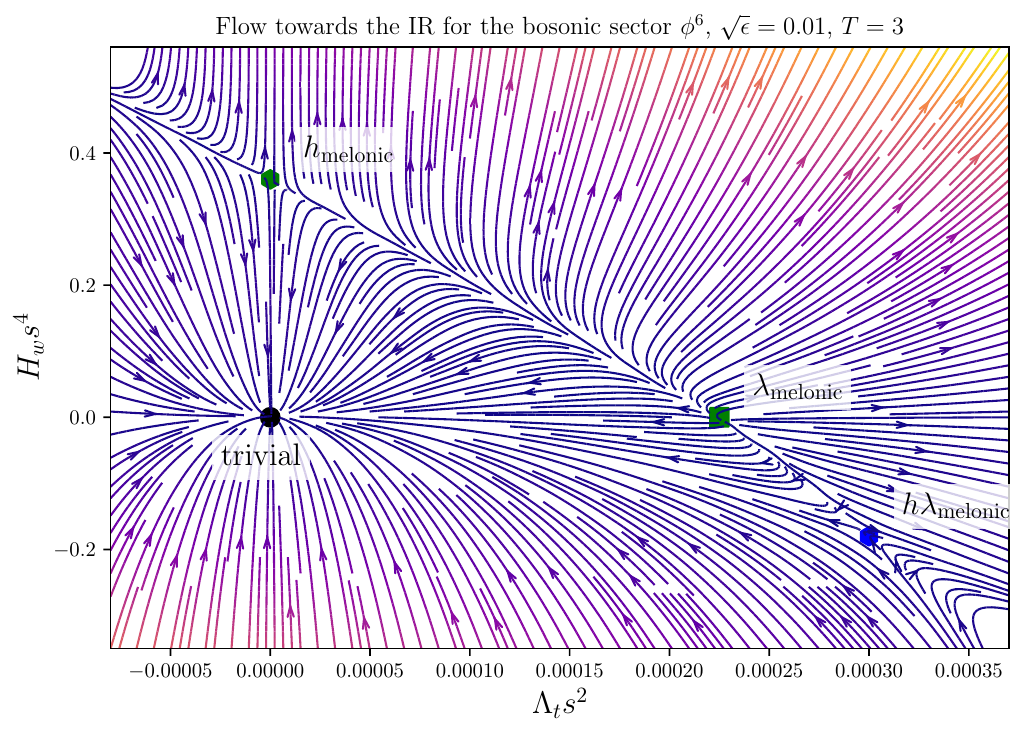}
    \caption{Flow towards the IR for the melonic-type theories, where we visualise the flow of the value of the coupling constants squared, to avoid the imaginary nature of $h_w$ for the \hlammelonic theory.}
    \label{fig:flowSystemForSquaredCCs}
\end{figure}

Tuning $T$ moves both $\lammelonic$ and $h\lammelonic$, with an apparent collision occurring for $T \to 2$; a collision that, as mentioned above, we will discover to be an artifact of perturbation theory in \cref{sec:hlamprismaticSDEfound}. Increasing $T$ seems to move \lammelonic{} along the line separating \hmelonic and \hlammelonic, towards the latter.

In the next section we look at these same fixed points from a complementary point of view, via the Schwinger-Dyson equations at large-$N$, which can calculate the scaling dimensions of the fundamental fields and singlet bilinears to all orders. However, the price we pay is having to sit precisely at the fixed points, therefore losing any information about the RG flow, or the values of the non-dominant coupling constants.

\section{Schwinger-Dyson equation analysis} \label{sec:SDEanalysis}

In this section, we use the Schwinger-Dyson equations (SDEs) to investigate the fixed points, assuming that the scaling symmetry at the fixed points is promoted to full conformal symmetry. This enhancement of symmetry is common, but not guaranteed. Various assumptions are required for a proof of this, including unitarity and locality \cite{Nakayama:2013is}, and that the conformal limit is taken before the large-$N$ limit \cite{Benedetti:2020yvb}.

\begin{figure}[H]
\centering
\begin{tikzpicture}[node distance=3cm, auto]
    \node (theory) [rectangle, draw, text width=4cm, text centered] {$\mathrm{O}(N)^3$ QFT};
    \node (prismatic) [rectangle, draw, text width=4cm, text centered, below right of=theory, xshift=2cm, yshift=0cm] {Prismatic QFT};
    \node (melonic) [rectangle, draw, text width=4cm, text centered, below left of=theory, xshift=-2cm, yshift=0cm] {Melonic QFT};
    \node (lamMelonic1) [rectangle, draw, text width=3cm, text centered, below left of=melonic, yshift=0cm] {Precursor finite-$N$ CFT};
    \node (hlamMelonic1) [rectangle, draw, text width=3cm, text centered, below right of=melonic, yshift=0cm] {Precursor finite-$N$ CFT};
    \node (hlprismatic1) [rectangle, draw, text width=6cm, text centered, below of=prismatic, yshift=0cm] {Precursor finite-$N$ CFT};
    \node (lamMelonic2) [rectangle, draw, text width=3cm, text centered, below of=lamMelonic1, yshift=0cm] {\lammelonic CFT \S\ref{sec:lammelonicSDEfound}\\CFT data \S\ref{sec:bilinears}};
    \node (hlamMelonic2) [rectangle, draw, text width=3cm, text centered, below of= hlamMelonic1, yshift=0cm] {\hlammelonic CFT, \S\ref{sec:lammelonicSDEfound}};
    \node (hlprismatic2) [rectangle, draw, text width=6cm, text centered, below of=hlprismatic1, yshift=0cm] {\hlamprismatic  CFT \S\ref{sec:hlamprismaticSDEfound}};
    \node (hlSUSYic) [rectangle, draw, text width=6cm, text centered, below of=hlprismatic2, yshift=0cm] {Supersymmetric \hlamprismatic\\(balanced DOFs) \S\ref{sec:qGenAndSUSYprismatic}};

    \path (theory) edge node {Add auxiliary field $X_{abc}$} (prismatic);
    \path (theory) edge node[above left] {No auxiliary field} (melonic);
    \path (melonic) edge[red, ->, thick] node {IR, $h=0$} (lamMelonic1);
    \path (melonic) edge[red,->, thick] node {IR, $h\neq 0$} (hlamMelonic1);
    \path (prismatic) edge[red, ->,thick] node {IR} (hlprismatic1);
    \path (lamMelonic1) edge node {$N\to\infty$} (lamMelonic2);
    \path (hlamMelonic1) edge node {$N\to\infty$} (hlamMelonic2);
    \path (hlprismatic1) edge node {$N\to\infty$} (hlprismatic2);
    \path (prismatic) edge[red, ->,thick] node {IR} (hlprismatic1);
    \path (hlprismatic2) edge node {Tune $r=2$ ($T=1$)} (hlSUSYic);
\end{tikzpicture}
\caption{Map of interacting fixed points in the theory in generic dimension (that is, ignoring the theories with free fermions, \hmelonic{} and \hprismatic{}). The CFT we obtain after the IR limit is determined by the initial coupling constants. We also ignore all symmetry breaking possibilities.} \label{fig:FPstructure} %
\end{figure}
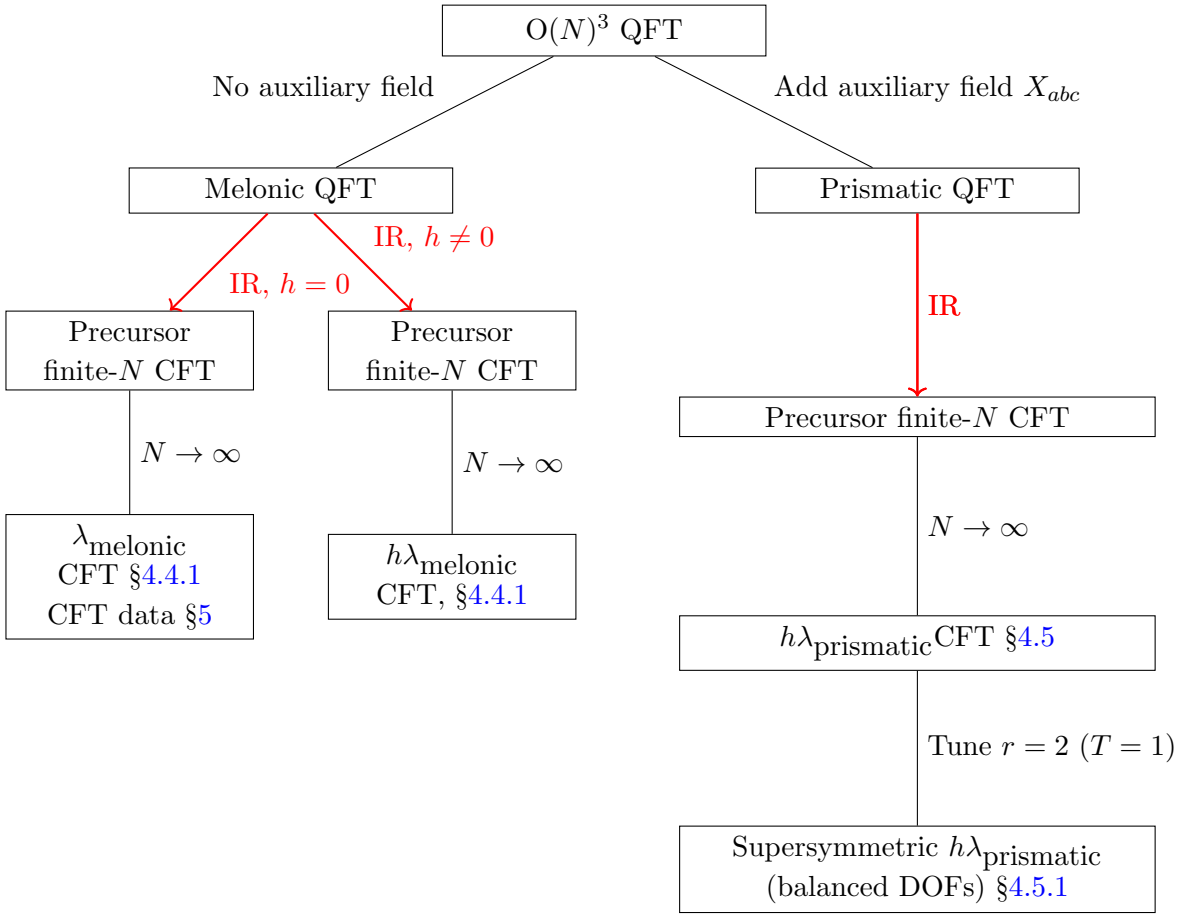

A similar analysis to what will follow is provided for the short-range and long-range $\mathrm{O}(N)^3$ bosonic tensor models in the series of papers \cite{Benedetti:2019eyl, Benedetti:2019ikb, Benedetti:2020yvb}. However, in the present case we will observe a more complex structure of fixed points, of which we give a cartoon in \cref{fig:FPstructure}. These exact solutions will match precisely the Wilson-Fisher-like fixed points found perturbatively in $D=3-\epsilon$ in \cref{sec:betas}. 

By ignoring the auxiliary field in the SDEs, we are assuming that of the dominant couplings, $h_w \neq 0$ and $h_p =0$. On the other hand, with the auxiliary field, we allow for $h_p\neq 0$: this gives us access to the prismatic-type fixed points via the SDEs.

\begin{figure}[H]
  \begin{align*}%
\vcenter{\hbox{\begin{tikzpicture}
  \begin{feynman}[every blob={/tikz/fill=gray!30,/tikz/inner sep=2pt}]
    \vertex[small, blob] (m) at (0,0) {};
    \vertex (a) at (-1,0) ;
    \vertex (b) at ( 1,0);
    \diagram* {
      (a) --[scalar] (m) --[scalar] (b),
      };
  \end{feynman}
\end{tikzpicture}}}
\quad &= \quad \vcenter{\hbox{\begin{tikzpicture}
  \begin{feynman}
    \vertex (a) at (-1,0) ;
    \vertex (b) at ( 1,0);
    \diagram* {
      (a) --[scalar] (b),
      };
  \end{feynman}
\end{tikzpicture}}} \quad+ \quad\vcenter{\hbox{\begin{tikzpicture}
    \begin{feynman}[every blob={/tikz/fill=gray!30,/tikz/inner sep=2pt,/tikzfeynman/small}, every edge=scalar]
   \vertex[small,blob] (m) at (0,0) {};
    \vertex (a) at (-1,0) ;
    \vertex (start) at ( 1,0);
    \vertex[small, blob] (centreblob0) at ( 2,1.2) {};
    \vertex[small, blob] (centreblob1) at ( 2,0.6) {};
    \vertex[small, blob] (centreblob2) at ( 2,0) {};
    \vertex[small, blob] (centreblob3) at ( 2,-0.6) {};
    \vertex[small, blob] (centreblob4) at ( 2,-1.2) {};
    \vertex (end) at ( 3,0);
    \vertex (e) at ( 4,0);
    \diagram* {
      (a) -- (m) -- (start) -- (centreblob2) -- (end) -- (e),
      (start) --[quarter left, out=30] (centreblob0) --[quarter left,in=150] (end),
      (start) --[quarter left, out=30] (centreblob1) --[quarter left,in=150] (end),
      (start) --[quarter right, out=-30] (centreblob3) --[quarter right,in=210] (end),
      (start) --[quarter right, out=-30] (centreblob4) --[quarter right,in=210] (end)
      };
    \end{feynman}
  \end{tikzpicture}}}\\ %
  &+\quad \vcenter{\hbox{\begin{tikzpicture}
    \begin{feynman}[every blob={/tikz/fill=gray!30,/tikz/inner sep=2pt,/tikzfeynman/small}]
   \vertex[small,blob] (m) at (0,0) {};
    \vertex (a) at (-1,0) ;
    \vertex (start) at ( 1,0);
    \vertex[small, blob,black] (centreblob1) at ( 2,0.8) {};
    \vertex[small, blob] (centreblob2) at ( 2,0) {};
    \vertex[small, blob,black] (centreblob3) at ( 2,-0.8) {};
    \vertex (end) at ( 3,0);
    \vertex (e) at ( 4,0);
    \diagram* {
      {[edges={scalar}] (a) -- (m) -- (start) -- (centreblob2) -- (end) -- (e)},
      (start) --[fermion, quarter left, out=30] (centreblob1) --[fermion, quarter left,in=150] (end),
      (start) --[anti fermion, quarter right, out=-30] (centreblob3) --[anti fermion,quarter right,in=210] (end)
      };
    \end{feynman}
  \end{tikzpicture}}} \quad +\quad \vcenter{\hbox{\begin{tikzpicture}
    \begin{feynman}[every blob={/tikz/fill=gray!30,/tikz/inner sep=2pt,/tikzfeynman/small}, every edge=scalar]
   \vertex[small,blob] (m) at (0,0) {};
    \vertex (a) at (-1,0) ;
    \vertex (start) at ( 1,0);
    \vertex[small, blob] (centreblob1) at ( 2,0.8) {};
    \vertex[small, blob] (centreblob2) at ( 2,0) {};
    \vertex[small, blob] (centreblob3) at ( 2,-0.8) {};
    \vertex (end) at ( 3,0);
    \vertex (e) at ( 4,0);
    \diagram* {
      (a) -- (m) -- (start) -- (centreblob2) -- (end) -- (e),
      (start) --[quarter left, out=30] (centreblob1) --[quarter left,in=150] (end),
      (start) --[quarter right, out=-30] (centreblob3) --[quarter right,in=210] (end),
      };
    \end{feynman}
  \end{tikzpicture}}}\\
\vcenter{\hbox{\begin{tikzpicture}
  \begin{feynman}[every blob={/tikz/fill=gray!30,/tikz/inner sep=2pt}]
    \vertex[small, blob,black] (m) at (0,0) {};
    \vertex (a) at (-1,0) ;
    \vertex (b) at ( 1,0);
    \diagram* {
      (a) --[fermion] (m) --[fermion] (b),
      };
  \end{feynman}
\end{tikzpicture}}}
\quad &= \quad \vcenter{\hbox{\begin{tikzpicture}
  \begin{feynman}
    \vertex (a) at (-1,0) ;
    \vertex (b) at ( 1,0);
    \diagram* {
      (a) --[fermion] (b),
      };
  \end{feynman}
\end{tikzpicture}}} \quad+ \quad
\vcenter{\hbox{\begin{tikzpicture}
    \begin{feynman}[every blob={/tikz/fill=gray!30,/tikz/inner sep=2pt,/tikzfeynman/small}]
   \vertex[small,blob,black] (m) at (0,0) {};
    \vertex (a) at (-1,0) ;
    \vertex (start) at ( 1,0);
    \vertex[small, blob] (centreblob1) at ( 2,0.8) {};
    \vertex[small, blob,black] (centreblob2) at ( 2,0) {};
    \vertex[small, blob] (centreblob3) at ( 2,-0.8) {};
    \vertex (end) at ( 3,0);
    \vertex (e) at ( 4,0);
    \diagram* {
      {[edges={fermion}] (a) -- (m) -- (start) -- (centreblob2) -- (end) -- (e)},
      (start) --[scalar, quarter left, out=30] (centreblob1) --[scalar, quarter left,in=150] (end),
      (start) --[scalar, quarter right, out=-30] (centreblob3) --[scalar,quarter right,in=210] (end)
      };
    \end{feynman}
  \end{tikzpicture}}}
  \end{align*}
\caption{Graphical SDE for the two-point functions in $\phi ^2 \bar{\psi}\psi$ theory, the black blobs denote full fermion propagators $F(p)$, and the grey blobs denote full boson propagators $B(p)$. We omit diagrams that vanish in dimensional regularization, as the fields are massless.}
\label{on3-2pt-sde-3dyuk}
\end{figure}
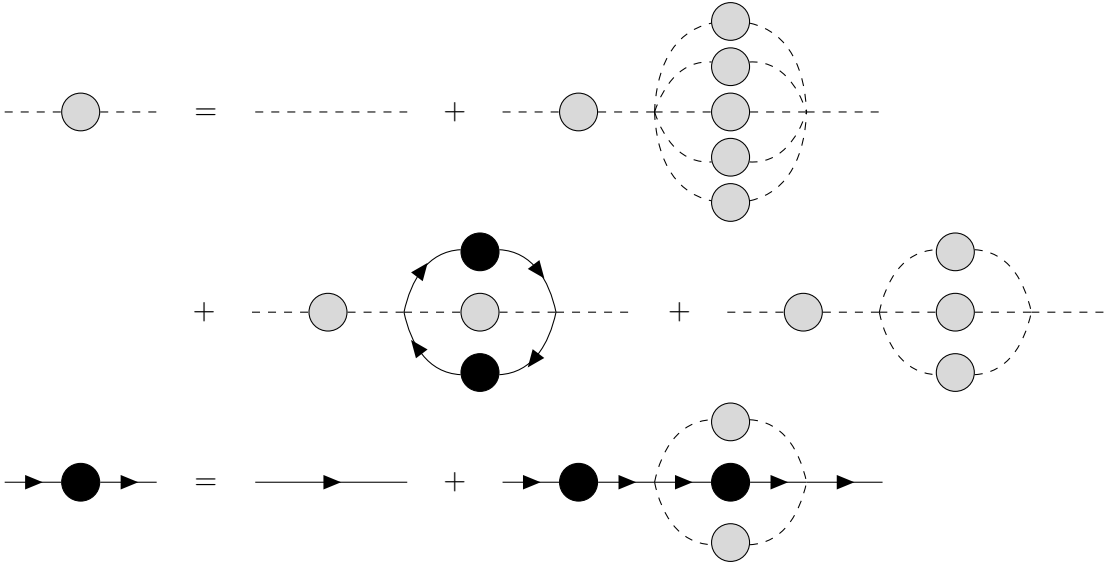

In the strict large-$N$ limit, melonic dominance enables the complete resummation of the SDEs to all orders in the coupling constants: indeed, for the purposes of this section, we could now forget about the tensorial origin of this theory, and imagine it as coming from an SYK-like theory with disorder, since we do not consider subleading corrections. Then we obtain the SDEs shown graphically in \cref{on3-2pt-sde-3dyuk}. 
\noindent To each vertex we assign a generic renormalised coupling to represent some combination of $\mathrm{O}(N)^3$-invariant coupling constants: $\lambda$ to the boson-fermion vertex, $h$ to the boson 6-point vertex, and $g$ to the boson 4-point vertex. For example, $\lambda^2 =\frac{\lambda_t^2}{6}$, though their precise forms will not matter.
Denoting by $F(p)$, $F_0(p)$, the full and bare fermion propagators, and by $B(p)$, $B_0(p)$, the full and bare boson propagators, we can divide through by $B_0(p)B(p)$ or $F_0(p)F(p)$ to obtain the Euclidean space SDEs

\begin{subequations}\label{eq:SDEsWithoutX}
\begin{align}
B_0(p)^{-1} & = B(p)^{-1} + \Sigma_B(p) \\
\Sigma_{B}(p) &= \frac{h^2}{5!} \int_{k,l,m,n} B(k+l+m+n+p)B(k)B(l)B(m)B(n) \notag\\
&+ \frac{\lambda^2}{1} \int_{k,l} (-1)\Tr\left[B(k+l+p)F(l)F(k)\right]\\
&+ \frac{g^2}{3!} \int_{k,l} B(k+l+p)B(l)B(k)\\ %
F_0(p)^{-1} & = F(p)^{-1} + \Sigma_F(p)  = F(p)^{-1} + \frac{\lambda^2}{1}  \int_{k,l} F(k+l+p)B(l)B(k) \label{eq:fermionSDEWithoutX}
\end{align}
\end{subequations}
By assuming the conformal form $\sim 1/\abs{x}^{2\Delta}$ for the full two-point functions, and taking an IR limit, where the free propagators are negligible, we exactly determine the scaling dimensions of the fundamental fields $\phi$ and $\psi$. We will find that all of our results depends only on the ratio of bosonic to fermionic degrees of freedom $r$. 
We note parenthetically that in the low energy/strong coupling limit (where we ignore the free propagator), the truncated SDEs have a large set of local symmetries, as described in \cite{Choudhury:2017tax,Benedetti:2018goh,Chang:2018sve}.

\subsection{Auxiliary field theory}

The motivation to introduce the auxiliary scalar $X_{abc}$ comes from the existence of perturbative solutions with $h\sim \epsilon$: \hprismatic and \hlamprismatic. These do not appear as solutions of \eqref{eq:SDEsWithoutX}; so, by comparison with the componentwise Schwinger-Dyson equations of the supersymmetric tensor model \cite{Popov:2019nja}, we conclude that the auxiliary field is necessary to obtain the SDEs for the prismatic-type QFTs.

This is simply another quadratic tensor interaction, and therefore does not modify the combinatorics of the large-$N$ limit: it simply adds a new SDE for $X_{abc}$ and an extra term for the $\phi$ SDE:
\begin{figure}[H]
  \begin{align*}%
\vcenter{\hbox{\begin{tikzpicture}
  \begin{feynman}[every blob={/tikz/pattern color=black,/tikz/inner sep=2pt}]
    \vertex[small, blob,pattern=north east lines,pattern color=black] (m) at (0,0) {};
    \vertex (a) at (-1,0) ;
    \vertex (b) at ( 1,0);
    \diagram* {
      (a) --[ghost] (m) --[ghost] (b),
      };
  \end{feynman}
\end{tikzpicture}}}
\quad &= \quad \vcenter{\hbox{\begin{tikzpicture}
  \begin{feynman}
    \vertex (a) at (-1,0) ;
    \vertex (b) at ( 1,0);
    \diagram* {
      (a) --[ghost] (b),
      };
  \end{feynman}
\end{tikzpicture}}} \quad+ \vcenter{\hbox{\begin{tikzpicture}
    \begin{feynman}[every blob={/tikz/fill=gray!30,/tikz/inner sep=2pt,/tikzfeynman/small}]
   \vertex[small,blob,pattern=north east lines] (m) at (0,0) {};
    \vertex (a) at (-1,0) ;
    \vertex (start) at ( 1,0);
    \vertex[small, blob] (centreblob1) at ( 2,0.8) {};
    \vertex[small, blob] (centreblob2) at ( 2,0) {};
    \vertex[small, blob] (centreblob3) at ( 2,-0.8) {};
    \vertex (end) at ( 3,0);
    \vertex (e) at ( 4,0);
    \diagram* {
      (a) --[ghost] (m) --[ghost] (start) --[scalar] (centreblob2) --[scalar] (end) --[ghost] (e),
      {[edges={scalar}]
      (start) --[quarter left, out=30] (centreblob1) --[quarter left,in=150] (end)},
      {[edges={scalar}]
      (start) --[quarter right, out=-30] (centreblob3) --[quarter right,in=210] (end)},
      };
    \end{feynman}
  \end{tikzpicture}}}\\
\Sigma_{B,\mathrm{aux}}(p) &= \Sigma_B(p) + \quad \vcenter{\hbox{\begin{tikzpicture}
    \begin{feynman}[every blob={/tikz/fill=gray!30,/tikz/inner sep=2pt,/tikzfeynman/small}]
   \vertex[small,blob] (m) at (0,0) {};
    \vertex (a) at (-1,0) ;
    \vertex (start) at ( 1,0);
    \vertex[small, blob] (centreblob1) at ( 2,0.8) {};
    \vertex[small, blob,pattern=north east lines, pattern color=black] (centreblob2) at ( 2,0) {};
    \vertex[small, blob] (centreblob3) at ( 2,-0.8) {};
    \vertex (end) at ( 3,0);
    \vertex (e) at ( 4,0);
    \diagram* {
      (a) --[scalar] (m) --[scalar] (start) --[ghost] (centreblob2) --[ghost] (end) --[scalar] (e),
      {[edges={scalar}]
      (start) --[quarter left, out=30] (centreblob1) --[quarter left,in=150] (end)},
      {[edges={scalar}]
      (start) --[quarter right, out=-30] (centreblob3) --[quarter right,in=210] (end)},
      };
    \end{feynman}
  \end{tikzpicture}}}
  \end{align*}
\label{auxSDEsAdded}
\end{figure}
\begin{equation}\begin{aligned}\label{eq:auxSDEsAdded}
A_0(p)^{-1} &= A(p)^{-1} + \Sigma_A(p)=A(p)^{-1} + \frac{\rho^2}{3!} \int_{k,l} B(k+l+p) B(l)B(k)\\
\Sigma_{B,\mathrm{aux}}(p) &= \Sigma_B(p) + \frac{\rho^2}{2!} \int_{k,l} A(k+l+p) B(l)B(k)
\end{aligned}\end{equation}
\noindent $A(p),A_0(p)$ denote the full and bare auxiliary field propagators, $\rho^2$ here stands for some quadratic combination of the $\mathrm{O}(N)^3$ invariant coupling constants that make up the 12-index coupling constant $\rho_{I(JKL)}$.

\subsection{Momentum scaling analysis of SDEs without auxiliary field}
 Assuming, as discussed above, that the fixed point possesses full conformal symmetry, and so is a CFT, we make the ansatz  
\begin{align}
B(p) &= \frac{B}{(p^2)^b}, \qquad  F(p) = \frac{F \not{p}}{(p^2)^{f + \half}} %
\end{align}
for the momentum space two-point functions. The bare two-point functions are
\begin{align}
 B_0(p) = \frac{B_0}{p^2 + m_0^2}\,,\qquad
F_0(p) = \frac{F_0}{\not{p} +M_0}\,.
\end{align}
$M_0$ plays no role in the subsequent analysis (due to parity symmetry) and can consistently be set to zero. The SDEs then take the form
\begin{align}
    (p^2+m_0^2) BB_0^{-1} &= p^{2b}+\lambda^2F^{-2}c_1\,p^{2(D-b-2f)}+h^2B^{-4}c_2\, p^{2(2D-5b)}+g^2B^{-2}c_3\,p^{2(D-3b)}\,,\label{eqn:SDEsubsB}\\
    (p^2)^\half FF_0^{-1}&= p^{2f}+\lambda^2B^{-2}c_4\,p^{2(D-2b-f)}\,,\label{eqn:SDEsubsF}
\end{align}
where $c_i$, $i=1\ldots 4$ are computable coefficients (see \cref{app:loopIntegrals}, or in more generality \cite{Fraser-Taliente:2024prep}). From demanding non-trivial IR scaling ($f<\frac{1}{2}$ and/or $b<1$), where we drop the free propagators, we conclude that:
\begin{enumerate}
    \item \label{SDEhmelonic}$\lambda =0$ gives the real sextic bosonic tensor model plus a free tensor fermion; we identify this theory as \hmelonic \cite{Benedetti:2019rja}:
\begin{equation}\label{eq:stdBosonicPhi6Sol}
    \gamma_\phi = \frac{\epsilon}{3}, \quad \gamma_\psi = 0, \quad \Leftrightarrow \quad \Delta_\phi = \frac{D}{6}, \quad \Delta_\psi = \frac{D-1}{2}
\end{equation}
This requires $D<3$ for validity of the IR solution. If $D>3$, we have a UV solution that breaks the unitarity bound for a scalar.
\item $\lambda\ne 0$ implies, from \eqref{eqn:SDEsubsF}, that  \begin{align} 2b+2f = D\,. \label{eqn:fbconstraint}
\end{align}
Consequently $g$ can be ignored and set to zero. To see this, note that if $g$ is relevant then, from \eqref{eqn:SDEsubsB},  $b=D/4$ and hence $f=D/4$ which is not a consistent IR solution if $D>2$. There are then two possibilities: \begin{enumerate}
    \item \label{SDElmelonic}$\lambda\ne 0$, $h=0$, gives solutions satisfying \eqref{eqn:fbconstraint} if and only if $D<3$; solving for the coefficients will then pick out a particular solution.
    \item \label{SDEhlmelonic}$\lambda\ne 0$, $h\ne 0$ implies from \eqref{eqn:SDEsubsB} that  $b \le D/3$ and therefore from  \eqref{eqn:fbconstraint} that $f \ge D/6$, so these solutions exist only if $D\le 3$. Note that the  $b=D/3$, $f=D/6$ solution has qualitatively different behaviour from the others as there are three terms of the same order on the r.h.s. of \eqref{eqn:SDEsubsB}. %
\end{enumerate}
\end{enumerate}
Note that unitarity has not entered into these considerations; we have only demanded consistent IR scaling of the equations.

As a warm-up, we now locate the bosonic fixed points that we found perturbatively earlier (in \eqref{eq:bosonicFixedPoints}): both of these are known, and correspond to the sextic bosonic tensor model and the prismatic tensor model. For convenience, we now switch to scaling dimensions instead of the momentum space powers $b,f$: $b=1-\gamma_\phi$ and $f=1/2 - \gamma_\psi$; $\Delta_\phi = \frac{D-2}{2} + \gamma_\phi$ and $\Delta_\psi = \frac{D-1}{2} + \gamma_\psi$. 

\subsection{Bosonic fixed points; or, \texorpdfstring{\hprismatic}{h-prismatic} and an introduction to the general characteristics of a multi-field melonic CFT} \label{sec:generalCharacteristicsPrismatic}

The first bosonic fixed point that we found perturbatively, $h_{\text{melonic}}$, with $\gamma_\phi= \epsilon/3$, is well understood as the \textit{sextic melonic CFT}\footnote{See \cite{Benedetti:2019rja} for a complex version, which is identical up to a modification to the spectrum of bilinears: in the real case, the odd-spin bilinears do not exist.}, and is trivially identified in the SDEs analysis without auxiliary field above as case \ref{SDEhmelonic}, \eqref{eq:stdBosonicPhi6Sol}. Thus, the non-perturbative value of the scaling dimension is linear in $D$: $\Delta_\phi = D/6$; so this theory with a single field has only a single solution, which is straightforward.

To identify the second bosonic fixed point, \hprismatic, we need the SDEs with auxiliary field \eqref{eq:auxSDEsAdded}\footnote{We note again that the theory with an auxiliary field is identical at the quantum level to the theory without auxiliary field, except for different values of the coupling constants $\lambda_i, h_i$. Due to these, different terms dominate in the SDEs.}. %
Indeed, if the above analysis is performed without fermions, then by solving the SDEs about $D=3-\epsilon$, we obtain:
\begin{align}
  &\frac{\Gamma (\Delta \phi ) \Gamma (D-\Delta \phi )}{\Gamma \left(\frac{D}{2}-\Delta \phi \right) \Gamma \left(\Delta \phi -\frac{D}{2}\right)}-\frac{3 \Gamma (3 \Delta \phi ) \Gamma (D-3 \Delta \phi )}{\Gamma \left(\frac{D}{2}-3 \Delta \phi \right) \Gamma \left(3 \Delta \phi -\frac{D}{2}\right)} = 0 \label{eq:prismatic2Pt}\\
  \implies &\gamma _{\phi } = \epsilon ^2-\frac{20 \epsilon ^3}{3}+\frac{1}{9} \left(472+3 \pi ^2\right) \epsilon ^4+O\left(\epsilon ^5\right). \label{eq:hprismaticPerturbative3meps}
\end{align}
\begin{figure}[h]
\centering
\includegraphics[width=0.9\textwidth]{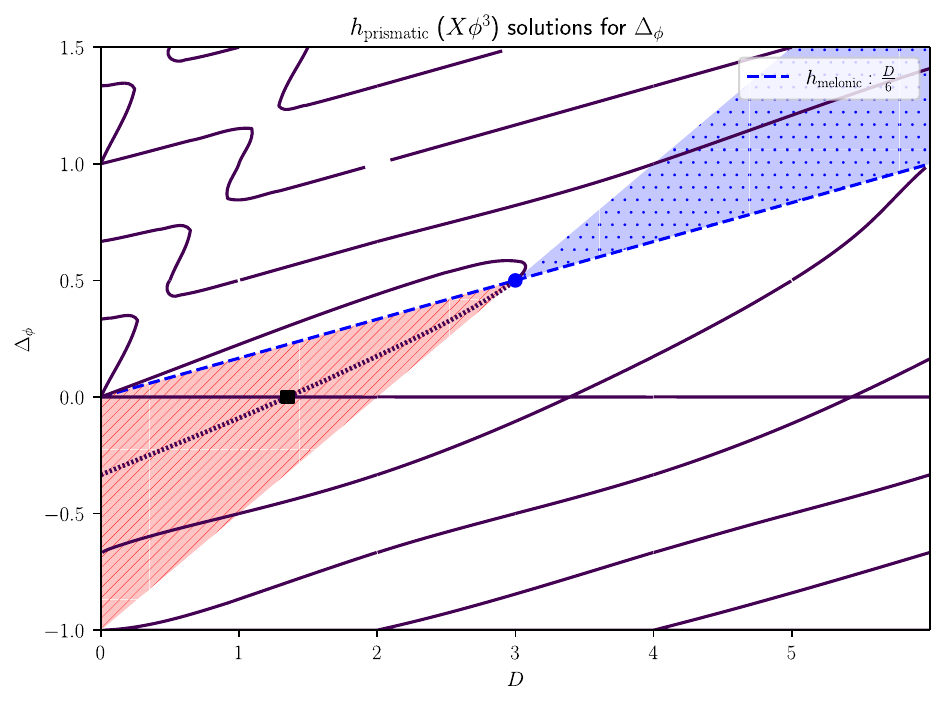}
\caption{Adding an auxiliary field $X$ to the sextic bosonic theory, we solve for \hprismatic{}, the sextic prismatic theory. The contours indicate the $\Delta_\phi$s which solve the SDEs, and $\Delta_X = D-3\Delta_\phi$ \cite{Giombi:2018qgp}. The region of IR validity \eqref{eq:IRscalingPrismatic} is shaded in ruled red; this also corresponds to the region where the anomalous dimensions of both $\phi$ and $X$ are positive (satisfying unitarity bounds). For $D>3$, we have the unitarity-bound-violating UV fixed point region in dotted blue. The standard \hmelonic{} solution is shown for comparison as a dashed blue line, $\Delta_\phi = D/6$. The free theory in $D=3$ is marked with a blue dot; the line of solutions \eqref{eq:hprismaticPerturbative3meps} descending from it in $D=3-\epsilon$ is indicated with a finely dashed line. For $D \simeq 1.353$, we have $\Delta_\phi=0$, marked with a black square; this will be discussed in \cref{sec:zeroScaling}.}
\label{fig:Bosonic6SDEdimPhi}
\end{figure}
This is precisely identical to the results of the \textit{prismatic bosonic tensor model} of \cite{Giombi:2018qgp} (and the combinatorics are identical to the $D=0$ theory studied in \cite{Krajewski:2023tbv}). The full solutions of the allowed $\Delta_\phi$ for every $D$ are plotted below in figure \ref{fig:Bosonic6SDEdimPhi}. For comparison, we indicate with the dashed blue line the \hmelonic theory.

Now, the solutions here demonstrate a number of features which we expect to be generic for multi-field melonic CFTs. Therefore, for emphasis, we provide these as a list:
\begin{itemize}
\item \textbf{The IR/UV wedges} The range of validity for an IR solution to \eqref{eq:prismatic2Pt} is 
\begin{equation}\label{eq:IRscalingPrismatic}
\gamma_\phi >0\text{ and }\gamma_A = \epsilon - 3\gamma_\phi  >0,
\end{equation}
indicated with the left red striped region, which we term the \textit{IR wedge}. Alternatively, swapping the signs of both of these conditions gives the non-unitary UV fixed point indicated with the second blue dotted region, the UV wedge. Note that positivity of all anomalous dimensions (for canonical kinetic terms) is also a necessary condition for unitarity of the theory. Hence, any solution outside the IR wedge must describe a non-unitary CFT (regardless of any concerns about evanescent operators).
\item \textbf{Infinite branches of solutions} We observe multiple different branches for almost every value of $D$ -- in fact, an infinite number. Those within the two wedges signal distinct vacua of the theory \cite{Chang:2021wbx}; those outside the wedges might seem to have no interpretation at all. However, this is not correct. The seemingly inaccessible lines of non-unitary fixed points can in fact become accessible, if we switch to so-called long-range kinetic terms for the scalar fields, which are the non-local expressions $\sim \int \phi (-\partial^{2})^{\zeta} \phi$ for arbitrary $\zeta$ \cite{Benedetti:2019eyl}. This allows us to modify the range of IR/UV validity, and so we can interpret these as non-local, manifestly non-unitary CFTs.
\item \textbf{Collision of two lines of fixed points} Occasionally, we appear to have fewer solutions than we might expect. For example, increasing $D$ from around $D=3$, two solutions collide and disappear at around $D \simeq 3.074$. $\Gamma(z)^\star = \Gamma(z^\star)$ then means that we have a pair of conjugate $\Delta_\phi$s; these will be illustrated graphically for \lammelonic in \cref{fig:Yuk46SDEdimPhiR4-3D}. This represents an instability of the theory \cite{Benedetti:2021qyk, Fraser-Taliente:2024prep}, and we will comment on it further in \cref{sec:windowsOfStability}.
\item \textbf{Disappearing solutions at exceptional values of the dimension $D$ and scaling dimension $\Delta_\phi$} We see breaks in the scaling dimension at certain \textit{exceptional values of the dimension}. For example, in even dimensions we find only a finite number of solutions to \eqref{eq:prismatic2Pt}, because the ratios of gamma functions become a rational function of $\Delta_\phi$. For example, in $D=2$ the infinite number of SDE solutions truncates to only two, being 
 \be
 \Delta_\phi = \frac{1}{13}(4 \pm \sqrt{3}).
 \ee
 Likewise, in even dimensions, we find $\lfloor \frac{2D +3}{3} \rfloor$ solutions to the scaling dimension equation, some of which may be complex. However, for all of these $D$s there are still perturbative fixed points in $D=2n\pm \epsilon$, which is why the contours only appear to have a small break in them. Another exceptional value is $\Delta_\phi=\half$ for $D=1$, where similarly we observe a perturbative solution $\Delta_\phi = \half + \frac{\epsilon}{6} + O(\epsilon^2)$, but no solution for exactly $\epsilon =0$. The nature of these exceptional fixed points is not understood.
\item \textbf{Zero scaling dimensions} The analysis that lead to \eqref{eq:hprismaticPerturbative3meps} must fail for certain values of $D$. For example, $\Delta_\phi=0$ implies a logarithmic two-point function; however, blindly evaluating \eqref{eq:hprismaticPerturbative3meps} with this value, we find that this is a solution for all values of $D$. In the IR region, this occurs for the branch descending from $D=3$ at $D_c \simeq 1.35$, as mentioned in \cite{Giombi:2018qgp}, as well as for the branch that has $\Delta_\phi=0$ always. This produces singularities in the dimensions of scalar bilinears, as we will see in \cref{sec:zeroScaling}, indicating that our approach to this theory breaks down. 
\end{itemize}
 We have now established in a simple multi-field model the various elements we will require: the IR wedge, the infinite numbers of solutions in general dimension, the breakdown of the conformal analysis at exceptional values of $D$ and $\Delta_\phi$, and the associated breaks in the contours; we are now ready to introduce fermions.

\subsection{Fermionic fixed points: SDEs without auxiliary field}

We now turn to the case where the fermions are coupled, where $\lambda \neq 0$, but for the moment we neglect the auxiliary field.

\subsubsection{\texorpdfstring{$\lambda_{\text{melonic}}$}{lambda-melonic} and \texorpdfstring{$h\lambda_{\text{melonic}}$}{hlambda-melonic} fixed points} \label{sec:lammelonicSDEfound}

\begin{figure}[H]
\centering
\begin{subfigure}{\textwidth}
  \centering
\includegraphics[width=0.8\textwidth]{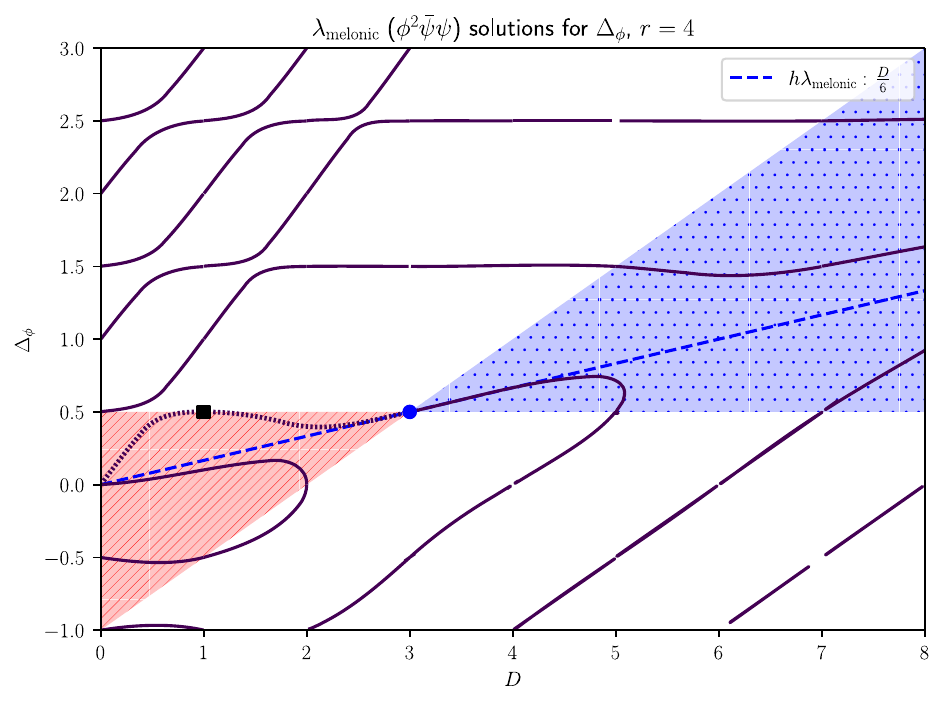}
  \caption{$r=4$, the branch used in the scaling dimensions analysis below.}
  \label{fig:Yuk46SDEdimPhiR4}
\end{subfigure}
\begin{subfigure}{\textwidth}
  \centering
\includegraphics[width=0.8\textwidth]{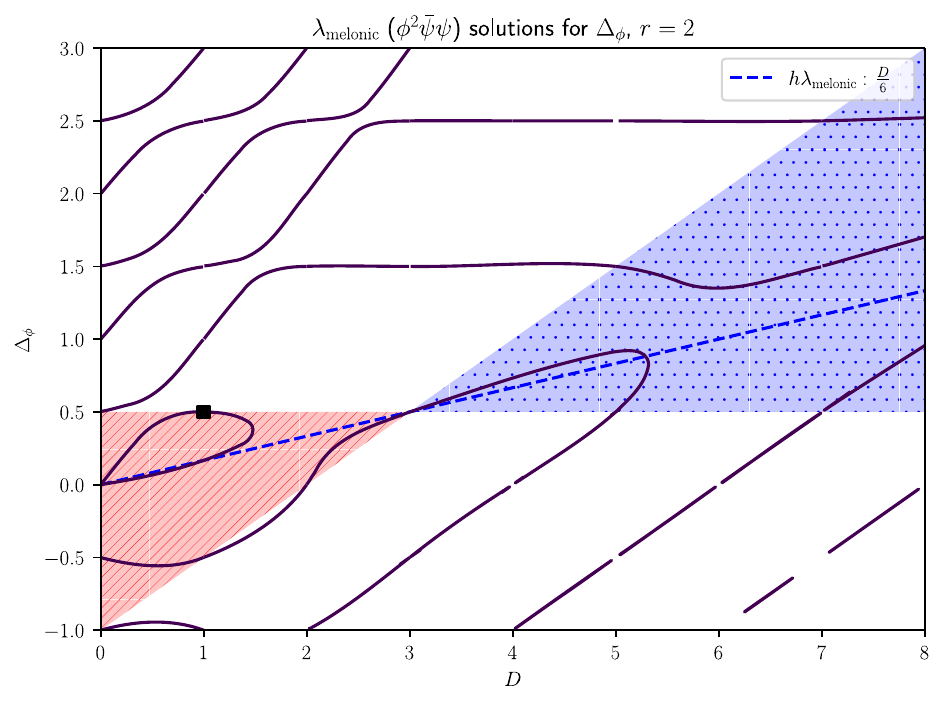}
  \caption{$r=2$ presented for comparison.}
  \label{fig:Yuk46SDEdimPhiR2}
\end{subfigure}
  \caption{$\phi$ scaling dimension for \lammelonic for $r=4,2$. The $r=4$ branch that we will use in the \lammelonic{} bilinear analysis of \cref{sec:bilinearsCalculationResults} finely dotted; it descends from the $D=3$ free theory, which is marked with a blue dot. The wedge of validity of the IR scaling solution to the SDEs is shaded in ruled red, and also corresponds to the regions where the scaling dimensions of $\phi,\psi$ satisfy the unitarity bounds. We give for comparison the \hlammelonic scaling dimension, $\Delta_\phi = \frac{D}{6}$ as a blue dashed line.  Also, note that in $D=1+ \epsilon$ for all values of $r$ we find $\Delta_\phi = \half- O(\epsilon^2)$, $\Delta_\psi =0 + O(\epsilon)$, marked with a black square. This solution does not exist in $D=1$ exactly, however. The IR wedge of \cref{fig:Yuk46SDEdimPhiR4} is shown in more detail and including the solutions for complex $\Delta_\phi$ in \cref{fig:Yuk46SDEdimPhiR4-3D}.}
  \label{fig:Yuk46SDEdimPhiBothRs}
\end{figure}

Setting $2b+2f=D$, the fermion SDE becomes
\begin{equation}\begin{aligned}
    &1+\frac{\lambda^2 B^2 F^2}{2  (4\pi)^D}\frac{\Gamma \left(\frac{D}{2}-b\right)^2}{\Gamma (b)^2} \frac{  \Gamma \left(b-\frac{D}{2}+\frac{1}{2}\right) \Gamma \left(b+\frac{1}{2}\right) }{\Gamma \left(\frac{D}{2} - b + \half\right) \Gamma \left(D-b+\frac{1}{2}\right)} = 0
\end{aligned}\end{equation}
This can be substituted into the boson SDE, which yields:
\begin{equation}\begin{aligned}
&\frac{B^6 h^2\Gamma (5 b-2 D)\Gamma \left(\frac{D}{2}-b\right)^5 }{ 5! (4\pi)^{2 D} \Gamma (b)^5 \Gamma \left(\frac{5 D}{2}-5 b\right)}(p^2)^{2 D-6 b}+\frac{2^{D-4 b} r \Gamma (-b) \Gamma (2 b) \Gamma \left(D-b+\frac{1}{2}\right)}{\Gamma \left(b-\frac{D}{2}+\frac{1}{2}\right) \Gamma \left(b+\frac{D}{2}\right) \Gamma (D-2 b)}=-1 \label{eq:SDEsForFermionic}%
\end{aligned}\end{equation}
A consistent solution requires either $h=0$ or  $2D-6b=0$.  We will discover that the former gives precisely the \lammelonic{} fixed point identified above. We plot the solutions and regions of validity for $h=0$ in \cref{fig:Yuk46SDEdimPhiR4,fig:Yuk46SDEdimPhiR2} for This can be substituted into the boson SDE, which yields:two different values of $r$. In \cref{sec:bilinears}, we will focus on the $r=4$ line of $\Delta_\phi$ solutions descending from the free theory in $D=3$, indicated in \cref{fig:Yuk46SDEdimPhiR4}.

\begin{figure}
  \centering
\includegraphics[width=0.8\textwidth]{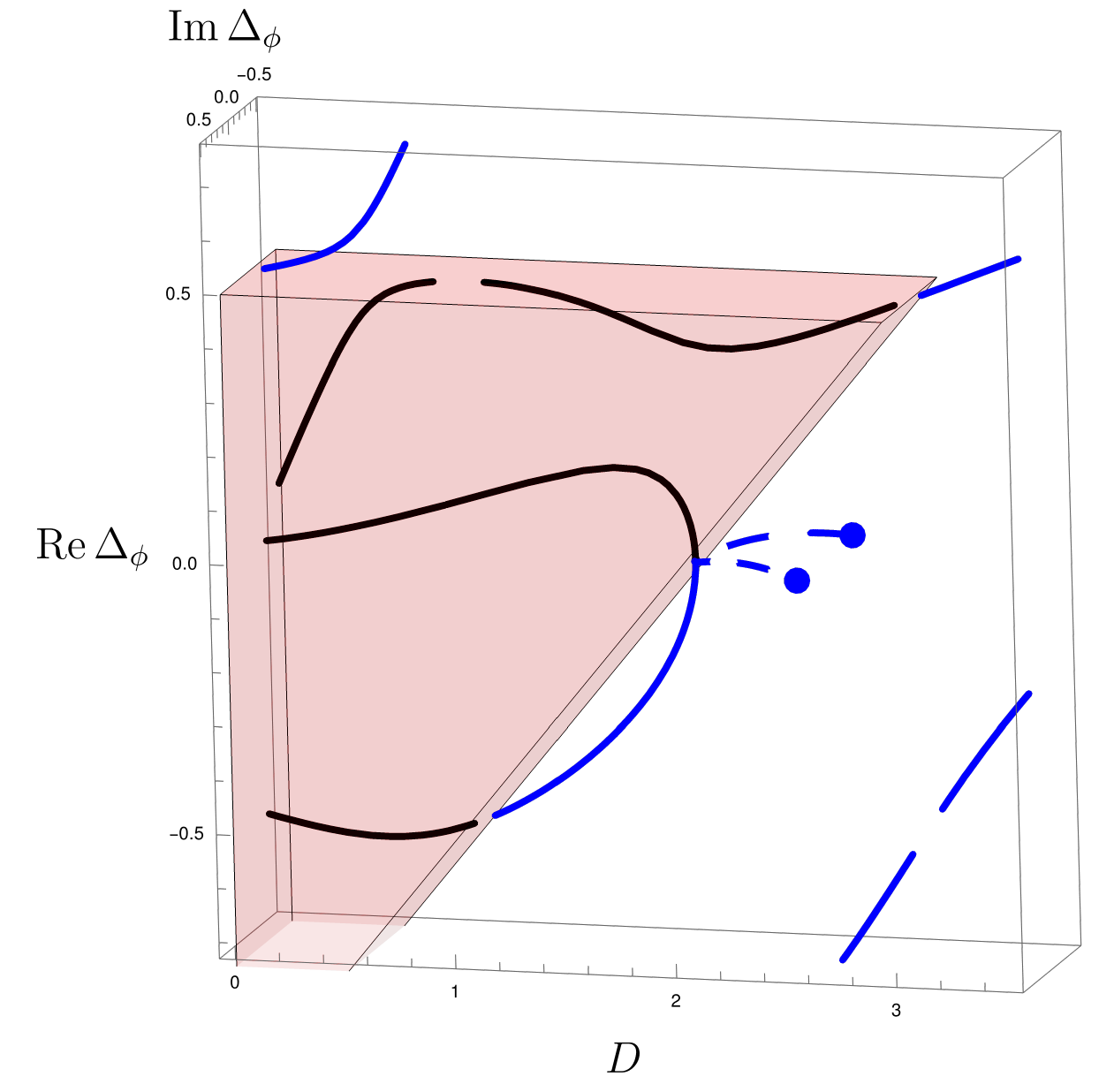}
  \caption{$\phi$ scaling dimension for \lammelonic for $r=4$; this is a reproduction of the IR wedge of \cref{fig:Yuk46SDEdimPhiR4}, but including complex $\Delta_\phi$s. Black lines lie inside the IR wedge; blue lines lie outside it. The two dashed lines in the centre are complex, and leave the plotting region at points marked with blue dots.}
  \label{fig:Yuk46SDEdimPhiR4-3D}
\end{figure}

Expanding $\gamma_\phi$ for $D=3-\epsilon, \Delta_\phi = \frac{D-2}{\epsilon}+\gamma_\phi$ in $\epsilon$ gives
\begin{align}\label{eq:gammaphiSDEsol}
\begin{split}
\gamma _{\phi } &= \frac{r \epsilon }{2(r+2)}+\frac{r (5 r-16) \epsilon ^2}{6 (r+2)^3}+\frac{r \left(2 r (r (17 r+164)-472)-224-3 \pi ^2 (r-2) (r+2)^2\right) \epsilon ^3}{36 (r+2)^5}\\
&+\frac{r \left[\splitfrac{3 \pi ^2 (r (r (5 r-82)+200)-64) (r+2)^2-378 (r-2) \zeta(3) (r+2)^4}{+4 r (r (r (r (62 r+1069)+5839)-17384)-7136)-4096}\right] \epsilon ^4}{216 (r+2)^7}+O(\epsilon ^5),
\end{split}
\end{align}
which exactly matches the $O(\epsilon^2)$ $\phi_{\mathbb{R}}\psi_{\mathbb{C}}$ perturbative result in \eqref{eq:fermionicFixedPoints} for $r=2T$. For later reference, for $r=4$ at fourth order 
\begin{equation}\label{eq:lamOnlyFPGammaPhiEp4}
    \gamma_\phi = \frac{\epsilon }{3}+\frac{\epsilon ^2}{81}+\frac{\left(428-27 \pi ^2\right) \epsilon ^3}{8748}+\left(-\frac{7 \zeta (3)}{108}-\frac{4 \pi ^2}{2187}+\frac{6299}{59049}\right) \epsilon ^4+O\left(\epsilon ^5\right).
\end{equation}
Note that this $D=3-\epsilon$ fixed point path, shown with a finely dotted line in \cref{fig:Yuk46SDEdimPhiR4}, leaves the IR wedge for $D>3$. However, as mentioned above, we can modify this by changing the UV kinetic term to some non-local $\int \phi (-\partial^{2})^\zeta \phi$, as in \cite{Benedetti:2019eyl}, at the cost of locality.
We note the presence of various features that first appeared in \hprismatic in \cref{sec:generalCharacteristicsPrismatic}. We have already mentioned the IR/UV wedges, and the infinite number of solutions in generic dimension are manifest in \cref{fig:Yuk46SDEdimPhiR4,fig:Yuk46SDEdimPhiR2}.
\begin{itemize}
\item  \textbf{Collision of two lines of fixed points}: once again, we have a collision of two lines of fixed points, that occurs, for example, around $D\sim 5$ for $r=2,4$, and at $D\simeq 1.48$ for $r=2$. At these points the actual solutions complexify, as is demonstrated for the latter in \cref{fig:Yuk46SDEdimPhiR4-3D}.
\item \textbf{Disappearing solutions at exceptional values of the dimension} In integer dimensions, the ratio of gamma functions in \eqref{eq:SDEsForFermionic} becomes a rational function of $\Delta$, and so we obtain $D$ solutions for even $D$, and $(D+1)/2$ solutions in odd $D$.
\item We have in $D=1$ a combination of both the \textbf{disappearing solutions at exceptional values of $\Delta_\phi$} and the \textbf{zero scaling dimensions}. That is, we have perturbative solutions $\Delta_\phi = 1/2 - O(\epsilon^2)$ and so $\Delta_\psi = 0 + O(\epsilon)$ in $D=1+\epsilon$, but no such solution for $\epsilon =0$. This will lead to the singularities in the dimensions of the scalar bilinears, as we will see in section \cref{sec:zeroScaling}; however, as we know from \hprismatic, these two phenomena should not be confused, just because they occur together here. In fact, for example, in the $\phi^p \bar\psi \psi$ theory, we always have a break in the contour for $\Delta_\phi = 1/2 - O(\epsilon^2)$, even though $\Delta_\psi = \half(D-p\Delta_\phi) \neq 0$ \cite{Fraser-Taliente:2024prep}.
\end{itemize}

However, we now also have a second melonic coupling: as discussed above, $b=D/3$ also gives a solution for $h\neq 0$. If the fermions are not free, i.e. $\lambda \neq 0$, then $2f+2b=D$ enforces $f=D/6$, which is indeed precisely the \hlammelonic fixed point that we found perturbatively in \cref{sec:betas}. At this fixed point, the $b=D/3,\Delta_\phi = D/6$ scaling dimension matches up exactly with the scaling dimension found in the \textit{complex} sextic bosonic tensor model \cite{Benedetti:2019rja}, despite the presence of the interacting fermions. Of course, the scaling behaviours of the respective SDEs are identical regardless of the fixed point, so the complex nature of that field, and the presence of fermions, are irrelevant. 

We also note that here that \lammelonic and \hlammelonic have manifestly different scaling dimensions beyond leading order in $\epsilon$. Though a perturbative analysis is not feasible here, it is likely simply that the value of the $h$s at the fixed point $h_i^* \propto \half (r-4)( \epsilon + g(r) \epsilon^2) + f(r) \epsilon^3$, where $f(4)\neq 0$. Thus, the $h^2$ term in $\gamma_\phi$ contributes only at $\epsilon^4$ order when $r=4$, so these fixed points which appeared to be identical at leading order in $\epsilon$ are actually different. This resolves the apparent collision of \lammelonic and \hlammelonic{} noticed perturbatively in \cref{sec:othercouplingconstants}. It still remains to deal with \hlamprismatic, but we will do so in \cref{sec:hlamprismaticSDEfound}.

\subsubsection{Majorana fermions, or other variations} \label{sec:rparameter}

It turns out that the seemingly arbitrary choice of a real scalar and complex fermion is not relevant, as any other choice can simply be subsumed (in the strict large-$N$ limit) by appropriate variations of $r$. For example, we can repeat the SDE analysis above with Majorana fermions, or other alternative fermions. The Feynman rules \cite{Denner:1992me} are straightforward in the context of our limited set of diagrams; we simply modify the symmetry factors. If we label the real or complex scalar field by $\phi_{\mathbb{R}/\mathbb{C}}$, and the Majorana or Dirac fermion by $\psi_{\mathbb{R}/\mathbb{C}}$, then the symmetry factors associated with the diagrams of \cref{on3-2pt-sde-3dyuk} (excluding the $\phi^4$ melon) are
\begin{center}
\begin{tabular}{L|LLL|L}
  & S_{FBB} & S_{BFF} & S_{B^5}  & \text{equivalent to }$r=$\\ \hline
 \phi_{\mathbb{R}} \psi_{\mathbb{{R}}} & 2 & 2 & 5! & T \\
 \phi_{\mathbb{R}} \psi_{\mathbb{{C}}} & 2 & 1 & 5! & 2T\\
 \phi_{\mathbb{C}} \psi_{\mathbb{{R}}} & 1 & 2 & 2! \cdot 3! & T/2\\
 \phi_{\mathbb{C}} \psi_{\mathbb{{C}}} & 1 & 1 & 2! \cdot 3! & T
\end{tabular}
\end{center}
If the $h$ couplings are irrelevant, i.e. we are at the \lammelonic{} fixed point, then the solution depends only on the ratio $r=T=S_{FBB}/S_{BFF}$, and the result is the same as above in \eqref{eq:gammaphiSDEsol}. This case shows no particularly interesting features, unlike the situation when the auxiliary field is included, analysed below. If $h$ is relevant, then we know that we have the completely constrained $b = D/3$ fixed point. Changing the value of $r$ does not modify this at all, and the solution always exists.
If $\lambda$ is irrelevant, the fermions are free, and the $r$ parameter has no meaning.

In fact, we will see in the $\lambda_{\text{melonic}}$ model that we can interpret $r=2T$ as the ratio of the number of real fermionic degrees of freedom in the field $\psi$ to the number of real bosonic degrees of freedom in $\phi$.  This can be made obvious by the fact that if we add a further $\mathrm{U}(N_f)$ fermion symmetry, i.e. have $N_f$ tensor fermion fields $\psi_{abc,K}, K=1,...N_f$, and likewise for $N_b$ tensor boson fields, the effect in the equations of the model is exactly the same\footnote{The Feynman rules say each fermion loop gives a factor of $(-1)\times T$; if we have a vector of $N_f$ fermions, which only interact via $\mathrm{U}(N_f)$-symmetric terms like $\sum_K \lambda_t \delta^{t,p,dt}_{abc,def,hij,klm} \phi_{abc} \phi_{def} \bar\psi_{hij,K} \psi_{klm, K}$, then each loop also contributes a factor of $N_f$. } as modifying $r=\frac{2T N_f}{N_b}$; for this reason we can at least formally modify $r$ to whatever value we like. 

This includes both negative and non-integer values of $r$. Combined with the fact that $T$ always comes with a factor of $(-1)$ due to fermion loops, we can interpret Dirac fermions of negative $T$ as $\Sp(\abs{T}/T_d)$-symmetric commuting fermions, via the well-known formal relation $\SO(-N) \simeq \Sp(N)$\footnote{At least up to ordering of operators in scattering, which does not concern us here; see \cite{Gurau:2022dbx,Keppler:2023lkb}.}, where $\abs{T}$ is the free parameter that we vary, and $T_D$ is the dimension of the minimal complex Dirac spinor in $D$ dimensions. A mathematical framework justifying non-integer values of $N_{f,b}$ and hence $r$ is presented in \cite{Binder:2019zqc}; see also \cite{Jepsen:2020czw}. Note that as $r$ is smoothly varied from $r=4$ to $r=2$, at $r\simeq 3.07$, the line of theories that we will use splits: this means that there is no valid IR solution for a dimensional window around $D \simeq 1.8$, as is shown in \cref{fig:Yuk46SDEdimPhiBothRs}. In the language of \cite{Kaplan:2009kr}, for $r=2$ the two IR fixed points annihilate with each other around $D=1.48$ as $D$ increases from $1$.

\subsection{The SDEs with auxiliary field} \label{sec:hlamprismaticSDEfound}

We can consider the set of SDEs with auxiliary field, \eqref{eq:auxSDEsAdded}; ansatzing the momentum-space IR propagator to be the following we require only $a>0$ for IR consistency:
\begin{equation}
    \expval{X_I(p)X_J(-p)}=A(p)\delta_{IJ} \equiv \frac{A}{p^{2a}} \delta_{IJ}
\end{equation}
This modifies only the bosonic sector of the SDEs, and in the deep IR we have:%
\begin{subequations}
\begin{align}
0 & = F(p)^{-1} +  \frac{\lambda^2}{2}  \int_{k,l} F(k+l+p)B(l)B(k),\\
0 & = B(p)^{-1} + \frac{h^2}{5!} \int_{k,l,m,n} B(k+l+m+n+p)B(k)B(l)B(m)B(n) \notag\\
&+\frac{\rho^2}{2} \int_{k,l} A(k+l+p)B(l)B(k) + \frac{\lambda^2}{1} \int_{k,l} (-1)\Tr\left[B(k+l+p)F(l)F(k)\right], \\
0 &= A(p)^{-1} + \frac{\rho}{3!} \int_{k,l} B(k+l+p)B(k)B(l).
\end{align}
\end{subequations}
Therefore, we obtain once again the usual $2f+2b=D$, along with $a+3b=D$. We again can either take $h\neq 0$, or $h=0$. In the former case, we force $\Delta_\phi =D/6$, which forces $\Delta_\chi=D/2$, which is the free field scaling dimension. This gives $\rho^2 A B^3 =0$, and so we find \hlammelonic again. In the latter, demanding consistency, we can solve for $B^2 F^2$ and $AB^3$, which yields the result
\begin{equation}
\frac{\csc (\pi  b) \Biggl[\splitfrac{3 (D-3 b) \sin (3 \pi  b-\pi  D) \Gamma \left(3 b-\frac{D}{2}\right) \Gamma \left(\frac{3 D}{2}-3 b\right)}{-r \Gamma \left(b+\frac{1}{2}\right) \cos \left(\pi  b-\frac{\pi  D}{2}\right) \Gamma \left(D-b+\frac{1}{2}\right)}\Biggr]}{b \Gamma \left(\frac{D}{2}-b\right) \Gamma \left(b+\frac{D}{2}\right)}=-1.
\end{equation}
This can be solved perturbatively to any order desired. Taking $b=1-\gamma_\phi$ and $D=3-\epsilon$, we obtain a match to the $\hlamprismatic$ perturbative analysis:
\begin{equation}\label{eq:auxFieldSRFC}
\gamma_\phi = \frac{r \epsilon }{2(r+2)}+ (r-2)\Biggl(\frac{4 (r-3) \epsilon ^2}{3 (r+2)^3}-\frac{\left(r \left(3 \pi ^2 (r+2)^2-8 r (5 r+99)+3056\right)-3840\right) \epsilon ^3}{36 (r+2)^5}+O(\epsilon^4)\Biggr).
\end{equation}
Note that for $r=4$, this equals the \lammelonic{} scaling dimension calculated in \eqref{eq:gammaphiSDEsol} up to order $\epsilon^3$. They then end up differing at order $\epsilon^4$: note the 6326 here, compared to 6299 in equation \eqref{eq:lamOnlyFPGammaPhiEp4}:
\begin{equation} \gamma_\phi = \frac{\epsilon }{3}+\frac{\epsilon ^2}{81}+\frac{\left(428-27 \pi ^2\right) \epsilon ^3}{8748}+\left(-\frac{7 \zeta (3)}{108}-\frac{4 \pi ^2}{2187}+\frac{\mathbf{6326}}{59049}\right) \epsilon ^4+O\left(\epsilon ^5\right).
\end{equation}
So, the auxiliary field solution is again different at order $\epsilon^4$, for the same reason as before. So all three of the fixed points which apparently collide at $r=2T=4$ in the perturbative analysis of \eqref{eq:fermionicFixedPoints} are in fact distinct. 

\begin{figure}
\begin{subfigure}{\textwidth}
  \centering
\includegraphics[width=0.8\textwidth]{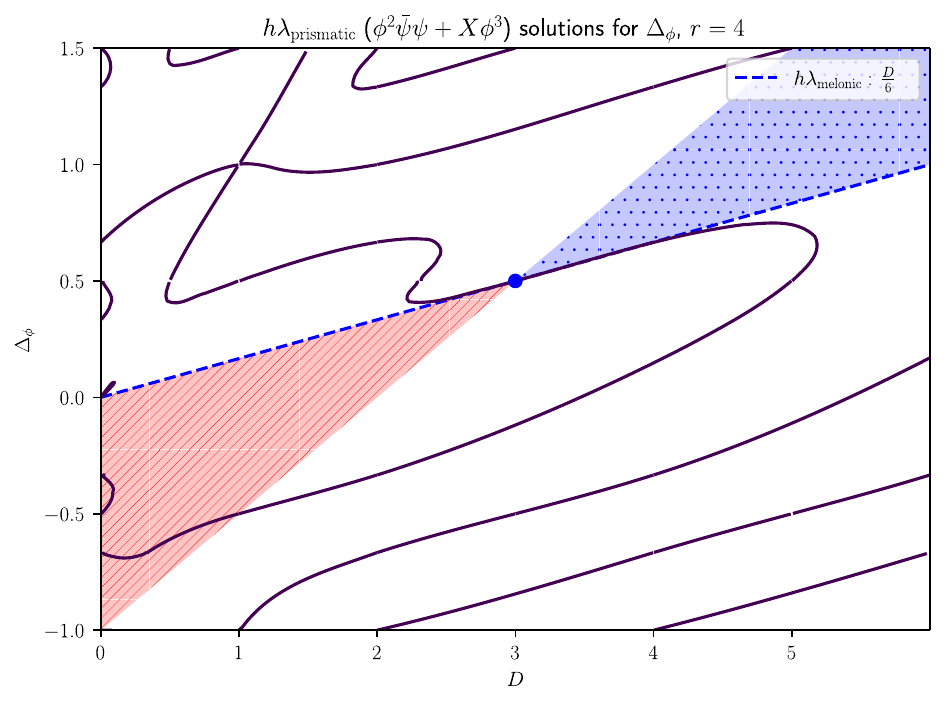}
  \caption{$r=4$}
  \label{fig:hlamprismatic2ptSDEs-r4}
\end{subfigure}
\begin{subfigure}{\textwidth}
  \centering
\includegraphics[width=0.8\textwidth]{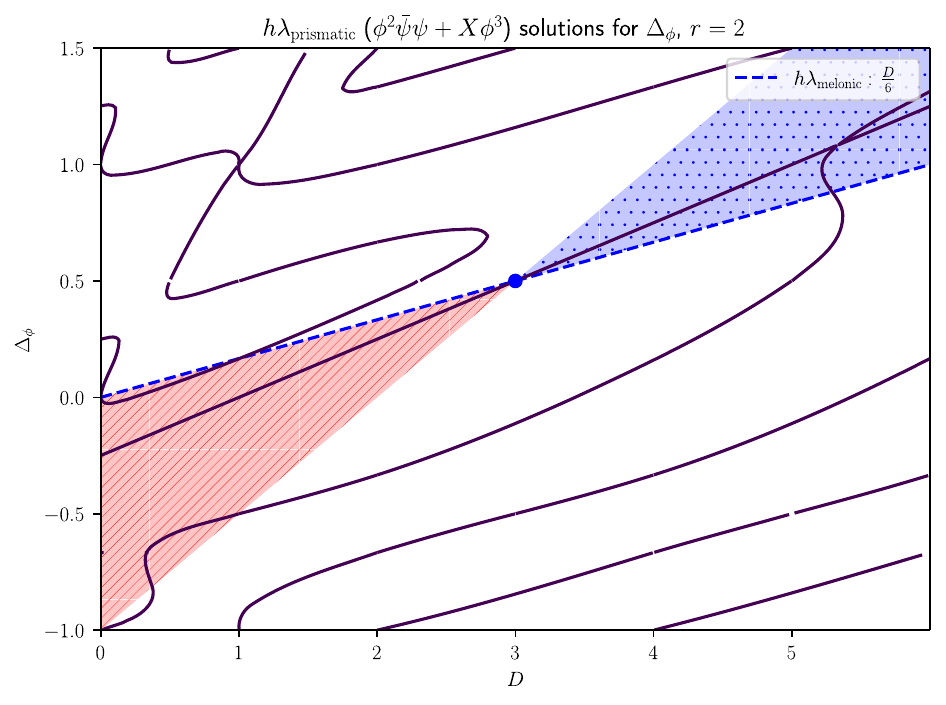}
  \caption{$r=2$, which contains the putative supersymmetric theory with $\Delta_\phi = \frac{D-1}{4}$ referred to in \cref{sec:qGenAndSUSYprismatic}, as well as other non-supersymmetric IR solutions.}
  \label{fig:hlamprismatic2ptSDEs-r2}
\end{subfigure}
\caption{$\phi$ scaling dimension for the prismatic theory, \hlamprismatic{}, with $r=4,2$ again. The unitary region, which is also the regime of validity of the IR scaling solution to the SDEs, is shaded in dashed red. The dotted blue region corresponds to the non-unitary UV solution to the SDEs.} \label{fig:prismaticScalingDimensions}
\end{figure}

In \cref{fig:prismaticScalingDimensions} we demonstrate the space of fixed points. Once again, red indicates the region of validity, and \hlammelonic is again indicated with the blue dashed line. For $r=0$, clearly \hlamprismatic{} reduces to \hprismatic{} (plus a free fermion CFT).

\subsubsection{Variation of \texorpdfstring{$r$}{r}, \texorpdfstring{$q$}{q}-generalisations and the supersymmetric solution} \label{sec:qGenAndSUSYprismatic}

As before, it is trivial to solve the above for general symmetry factors, which correspond to the following theories. Note that $\psi_{\mathbb{R}}$ corresponds to Majorana and $\psi_{\mathbb{C}}$ corresponds to Dirac fermions, although we leave $T$ general.

Note that if $r=2$, we obtain
\be
\gamma_\phi = \gamma_\psi = \gamma_F = \epsilon/4
\ee
exactly, to all orders.  This corresponds to the scaling dimension of the supersymmetric solution in the $\phi_{\mathbb{R}} \psi_{\mathbb{R}}$  case ($\mathcal{N}=1$ scalar superfield) and $\phi_{\mathbb{C}} \psi_{\mathbb{C}}$  cases ($\mathcal{N}=2$ scalar superfield); both of these have $\Tr[\mathbb{I}_s]=2$, the expected dimension of the spinor space in 3D. For this value of $r$, if we include the auxiliary field, the number of off-shell fermionic degrees of freedom then matches the number of off-shell bosonic degrees of freedom $N^3+N^3$, as it should for a supersymmetric theory. Of course, for the theory $\phi_{\mathbb{R}} \psi_{\mathbb{C}}$, this would require $\Tr[\mathbb{I}_s]=1$, which is not a physically permitted dimension of the spinor space; nonetheless, we still appear to have supersymmetry.

Continuous dimension is in general incompatible with supersymmetry; however, we have performed a variant of the dimensional reduction scheme used in loop calculations of supersymmetric theories: we have kept $T$ fixed and continued in $D$, in order to consider a supersymmetric theory in general dimension \cite{Giombi:2014xxa}.

Assuming either real scalar fields and Majorana fermions, or complex scalar fields and Dirac fermions, adding in an auxiliary field, we can use the superspace formalism: that is, we combine the three fields into a (real or complex) superfield $\Phi$. The superfield formalism enforces that the dimensions of each of the three components of the superfield ($\phi$, $\psi$, $X$) differ by $1/2$, and hence the anomalous dimensions of each must be equal \cite{Popov:2019nja,Lettera:2020uay}. Thus, in a manner very similar to the standard 4D quartic bosonic model, we define the superfield propagator to be $P(p)$, and find the IR SDEs to be $P(p)^{-1} \propto \int_{k,l} \Tr(P(k+l+p) P(k) P(l))$. Therefore, simple dimensional analysis enforces $\gamma_\Phi = \epsilon/4 = \gamma_\phi = \gamma_\psi = \gamma_X$ -- precisely the result of \eqref{eq:auxFieldSRFC}. 
In fact, this can be done with general coupling $\Phi^q$ to find that the superfield gets scaling dimension $\Delta_\Phi = \frac{D-2}{2} + \gamma_\Phi = \frac{D-1}{q}$. Motivated by this, these SDEs with auxiliary field easily generalise, with or without fermions. We simply set the auxiliary field to be $X=\phi^{q-1}$, and perform the same set of integrals. If we include fermions this is a generalisation of the aforementioned $q$-generalised supersymmetric tensor models (\cite{Popov:2019nja,Lettera:2020uay}) to non-matching bosonic and fermionic degrees of freedom. Without the auxiliary field, and without fermions, this is the $q$-generalisation of the original tensor model studied in \cite{Giombi:2017dtl}. See \cite{Fraser-Taliente:2024prep} for more.

\section{Spectrum of bilinears for the \texorpdfstring{\lammelonic}{lambda-melonic} fixed point} \label{sec:bilinears}

The simplest of these new fermionic fixed points is \lammelonic{}, which we now investigate in more detail by computing the spectrum of the set of $\mathrm{O}(N)^3$ singlet bosonic operators that appear in the conformal OPEs $\phi_{abc} \times \phi_{abc}$ and $\bar\psi_{abc} \times \psi_{abc}$. For these melonic theories, this is a standard computation using the Schwinger-Dyson equation, which at large $N$ we can write down exactly \cite{Benedetti:2023mli}. In the conformal limit this Schwinger-Dyson equation can be integrated analytically, and the computation of the spectrum reduces to finding the scaling dimensions $\Delta$ such that a particular kernel matrix $k(\Delta)^a_{b}$ has a unit eigenvalue.

We shall find results that are analytic in spin; however, it is enlightening to consider explicitly the form of the spin-zero operators appearing in this OPE. They should be 
$\phi_{abc} (\partial^2)^n \phi_{abc}$, $\bar\psi_{abc} (\partial^2)^n \psi_{abc}$, and $\bar\psi_{abc} (\partial^2)^n \not{\partial} \psi_{abc}$; of course, operators of definite scaling dimensions will be a linear combination of these.

There are four main observations to be made about this part of the bilinear spectrum for the \lammelonic fixed point:
\begin{enumerate}
\item Near the free theory and near $D=0$ the operators in the spectrum exactly match the known scaling dimensions. Additionally, we have cross-checks via the perturbative computations of the mass operators.
\item There are windows of stability of the parameters $r$ and $D$, outside which operators have a complex scaling dimension; these appear to be characteristic of tensor models.
\item The spectrum contains conserved operators at spins $s=0,1,2$. These are, respectively, a redundant operator, the $\mathrm{U}(1)$ current, and the stress tensor.
\item Divergences appear in the spectrum when the scaling dimension of one of the fundamental fields goes to zero; in this model, this occurs at $D=1$, just as occurs for \hprismatic in $D\simeq 1.35$.
\end{enumerate}

We will restrict ourselves to the subset of the operator spectrum consisting of bosonic operators $\cO_{(\mu_1 \cdots \mu_s)}$ that are traceless symmetric tensors in $\SO(D)$. These transform in the traceless symmetric spin-$s$ representations $\rho$, and so are labelled by Dynkin label $(s,0,0,\ldots)$, and exist in any $D$.  They take the schematic form\footnote{We will not need the exact forms, which can be found in e.g. \cite{Hikida:2016cla}.}
\begin{equation}\begin{aligned}
\cO_{s} \sim [(\partial^2)^x \partial_{(\mu} \cdots \partial_{\mu_i} \phi][\partial_{\mu_{i+1}} \cdots \partial_{\mu_s)} (\partial^2)^y \phi] -\text{traces}
\end{aligned}\end{equation}
We will not consider any other representations (thus ignoring any fermionic operators), as in that case analytic continuation in the dimension and Dynkin label entries is less clear.

We will begin with an overview of the pure bosonic calculation in \cref{sec:bilinearCalculationOverview}. %
Moving to the \lammelonic{} theory, we demonstrate the elaborations necessary for a fermionic theory in \cref{sec:bilinearsCalculationDetails}, and end with observations and comparison to \hprismatic in \cref{sec:bilinearsCalculationResults}.

\subsection{Spectrum for the bosonic tensor model with \texorpdfstring{$V(\phi)=\phi^q$}{V(phi)=phi\string^ q}} \label{sec:bilinearCalculationOverview}

First, we recall the determination of the spectrum  for a purely bosonic CFT, recapping the salient points of \cite{Giombi:2017dtl}. It is informative to see the dependence of the melonic spectrum on the degree of the potential, $q$, which we keep general. 
To identify the operators $\cO_{\Delta,\rho}$, of scaling dimension $\Delta$ and $\SO(D)$ representation $\rho$, appearing in the OPE $\phi\times\phi$, we consider the Schwinger-Dyson equation for the three-point function $v(\cO)=\expval{\phi(x)\phi(y)\cO_{\Delta,\rho}}$. %
The spectrum consists of those operators $\cO_{\Delta,\rho}$ for which this Schwinger-Dyson equation is self-consistent.

\subsubsection{The Schwinger-Dyson equation for the three-point function}

The self-consistent Schwinger-Dyson equation for the three-point function $v(\cO)$ takes the form \cite{Gross:2016kjj,Liu:2018jhs}
\begin{figure}[H]
  \begin{align*}%
  \vcenter{\hbox{
\begin{tikzpicture}
    \begin{feynman}
        \vertex [blob] (a) at (0,0) {$v$};
        \vertex (b) at (-1,0.75) {$x$};
        \vertex (c) at (-1,-0.75) {$y$};
        \vertex (d) at (1.5,0) {$z$};
        \diagram* {
            (a) -- (b),
            (a) -- (c),
            (a) -- (d),
        };
    \end{feynman}
\end{tikzpicture}
}}
\quad &= \quad
\vcenter{\hbox{
\begin{tikzpicture}
    \begin{feynman}
        \vertex [blob] (a) at (0,0) {$v$};
        \vertex (y1) at (-1,0.75);
        \vertex (y2) at (-1,-0.75);
        \vertex (d) at (1.5,0) {$z$};
        \vertex (x) at (-3.0, 0.75) {$x$};
        \vertex (y) at (-3.0,-0.75) {$y$};
        \diagram* {
            (a) -- (y1) -- (x),
            (a) -- (y2)--(y),
            (a) -- (d),
        };
    \end{feynman}
\end{tikzpicture}
}}
\quad + \quad \vcenter{\hbox{
\begin{tikzpicture}
    \begin{feynman}
        \vertex [blob] (a) at (0,0) {$v$};
        \vertex (y1) at (-1,0.75);
        \vertex (y2) at (-1,-0.75);
        \vertex (d) at (1.5,0) {$z$};
        \vertex (x) at (-3.0, 0.75) {$x$};
        \vertex (y) at (-3.0,-0.75) {$y$};
        \diagram* {
            (a) -- (y1) -- (x),
            (a) -- (y2)--(y),
            (a) -- (d),
        };
        \draw[draw=gray,fill=gray] (-1,0.75) rectangle ++(-1.0,-1.5);
    \end{feynman}
\end{tikzpicture}
}}
  \end{align*}
\label{fig:3ptFuncVisualSDE}
\end{figure}
\noindent where the grey block represents the 2PI connected four-point Bethe-Salpeter kernel. %
The first term on the r.h.s., which is the free-field contribution to the kernel, %
drops out in the conformal IR. The tensor structure at the \lammelonic fixed point is identical to that of the quartic model, so the argument of \cite{Giombi:2017dtl} transfers exactly. %
The complete non-perturbative connected Bethe-Salpeter kernel is (illustrated for $q=4$):
\begin{figure}[H]
  \begin{align*}%
\vcenter{\hbox{\begin{tikzpicture}
    \begin{feynman}[every blob={/tikz/pattern color=black,/tikz/inner sep=2pt}]
    \vertex (a) at (-1,0) ;
    \vertex (b) at ( 1,0);
    \vertex (c) at (-1,1);
    \vertex (d) at ( 1,1);
    \draw[draw=gray,fill=gray] (-0.5,0) rectangle ++(1.0,1.0);
    \diagram* {
      (a) -- (b),
      (c) -- (d)
      };
  \end{feynman}
\end{tikzpicture}}}
\quad &= 
\quad \half \quad \vcenter{\hbox{\begin{tikzpicture}
    \begin{feynman}[every blob={/tikz/pattern color=black,/tikz/inner sep=2pt}]
    \vertex (a) at (-1,0) ;
    \vertex (b) at (1.5,0);
    \vertex (m1) at (0,0);
    \vertex (m2) at (0,1);
    \vertex (c) at (-1,1);
    \vertex (d) at (1.5,1);
    \foreach \point in {(-0.7,0), (-0.7,1), (0.22,0.5), (-0.22, 0.5)} {
        \node [circle, fill, inner sep=3pt] at \point {};
    }
    \diagram* {
      (a) -- (m1)-- (b),
      (m1)-- [quarter left](m2),
      (m1)-- [quarter right](m2),
      (c) -- (m2)-- (d),
      };
  \end{feynman}
\end{tikzpicture}}} + \half \quad \vcenter{\hbox{\begin{tikzpicture}
    \begin{feynman}[every blob={/tikz/pattern color=black,/tikz/inner sep=2pt}]
    \vertex (a) at (-1,0) ;
    \vertex (b) at (1.5,0);
    \vertex (b1) at (2.5,1);
    \vertex (m1) at (0,0);
    \vertex (m2) at (0,1);
    \vertex (c) at (-1,1);
    \vertex (d) at (1.5,1);
    \vertex (d1) at (2.5,0);
    \foreach \point in {(-0.7,0), (-0.7,1), (0.22,0.5), (-0.22, 0.5)} {
        \node [circle, fill, inner sep=3pt] at \point {};
    }
    \diagram* {
      (a) -- (m1)-- (b)--(b1),
      (m1)-- [quarter left](m2),
      (m1)-- [quarter right](m2),
      (c) -- (m2)-- (d)--(d1),
      };
  \end{feynman}
\end{tikzpicture}}}
  \end{align*}
\label{fig:3ptFuncVisualSDEMelonic}
\end{figure} %
\noindent where, as before, the blobs on lines indicate full resummed propagators. %
In the IR limit, and in position space, we get
\begin{equation}\begin{aligned}\label{eq:3ptSDEnobasis}
v(\cO)(x,y,z) = %
\int_{w_1,w_2} K((x,y),(w_1,w_2))\,  v(\cO)(w_1,w_2, z)\,,
\end{aligned}\end{equation}
where, for general $q$,
\begin{align}
    K((x,y),(x_a,x_b)) = (q-1)\lambda^2 \times \half & \left[G(x,x_a)  G(y,x_b) G(x_a,x_b)^{q-2} \right. \nonumber \\ &\qquad\qquad\left. + G(x,x_b) G(y,x_a) G(x_b,x_a)^{q-2}\right]\, . \label{eqn:4ptKernel}
\end{align}
$K$
is, %
in general, a complicated function but in the conformal limit, we know exactly the form of the two-point function $G$. The resulting integral in \eqref{eqn:4ptKernel} is tractable, allowing us to determine precisely the three-point functions $v$, up to overall factors; this gives us the scaling dimensions.

\subsubsection{Conformal structures}

In a conformal theory, $v$ is simply determined by the three-point coefficients $c_{\phi\phi\cO}^a$:
\begin{equation}\begin{aligned}
\expval{\phi(x)\phi(y) \cO(z)} = \sum_a c_{\phi\phi\cO}^a \expval{\phi(x)\phi(y) \cO(z)}^a \equiv  \sum_a c_{\phi\phi\cO}^a \, v^a(\Delta,\rho)(x,y,z)\, . \label{eqn:three-pt}
\end{aligned}\end{equation}
Here the index $a$ runs over the possible conformal structures that can appear in the OPE of two scalars and an operator $\cO$ in $SO(D)$ representation $\rho$; it is a group-theoretical exercise to enumerate these \cite{Kravchuk:2016qvl}. %
The action of the kernel on each of these structures is restricted %
by conformal invariance to take the form
\begin{equation}\begin{aligned}\label{eq:3ptSDE}
\int_{w_1,w_2} K((x,y),(w_1,w_2))\,  v^a(\cO)(w_1,w_2,z) \equiv k^a{}_b\,  v^b(\cO)(x,y,z),
\end{aligned}\end{equation}
defining the matrix $k^a{}_b$ %
which is a function only of the conformal representations of $\phi$ ($\Delta_\phi$, scalar) and $\cO$ ($\Delta$, $\rho_s$).
Finding the scaling dimensions in the IR limit for the melonic theory from \eqref{eq:3ptSDEnobasis} is then identical to solving the matrix eigenvalue problem
\begin{equation}\begin{aligned}
\det(k^a{}_b - \delta^a{}_b) = 0\,.
\end{aligned}\end{equation}

For the traceless bosonic spin-$s$ operators $[\cO_{\Delta,\rho_s}]_{\mu_1 \ldots \mu_s}$ %
only one conformal structure is allowed  in \eqref{eq:3ptSDE}.
Therefore $k^a{}_b=k\,\delta^a{}_b$, and, introducing a generic null vector $\xi^\mu$,
we can project out the traceless component by considering the operator $\cO_{\Delta,\rho_s} (x;\xi)= [\cO_{\Delta,\rho_s}(x)]_{\mu_1 \ldots \mu_s} \xi^{\mu_1} \ldots \xi^{\mu_s}$, and the corresponding correlator
\begin{equation}\begin{aligned}
v^1_{\phi_1\phi_2\cO_{s}}(x_1,x_2,x_3;\xi) &=  \expval{\phi_1(x_1) \phi_2(x_2) \cO_{\Delta,\rho_s}(x_3; \xi)}  \\ &=\frac{(\mathcal{X}_3 \cdot \xi)^{s}}{(x_{12}{}^2)^{\frac{1}{2}(\Delta_1+\Delta_2-\Delta+s)}(x_{23}{ }^2)^{\frac{1}{2}(\Delta+\Delta_2-\Delta_1-s)}(x_{31}{ }^2)^{\frac{1}{2}(\Delta+\Delta_1-\Delta_2-s)}}
\end{aligned}\end{equation}
where
\begin{equation}\begin{aligned}
x_{ij} = x_i - x_j, \quad \mathcal{X}_{3 \mu}=\frac{(x_{31})_\mu}{x_{31}{ }^2} -\frac{(x_{32})_\mu}{x_{32}{ }^2}  .
\end{aligned}\end{equation}
Then we exploit the fact that \eqref{eq:3ptSDE} applies for any $z$ to take  $z\to\infty$ and work with the index-free eigenvector %
\begin{equation}\begin{aligned}
v^1_{\phi\phi\cO_{s}}(x_{12};\xi)=\lim_{x_3 \to \infty} (x_3{}^2)^{\Delta} v^1_{\phi\phi\cO_{s}}(x_1,x_2,x_3;\xi) = \frac{(\xi \cdot x_{12})^s}{\abs{x_{12}}^{2 \Delta_\phi - \Delta + s}}\, ,
\end{aligned}\end{equation}
where $\Delta_1=\Delta_2=\Delta_\phi$.

The conformal scalar IR propagator is known from the two-point function SDE \cite{Giombi:2017dtl}
\begin{equation}\begin{aligned}
 G(x)= \frac{B}{\abs{x}^{2\Delta_\phi}},\quad \Delta_\phi = \frac{D}{q}, \quad \lambda^2 B^q \equiv - \frac{1}{X_\phi X_{\tilde{\phi}}},
\end{aligned}\end{equation}
 where, for scalar operators such as $\phi$ and its shadow $\tilde{\phi}$, we define
 \begin{equation}\begin{aligned}
X_\phi &= \pi^{D/2} \frac{\Gamma\left(\frac{D}{2}-\Delta_\phi\right)}{\Gamma(\Delta_\phi)} \quad \implies \quad X_{\tilde\phi} = \pi^{D/2} \frac{\Gamma\left(\frac{D}{2}-\tilde\Delta_\phi\right)}{\Gamma(\tilde\Delta_\phi)},\\
\end{aligned}\end{equation}
is convenient, and we have defined the shadow dimension, the dimension of the shadowed operator, by $\Delta_{\tilde{\cO}} \equiv \tilde{\Delta}_\cO \equiv D-\Delta_{\cO}$; see \cite{Fraser-Taliente:2024prep} for more on shadow operators and these computations.
Using the kernel \eqref{eqn:4ptKernel} and $v^1_{\phi\phi\cO_{s}}(x_{12};\xi)$, we can then compute \eqref{eq:3ptSDE} to find that the eigenvalue condition, $\det(k^1{}_1-1) =0$,  is 

\begin{equation}\begin{aligned}\label{eqn:implicitdelta}
 1+(q-1) P_s \frac{Y_\phi(s,\Delta)}{Y_{\tilde\phi}(s,\Delta)} =0\, .%
\end{aligned}\end{equation}
Here we have defined the useful function $Y_\cO(s,\Delta)$ and the projector onto even spin:
\begin{equation}\begin{aligned}\label{eq:Yfuncdef}
Y_\cO (s,\Delta) &= X_{\cO} \, \Gamma\left(\Delta_\cO +\frac{s}{2} - \frac{\Delta}{2}\right) \Gamma\left(\Delta_\cO + \frac{s}{2} - \frac{\tilde\Delta}{2}\right)\\
P_s &\equiv \frac{1+(-1)^s}{2}\,.
\end{aligned}\end{equation}
We will often suppress the $\Delta$ dependence, writing $Y_\cO(s)$. The implicit equation \eqref{eqn:implicitdelta} has an infinite number of $\Delta$  solutions for fixed $D,s$; these can be found either perturbatively around the free theory (to any order desired) or exactly numerically. 
Note that there are no solutions for odd $s$; this is because $\phi$ is a real field.

\subsection{Extension to the \texorpdfstring{\lammelonic}{lambda-melonic} fixed point} \label{sec:bilinearsCalculationDetails}

The essence of the analysis is unchanged, as the tensor structure at the \lammelonic fixed point is identical to that of the quartic model. 
However, the computations are now complicated by (1) multiple fields, including complex fermions, (2) multiple conformal structures appearing in a given conformal correlator, and (3) mixing between conformal correlators of different types. Note also that the results will in general now depend on the ratio of fermionic/bosonic degrees of freedom, $r=2T$.

The structures compatible with conformal invariance are now
\begin{equation}\begin{aligned}
\{v^a(\cO)\} \, \equiv \, \{ \langle \phi_{abc}(x_1) \phi_{abc}(x_2) \cO(x_3) \rangle^a\} \cup \{\langle \psi_{abc}(x_1) \bar\psi_{abc}(x_2) \cO(x_3)\rangle^a\}
\end{aligned}\end{equation}
 where the operators $\cO$ with non-zero $c^a_{\phi\phi\cO}$ and $c^a_{\psi\bar{\psi}\cO}$ are precisely those operators which could appear in the OPE of $\phi \times \phi$, $\psi\times\bar{\psi}$ respectively \eqref{eqn:three-pt}.
We do not here need to consider $\langle \psi \phi \cO\rangle$, which is non-zero only for fermionic $\cO$.

\subsubsection{Eigenvector candidates}

It is convenient to work in a $v^a$ basis labelled by the free field operators in $D=3$, shown in Table \ref{table:3Dfree} grouped by their parity. Recall that we are considering operators in the traceless symmetric $s$-tensor representation in the index-free representation, and sending $x_3\to\infty$ in all correlators. 

The purely bosonic operators take the same form as before. 
The spin-0 bilinear operators made of fermions take the schematic form $\bar{\psi} (\not{\partial})^n \psi$. In  $D=3$, for odd $n=2m+1$, this is the parity-even scalar $\bar\psi(\partial^2)^m \not{\partial}\psi$; for even $n=2m$ it is the parity-odd pseudoscalar $\bar{\psi} (\partial^2)^m \psi$ \cite{Giombi:2017rhm}.

The most general non-chiral (i.e. without $\gamma_5$) three point function of two fermions and an $s=0$ scalar operator must then take the form %
\begin{equation}\begin{aligned}\label{eq:threeptpsipsibarO}
\left\langle\psi(x_1)\bar{\psi}(x_2)\mathcal{O}_\Delta(x_3)\right\rangle=\frac{c_{\phi\bar{\phi}\cO}^{P+} \left(\frac{\not{x_{12}}}{\abs{x_{12}}^2}\right)+c_{\phi\bar{\phi}\cO}^{P-} \left(\frac{\not{x_{13}}\not{x_{32}}}{\abs{x_{12}}\abs{x_{31}}\abs{x_{23}}}\right)}{\left|x_{31}\right|^\Delta\left|x_{12}\right|^{2 \Delta_\psi-1-\Delta}\left|x_{23}\right|^\Delta}.
\end{aligned}\end{equation}
Taking $x_3\to\infty$ gives the two eigenvectors $ v_{s=0}^{F\bar{F}_{\not{x}}}$ and $v_{s=0}^{F\bar{F}_{1}}(x_1, x_2)$, shown in Table \ref{table:3Dfree}. Generalising to the case of $s>0$, we find  one $\expval{\phi\phi\cO}$ and four $\expval{\psi\psi\cO}$ structures.%

\begin{table}[h!]
\begin{tabular}{L|L|L|L|L}
  D=3 \text{ symmetry} &\text{name} & \text{schematic form} & \text{eigenvector candidate}\\\hline
  &BB & \phi (\xi \cdot \partial)^s  (\partial)^{2n} \phi & \frac{(\xi \cdot x_{12})^s}{\abs{x_{12}}^{2\Delta_\phi - \tau}} & \\
  P\text{-even} &F\bar{F}_{\not{\xi}} & \bar{\psi} (\xi \cdot \partial)^{s-1} (\partial)^{2n} \not{\xi} \psi & \frac{(\xi \cdot x_{12})^{s-1} \not{\xi}}{\abs{x_{12}}^{2\Delta_\psi -1 - \tau}}& \\
  & F\bar{F}_{\not{x}} & \bar{\psi}(\xi \cdot \partial)^s  (\partial)^{2n} \not{\partial}\psi & \frac{(\xi \cdot x_{12})^s \not{x_{12}}}{\abs{x_{12}}^{2\Delta_\psi + 1 - \tau}}& \\\hline
  P\text{-odd} &  F\bar{F}_{1} & \bar{\psi}(\xi \cdot \partial)^s (\partial)^{2n} \psi & \frac{(\xi \cdot x_{12})^{s}}{\abs{x_{12}}^{2\Delta_\psi - \tau}}& \\
   &  F\bar{F}_{\not{\xi}\not{x}} & \bar{\psi}(\xi \cdot \partial)^{s-1} (\partial)^{2n} ( \not{\xi}\not{\partial}- \not{\partial}\not{\xi}) \psi & \frac{ (\xi \cdot x_{12})^{s-1} \, (\not{x}_{12} \not{\xi}-\not{\xi}\not{x}_{12} ) }{\abs{x_{12}}^{2\Delta_\psi  - \tau}} & \\
\end{tabular}
\caption{The basis of conformal structures for bilinears.}
\label{table:3Dfree}
\end{table}
Note that the $P$ symmetry constrains there to be no mixing between odd and even operators. Additionally,  the $P$-odd kernel turns out to be diagonal in the basis of \cref{table:3Dfree} so %
the $5\times 5$ kernel $k^a{}_b$ is in fact block diagonal. 

\subsubsection{Kernel}

Writing down the melonic kernel corresponds to drawing the forward two-particle scattering diagrams allowed in the melonic limit, and we find
\newcommand\kernelDiagram[6]{\vcenter{\hbox{\scalebox{0.5}{\feynmandiagram[small, layered layout, horizontal=a to b] {%
i1 -- [#1] a -- [#2] b,
i2 -- [#3] c -- [#4] d,
{ [same layer] a -- [#5, quarter right] c},
{ [same layer] a -- [#6,quarter left] c},
};}}}}
\begin{align}\label{eq:kernelBBFFbPCons}
    K((x_1, x_2), (x_a, x_b)) = \begin{pmatrix}
    \kernelDiagram{scalar}{scalar}{scalar}{scalar}{anti fermion}{fermion} & \kernelDiagram{scalar}{anti fermion}{scalar}{fermion}{scalar}{fermion}\\
    \kernelDiagram{anti fermion}{scalar}{fermion}{scalar}{scalar}{anti fermion} & \kernelDiagram{anti fermion}{anti fermion}{fermion}{fermion}{scalar}{scalar}
    \end{pmatrix} \equiv \frac{\lambda_t^2}{6}\begin{pmatrix}
        (-T)K_{BB\gets BB} & \mathbf{2}(-T)K_{BB\gets F\bar{F}}\\
        K_{F\bar{F}\gets BB} & \frac{1}{2} K_{F\bar{F}\gets F\bar{F}}
    \end{pmatrix}
\end{align}
 Clearly, the off-diagonal components mean that a $BB$-type eigenvector will generically mix with an $F\bar{F}$-type, and vice versa. Note in addition that
 \begin{enumerate}
    \item The prefactor $\lambda_t^2/6$  comes from $\lim_{N\to\infty}$ of each of the tensor contractions of two $\lambda_{IJKL}$s. 
    \item It is necessary to keep track of closed fermion loops; each generates a factor of $-T$ in the evaluation of \eqref{eqn:three-pt}. These occur when the top row of $K$ is contracted with an eigenvector, either as an explicit fermion loop in the diagram for $K_{BB\gets F\bar{F}}$ or when the external fermion lines of $K$  cap off the fermionic component in an eigenvector. 
    \item The bold factor of $2$ in front of $K_{BB-F\bar{F}}$ could go on either of the off-diagonal entries; it comes from the two ways of placing an element of the kernel to complete a fermion loop in a ladder. The loop can either match or reverse the orientation of the previous fermion loop.
    \item Though there is only one boson structure, $K_{i\gets FF}$ generically will mix between the different possible fermion structures. Indeed, for example, $k^{BB}{}_{F\bar{F}_{\not{xi}}}\neq k^{BB}{}_{F\bar{F}_{\not{x}}}$. %
\end{enumerate}
We illustrate the calculation of these $k$s via \eqref{eq:3ptSDE} by calculating $k^{BB}_{F\bar{F}_i}$, which is only non-zero for $i=F\bar{F}_{\not{\xi}},F\bar{F}_{\not{x}}$:
\begin{equation}\begin{aligned}
&\kernelDiagram{anti fermion}{scalar}{fermion}{scalar}{scalar}{anti fermion}\cdot v^{BB}(x_{ab}) = \frac{\lambda_t^2}{6}\int K_{F\bar{F}\gets BB}\cdot v^{BB}(x_{ab}) \\
&=\frac{\lambda_t^2}{6}\int_{x_a, x_b} F(x_{1a}) B(x_{ab}) F(x_{ab}) F(x_{b2})\frac{v^{BB}(x_{ab}) + v^{BB}(x_{ba})}{2}\\
&=\frac{\lambda_t^2}{6}\int_{x_a, x_b} F(x_{1a}) B(x_{ab}) F(x_{ab}) F(x_{b2}) P_s v^{BB}(x_{ab}) \equiv \sum_i k^{BB}{}_{F\bar{F}_i}\, v^{F\bar{F}_i}(x_{12}),\\
\end{aligned}\end{equation}
where we have used the fact that $v^{BB}(x_{ba})=(-1)^s v^{BB}(x_{ab})$, and in the final step decomposed the result of the integral in the basis of Table \ref{table:3Dfree}. The other calculations proceed likewise; plugging these $k$s into $\det(k-1)=0$, we can eliminate $\lambda^2 B^2 F^2$ with the two-point function solution, and $T$ via  \eqref{eq:SDEsForFermionic}; this gives an implicit equation for the scaling dimension $\Delta$ of the bilinear.

\subsection{Bilinear spectrum results for \texorpdfstring{\lammelonic}{lambda-melonic} fixed point} \label{sec:bilinearsCalculationResults}

In the following, we solve for numerically and plot as a function of $D$ the scaling dimensions of these $\mathrm{O}(N)^3$-singlet, traceless spin-$s$, bosonic bilinears -- and then discuss the results. We select the branch of the \lammelonic{} theory that descends from the free theory in $D=3$ dimensions.

\subsubsection{Eigenvalue conditions}

\noindent \textbf{Parity-even sector} Using again the projector onto even spins $P_s$, we have:
\begin{equation}\begin{aligned}\label{eq:PevenGeneralsSDcondition}
0=&\left(1-\frac{Y_{\psi }(s-1)}{Y_{\tilde{\psi }}(s-1)}\right) \left(1+P_s \frac{Y_{\phi }(s)}{Y_{\tilde{\phi }}(s)}+\frac{Y_{\psi }(s+1)}{Y_{\tilde{\psi }}(s+1)} -3P_s\frac{Y_{\phi }(s)}{Y_{\tilde{\phi }}(s)}\frac{Y_{\psi }(s+1)}{Y_{\tilde{\psi }}(s+1)} \right)\\
&+2 s \frac{Y_{\psi }(s-1)}{Y_{\tilde{\psi }}(s+1)} \left(\Delta _{\phi }+\left(D+s-2-\Delta _{\phi }\right) P_s \frac{Y_{\phi }(s)}{Y_{\tilde{\phi }}(s)}\right),\\
&\text{where } X_\psi \equiv -i \pi^{D/2} \frac{\Gamma\left(\frac{D}{2}-\Delta_\psi + \half \right)}{\Gamma(\Delta_\psi + \half)}.\\
\end{aligned}\end{equation}
We stress the different definition of $X_\psi$ for a fermionic field.

\noindent \textbf{Parity-odd sector} This is in fact diagonal in the basis of Table \ref{table:3Dfree}. 
Recalling the definition \eqref{eq:Yfuncdef} of $Y_\cO(s)$, we find:
\be \label{eq:PoddGeneralsSDcondition}
F\bar{F}_1 : \quad 0&=1+\frac{Y_\psi(s)}{Y_{\tilde\psi}(s)}, \quad \quad F\bar{F}_{\not{\xi}\not{x}} : \quad 0= 1-\frac{Y_\psi(s)}{Y_{\tilde\psi}(s)}.
\end{aligned}\end{equation}
It is clear that when $s=0$, the sector corresponding to $F\bar{F}_{\not\xi}$ decouples as expected, since $F\bar{F}_{\not\xi}$-type operators are non-local for $s=0$.

In these results, we expect a \textit{shadow symmetry} of the spectrum. That is, every physical operator of dimension $\Delta$, has a corresponding non-physical, non-local operator of dimension $\tilde{\Delta}\equiv D- \Delta$, called the shadow operator \cite{Giombi:2018qgp}. In a given CFT, we can identify which of this pair of operators is physical by analytically continuing from the free theory. Mathematically, these arise due to the Schwinger-Dyson equations having a symmetry under $\Delta \leftrightarrow \tilde\Delta$. Since $Y_{i}(s,\Delta)=Y_{i}(s,\tilde\Delta)$ \eqref{eq:Yfuncdef} is also manifestly shadow symmetric, these eigenvalue conditions are also symmetric about the line $\Delta= \frac{D}{2}$, shown in the plots below in widely dashed blue. This is similar to the shadow symmetry observed in Chew-Frautschi plots of Regge trajectories \cite{Caron-Huot:2022eqs,Simmons-Duffin:2023cftNullPlane}.

To solve \eqref{eq:PevenGeneralsSDcondition} and \eqref{eq:PoddGeneralsSDcondition} numerically, we must input the scaling dimensions of the fundamental fields. We can eliminate $\Delta_\psi$ via $2\Delta_\phi + 2\Delta_\psi =D$; however, if we want to find the spectrum for a given $D$, we must first find $\Delta_\phi$ via the two-point SDEs, \eqref{eq:SDEsForFermionic}. Of course, there are multiple branches of $\Delta_\phi$ available at each $D$; here, we use the solution branch for $\Delta_\phi$ descending from the free theory in $D=3$, marked in \cref{fig:Yuk46SDEdimPhiR4}. As noted above, although this theory exits the regime of validity at $D < 1.46$, it is nonetheless interesting to look at the continuation of this path of theories down to $D=0$; although non-unitary, they could be accessed with a modified bare scaling dimension. %

Due to the presence of $P_s$, even and odd spins are part of different Regge trajectories, and so must be analytically continued separately \cite{Chang:2021wbx}. 

\subsubsection{Scaling dimensions} \label{sec:bilinearsObservations}

\begin{figure}[h]
\centering
\includegraphics[width=0.7\textwidth]{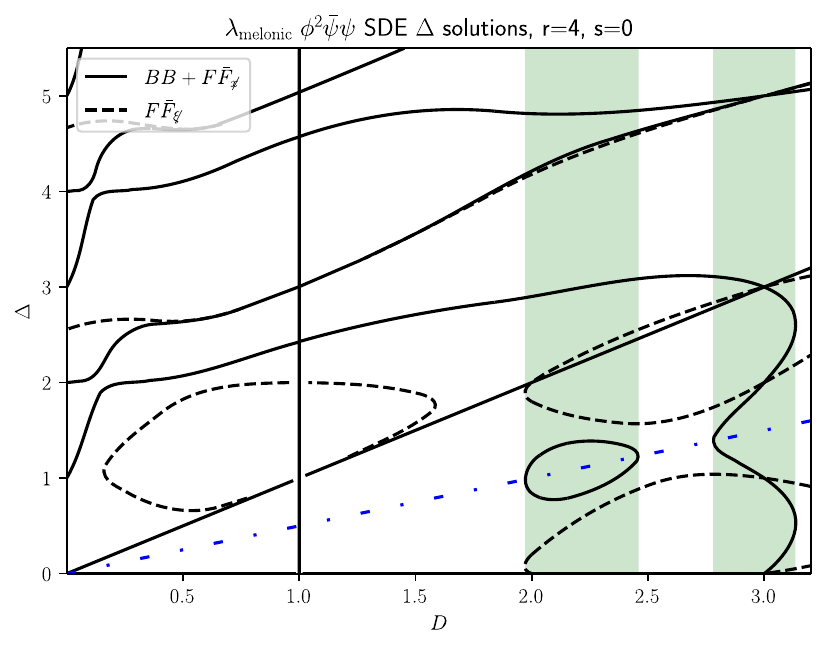}
\caption{P-even $s=0$ spectrum for $r=4$ \lammelonic.  The dimensional windows of stability, where all known local operators are real, are shown here in green. In the left-hand green window, the $F\bar{F}_{\not{\xi}\not{x}}$ (shown in \cref{fig:lmelonicSpecPoddr4S0}) has a complex scaling dimension; however, it is non-local, so  the implications for the fate of the theory are unclear. Note that the theory for $D>3$ automatically violates the unitarity bounds, since $\Delta_\phi < \frac{D-2}{2}$ (see \cref{fig:Yuk46SDEdimPhiR4}). The widely dashed blue line indicates the line of shadow symmetry. To see what happens to the disappearing solution, see the complex version of this plot in \cref{fig:lmelonicSpecPevenr4S0-3Dplot}.} %
\label{fig:lmelonicSpecPevenr4S0}
\end{figure}

\begin{figure}[h]
\centering
\includegraphics[width=0.7\textwidth]{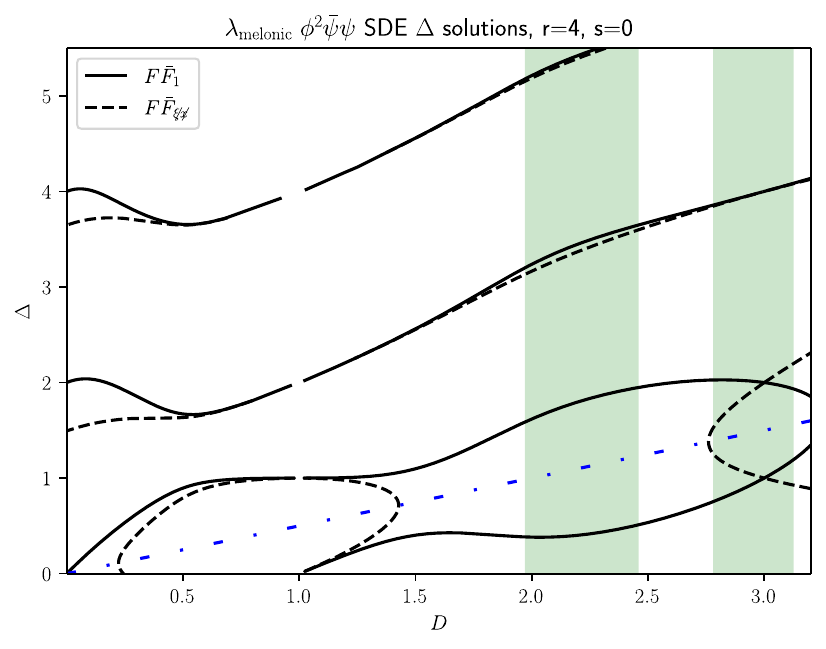}
\caption{P-odd $s=0$ spectrum for $r=4$ \lammelonic. Note that the dashed $F\bar{F}_{\not{\xi}}$ is again non-local for this value of $s$. There exist no solutions for $D=1$, hence the break in the line. We shade the $D$-range of stability given by \cref{fig:lmelonicSpecPevenr4S0} again in green.} %
\label{fig:lmelonicSpecPoddr4S0}
\end{figure}

\begin{figure}[h]
\centering
\begin{subfigure}{0.5\textwidth}
  \centering
\includegraphics[width=\textwidth]{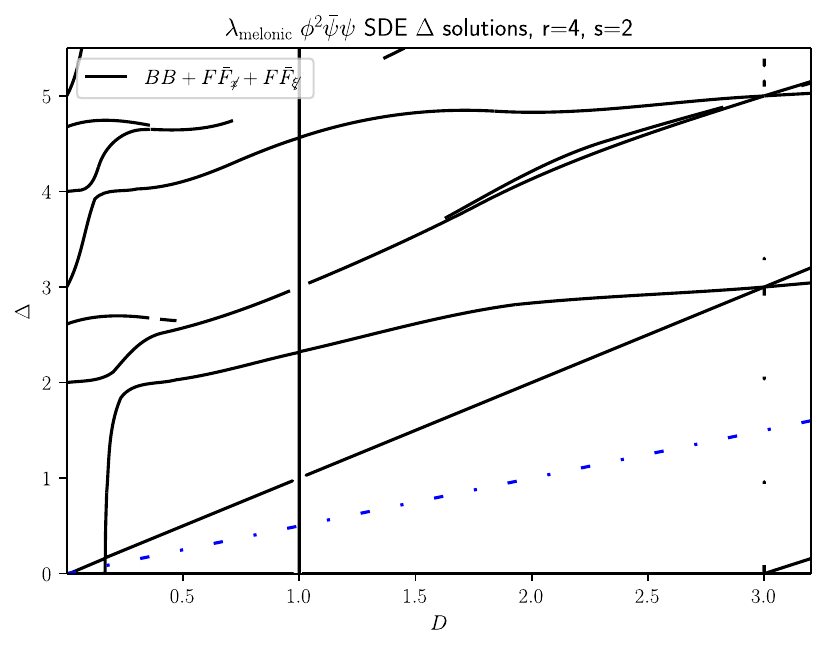}
  \caption{P-even $s=2$}
  \label{fig:lmelonicSRFCs4T2specPEven}
\end{subfigure}%
\begin{subfigure}{0.5\textwidth}
  \centering
\includegraphics[width=\textwidth]{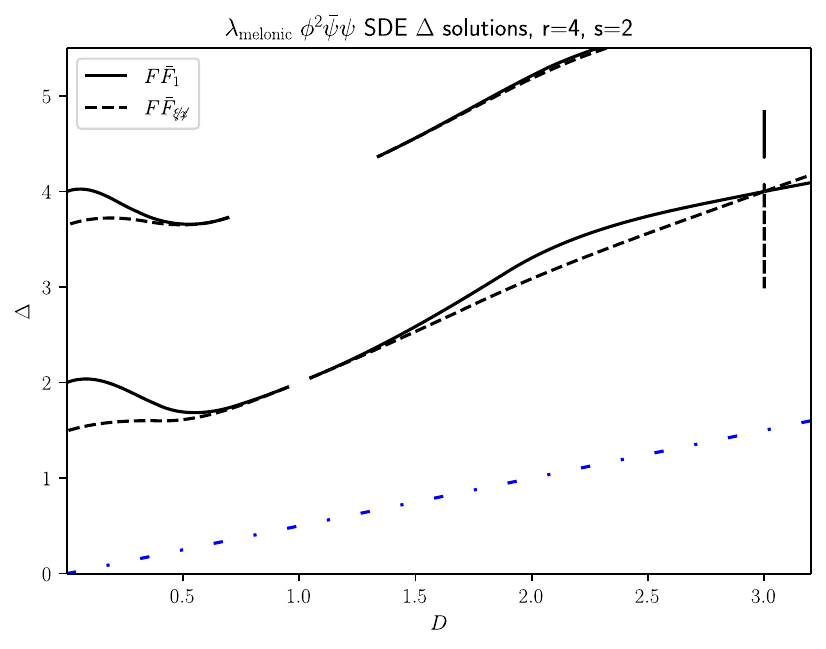}
  \caption{P-odd $s=2$}
  \label{fig:lmelonicSRFCs4T2specPOdd}
\end{subfigure}
  \caption{$s=2$ spin spectrum for $r=4$ \lammelonic{}. All operators shown here are local, including $F\bar{F}_{\xi}$, which is no longer decoupled. As usual, we show the line of shadow reflection. Note the presence of the $\Delta=D$ stress tensor in the P-even spectrum.} \label{fig:lmelonicSRFCs4T2spec}
\end{figure}

In \cref{fig:lmelonicSpecPevenr4S0,fig:lmelonicSpecPoddr4S0,fig:lmelonicSRFCs4T2spec} we display the spectrum of this theory as a function of continuous dimension $D<3$; the spectrum at a particular value of $D$ can be found by slicing the contour plot at that $D$. We restrict to the low-$\Delta$ region (around $\Delta <5$), as the high-$\Delta$ spectrum rapidly asymptotes to a known trivial form,
\begin{align}
&\text{for $BB$: }\quad \Delta=2\Delta_\phi + 2n + s+ O(1/n);\\
&\text{for $F\bar{F}_i$: }\quad \Delta = 2\Delta_\psi + 2n + s+ O(1/n).
\end{align}
 We postpone discussion of the $D=1$ region, where all $\Delta$s solve \eqref{eq:PevenGeneralsSDcondition}, and no $\Delta$s solve \eqref{eq:PoddGeneralsSDcondition}, to \cref{sec:zeroScaling}.

\subsubsection{Discussion of the spectrum}

We start by identifying the elements of the physical spectrum. This is not necessarily trivial, as discussed in appendix A of \cite{Benedetti:2020iku}; however, we can exploit the fact that near the free theory in $D=3$, the operators in the spectrum must exactly match the known scaling dimensions of the physical operators. In the P-even spectrum for $s=0$, we ignore the $F\bar{F}_{\not{\xi}}$ operator, as it is both non-local and decoupled. We can then identify in \cref{fig:lmelonicSpecPevenr4S0} that there are five physical operators. In $D=3$, we can identify them exactly: they have dimensions $\Delta=1,3,5$ and $\Delta=3,5$, and correspond to $\bar\phi (\partial^2)^{n=0,1,2} \phi$ and $\bar\psi (\partial^2)^{n=0,1} \not\partial \psi$. Hence, the solid contour leaving $\Delta=1$ is the physical scalar mass operator $\phi^2$, and the solid contour leaving $\Delta=2$ is its shadow. For $s=2$, \cref{fig:lmelonicSRFCs4T2specPEven}, all operators are physical.

In the P-odd spectrum, for $s=0$, we again ignore the non-local $F\bar{F}_{\not{\xi}\not{x}}$. We then have three physical local operators in \cref{fig:lmelonicSpecPoddr4S0}, which we can identify as $\bar\psi (\partial^2)^{n=0,1,2} \psi$. This makes it clear that, in this case, the second from bottom solid contour is the fermion mass operator $\bar\psi \psi$, with the bottom one being its non-physical shadow. For $s=2$, \cref{fig:lmelonicSRFCs4T2specPOdd}, all operators are again physical.

Note that the singlet mass operators of the two fundamental fields, $\phi^2$ and $\bar\psi\psi$, are present, so we can use them to cross-check our interpretation. Setting $D=3-\epsilon$ and expanding the eigenvalue condition for $s=0$, we find that for $\Delta=1+O(\epsilon)$ in the parity-even sector, $\Delta=2+O(\epsilon)$ for $F\bar{F}_1$ in the parity-odd sector, we exactly match the $O(\epsilon^2)$ perturbative computations of $\Delta_{\bar\phi \phi}^{\text{pert}} \equiv D+\dv{\beta_{m^2}}{m^2}|_{\lambda_*}$ and $\Delta_{\bar\psi \psi}^{\text{pert}} \equiv D+\dv{\beta_{M}}{M}|_{\lambda_*}$ at the \lammelonic fixed point \eqref{eq:perturbativeMbetaFunc}. Additionally, since this is a large-$N$ CFT, we expect the scaling dimensions of multi-trace operators to factorise. We calculated $\Delta_{h_8}  = [(\phi_{abc}\phi_{abc})^3]$ and $\Delta_{\lambda_{d_S}}  = [(\phi_{abc}\phi_{abc})(\bar\psi_{abc}\psi_{abc})]$ in \cref{sec:stabmats}, and indeed we find them to be equal to $3\Delta_{\phi^2}$ and $\Delta_{\phi^2} +\Delta_{\bar\psi \psi}$ respectively.

\subsubsection{Windows of stability} \label{sec:windowsOfStability}

\begin{figure}
\centering
\includegraphics[width=0.7\textwidth]{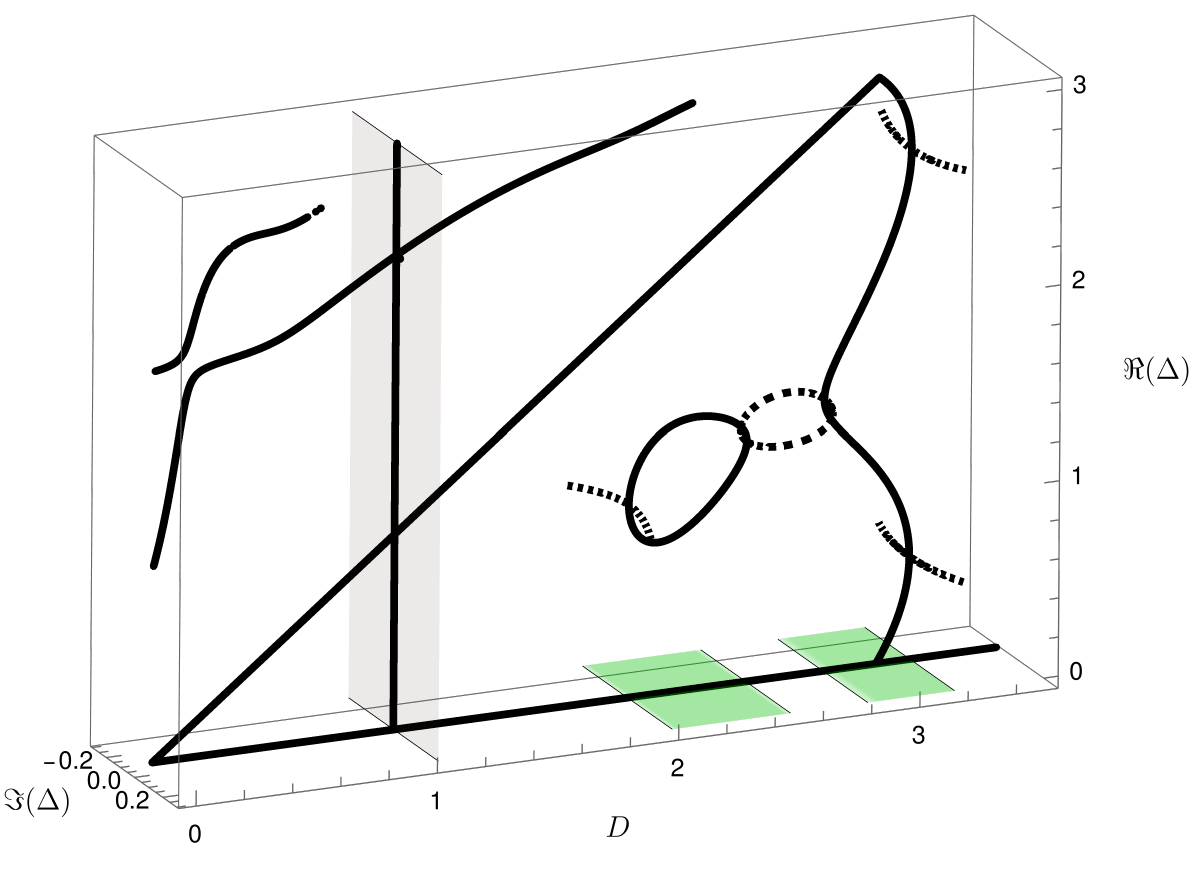}
\caption{P-even $s=0$ local spectrum for $r=4$ \lammelonic ($BB+F\bar{F}_{\not{x}}$), now including the complex solutions for $\Delta$. The real slice here precisely matches the lower part of \cref{fig:lmelonicSpecPevenr4S0}. Purely real solutions are solid, and complex solutions are dashed; all lines that seemingly end actually just continue on into the complex $\Delta$ plane. Once again, the windows of stability (with no imaginary solutions) are shown in green, and we shade $D=1$ to indicate that all complex values of $\Delta$ solve the equations. Note the collision of the $\phi^2$ operator with its shadow, giving a pair of operators with dimension $\Delta = \frac{D}{2} \pm i \alpha$.} 

\label{fig:lmelonicSpecPevenr4S0-3Dplot}
\end{figure}

It is worth first distinguishing the following three types of CFT:
\begin{enumerate}
    \item A unitary CFT with all local scaling dimensions real.
    \item A non-unitary CFT with all local scaling dimensions real.
    \item A CFT with some complex local scaling dimensions, which makes it automatically non-unitary.
\end{enumerate}
\noindent In non-integer dimension, from studies of the Wilson-Fisher fixed point, we expect all CFTs to be non-unitary due to the presence of evanescent operators \cite{Hogervorst:2015akt,DiPietro:2017vsp,Ji:2018yaf}. We see in \cref{fig:lmelonicSpecPevenr4S0} that for some ranges of dimension $D$, all the (local) scaling dimensions are real: namely, $D\in (1.97,2.46)$ and $(2.78, 3.14)$. On the boundaries of these ranges, the scaling dimensions of two operators collide, and we then obtain a pair of operators with complex conjugate scaling dimensions. This is manifest for the $D\in (1.97,2.46)$ window; in \cref{fig:lmelonicSpecPevenr4S0-3Dplot}, we see that the $\phi^2$ operator. Hence, inside the windows of stability: in integer dimension, the \lammelonic{} theory is type 1; in non-integer dimension, it is type 2. Outside the windows, it is type 3.

This phenomenon of windows of stability for the parameter values, only apparent non-perturbatively, appears to be characteristic of melonic models \cite{Giombi:2017dtl,Prakash:2017hwq,Giombi:2018qgp,Benedetti:2019eyl,Benedetti:2019rja,Kim:2019upg,Benedetti:2020iku,Chang:2021wbx,Prakash:2022gvb} (the windows in $D$ are demonstrated for \hprismatic{} in \cref{fig:Bosonic6SDEdimPhi}). However, this behaviour depends strongly on the details of the theory -- in this case, $r$ and $D$: for example, for $r=8$, the $D\in (1.97,2.46)$ window visible in \cref{fig:lmelonicSpecPevenr4S0} disappears entirely. Additionally, note that $F\bar{F}_{\not{\xi}\not{x}}$ remains complexified throughout the $(1.97,2.46)$ window; however, it is a non-local operator and its interpretation is unclear.

The complex scaling dimension of this form means that the critical points along this line are unstable -- the conformal vacuum is not the true vacuum of the precursor QFT; for this reason, we term these as windows of stability. It was proven in \cite{Benedetti:2021qyk} that the presence of an operator $\cO$ of the form $\Delta=D/2 + i \alpha$ in the OPE of two fundamental scalar fields means that the conformal solution is unstable; by the AdS/CFT duality, this is the CFT counterpart of the well-known Breitenlohner-Freedman bound in AdS: the dual field of the operator has a mass below the BF bound. It is conjectured that the fate of these theories after this instability that in the true vacuum, this operator acquires a VEV, $\expval{\cO}\neq 0$ \cite{Kim:2019upg,Benedetti:2021qyk}.

It is interesting to observe that in the standard bosonic $\mathrm{O}(N)^3$ quartic and sextic models (as well as in generalisations and extensions of them) it is also the operator that roughly corresponds to $\phi^2$ that receives a complex scaling dimension along with its shadow \cite{Giombi:2017dtl,Benedetti:2019rja} when descending from the free theory; note that in all these cases, this is the physical bilinear operator of lowest scaling dimension, with $\Delta < D/2$. %

Despite the non-unitarity of the CFTs, it is easy to check that for integer spins $s$, the physical operators identified here all satisfy the CFT unitarity bounds \cite{Benedetti:2023mli} for $D<3$:
\begin{equation}
    \Delta_{s=0} \ge \frac{D}{2}-1,\quad  \Delta_{s \neq 0}\ge D+s-2
\end{equation}
There is one exception: for $D=1$, every scaling dimension is a solution of \eqref{fig:lmelonicSpecPevenr4S0}. However, this dimension is pathological in any case, as we will discuss in \cref{sec:zeroScaling}. For $D>3$, of course, $\phi$ itself is non-unitary.

\subsubsection{Operators of protected dimension} \label{sec:protectedOperators}

We frequently observe operators of protected dimension; that is, operators which have scaling dimension fixed at $D$ or $D-1$. These are usually associated to some symmetry.

In the P-even $s=0,2$ sectors, there are always operators with dimension $\Delta=D$. The $s=2$ operator corresponds to the conserved stress tensor. The $s=0$ operator is a vanishing mixture of $\phi \, \partial^2\,  \phi$ and $\bar\psi \not{\partial}\psi$ -- this occurs because, by the equations of motion, both are proportional to $\phi^2 \bar\psi \psi$; hence, a linear combination of them vanishes as an operator \cite{Bulycheva:2017ilt}. Note that this operator is unrelated to the trace of the stress tensor (which is of course zero). In the case of the generalised SYK model, Gross interpreted this as coming from the IR rescaling symmetry \cite{Gross:2016kjj}; in our case, this is the IR local rescaling symmetry $\psi \to A \psi, \phi \to A^{-1} \phi$, for a local function $A(x)\neq 0$. However, in the generalised SYK model, its OPE coefficient with the fundamental fields was found to vanish; we suspect that the same would apply here. %

In the P-even spin-$1$ sector, we find a bilinear that involves only the fermions (because $P_{s=1}=0$), with $\Delta = D-1$; this corresponds to the conserved $\mathrm{U}(1)$ current that rotates the complex fermions, $\bar\psi \gamma^\mu \psi$. Like the rescaling symmetry, this generically conserved current is, in the deep IR, upgraded to a local $\mathrm{U}(1)$ symmetry, just as in the SYK model and melonic supertensor models \cite{Davison:2016ngz,Chang:2018sve}: $\psi(x) \to \psi(x) e^{i \alpha(x)}$. In the P-odd spin-$1$ sector, we also find an $F\bar{F}_{\not{\xi}\not{x}}$-type operator with $\Delta = D$; the associated symmetry is not clear. The $s\neq 0, 2$ sectors are otherwise uninteresting, and so not plotted.

\subsubsection{Spectrum divergences when fundamental scaling dimensions hit zero} \label{sec:zeroScaling}

In exactly $D=1$ dimensions, there is no solution giving $\Delta_\phi=1/2$. However, perturbatively, in $D=1+\epsilon$, we find a solution $\Delta_\phi = \half- O(\epsilon^2)$, $\Delta_\psi =0 + O(\epsilon^2)$, for arbitrary $r$. These are marked with black squares in \cref{fig:Yuk46SDEdimPhiBothRs}. If we ignore this, and take proceed anyway, taking $\epsilon \to 0$, we find that every value of $\Delta$ solves the P-even determinant condition, for any value of $s$. For the P-odd family, the determinant condition evaluates to $-1=0$ for $D=1$, and so there is a break visible in every contour along $D=1$. This missing tower of operators is just as in the $D=2$ case of the standard bosonic tensor model \cite{Giombi:2017dtl}: here we have perturbative solutions for $D=1+\epsilon$, $\Delta = 2n+ O(\epsilon)$, which do not exist at $D=1$.
 
\begin{figure}
\centering
\includegraphics[width=0.7\textwidth]{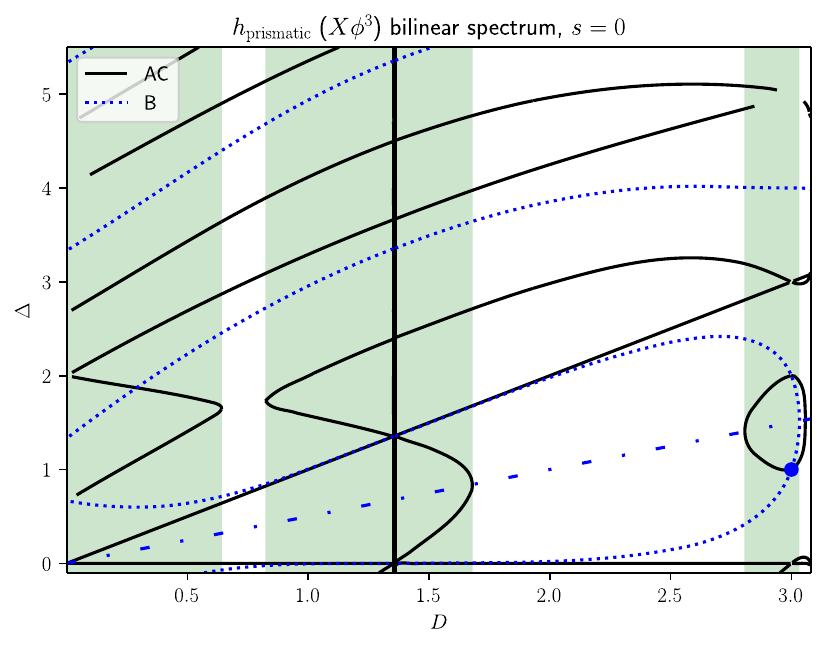}
\caption{Sextic prismatic model, \hprismatic, $s=0$ scaling dimensions \cite{Giombi:2018qgp} for the branch indicated in \cref{fig:Bosonic6SDEdimPhi}. Green shading again indicates the regions of stability; the line of shadow reflection is blue and dashed. For $D\simeq 1.353$, every $\Delta$ solves the eigenvalue equation for AC-type bilinears, and no $\Delta$s solve the B-type. The mass operator $\phi^2$ in the $D=3$ free theory is marked with a blue dot; just as with \lammelonic, it is this operator that obtains a complex scaling dimension at $D\simeq 2.81$. The bilinear $X\phi$, by contrast, is the first to obtain a complex scaling dimension if we increase $D$ to $D\simeq 3.03$.} 
\label{fig:hprismaticSpectrum}
\end{figure}

The same phenomenon also occurs for other theories when the scaling dimension of one of the fields $\phi$ hits certain integer values. Returning to the prismatic $\phi^6$ model \cite{Giombi:2018qgp}: following the family of CFTs descending from $D=3-\epsilon$ indicated in \cref{fig:Bosonic6SDEdimPhi}, while still within the IR wedge we reach $\Delta_\phi =0$ at $D_C \simeq 1.353$\footnote{$D_C$ solves $\psi ^{(0)}(D)+\gamma_E=\psi ^{(0)}(\frac{D}{2})-\psi ^{(0)}(-\frac{D}{2})$, with $\psi^{(0)}(x)=\mathrm{Polygamma}[0,x]$ and $\gamma_E$ Euler's constant.}. As shown in  \cref{fig:hprismaticSpectrum}, for any spin, every value of $\Delta$ solves the A/C-type bilinear equations there, and no values of $\Delta$ solve the B-type bilinear equations.

Likewise, consider the other branches of \lammelonic in \cref{fig:Yuk46SDEdimPhiBothRs} -- in particular, those that descend from the free theory in $D=2n+1-\epsilon$, when they hit $D=2n-1$, for $n \in \mathbb{N}$. At those points\footnote{It is worth noting, however, that the $D=1$ evaluation of the $\Delta_\phi=+1/2$ solution branch was doubly incorrect, as at precisely this point $\Delta_\psi =0$. This is the same as the free scaling dimension $\frac{D-1}{2}$ in the two-point SDE \eqref{eq:fermionSDEWithoutX}, so in exactly $D=1$ we were not permitted to drop it in the IR.}, we have $\Delta_\phi = (2n-1)/2 + O(\epsilon^2)$ exactly, and hence $\Delta_\psi = 0+ O(\epsilon^2)$. Likewise, along the line $\Delta_\phi =0$, we also see breaks in the contours. 
Indeed, we expect this behaviour to be generic for any melonic-type CFT with multiple fundamental fields -- of course, if there is only one field, $\Delta_\phi =0$ only in $D=0$. One exception to this, in the case of only one field, is in the supersymmetric $\cN=1$ quartic $\mathrm{O}(N)^q$ model, which, as described in \cref{sec:qGenAndSUSYprismatic}, has $\Delta_\Phi = (D-1)/q$; thus, taking $\cN=1$ theory \cite{Popov:2019nja} we do indeed observe in $D=1$: a missing tower of operators for the BB and FF-type operators; every value of $\Delta$ satisfies the eigenvalue condition for BF-type operators. %

Despite the presence of these spectrum divergences, there is evidence to suggest that the theories considered in the divergent dimensions may still have the missing states in the spectrum. \cite{Benedetti:2019ikb} used a long-range tensorial $g \phi^4$ model, which exhibits a break in the scaling dimension solutions as $D\to 2$ for arbitrary values of the marginal coupling: a character decomposition derivation of the spectrum for the free theory suggested that those states which are seem to be absent are in fact still present. By continuity of the spectrum in $g$, this would also hold for the short-range models. If true, this would also apply to the other $\phi^q$ models, each of which have no solutions to the kernel for $D= \mathbb{N}\frac{2q}{q-2}$.

\section{Outlook} \label{sec:outlook}

We have obtained a set of novel interacting conformal field theories, which generalise and extend the known set of higher-dimensional tensor field theories to include fermions, and have considered the RG flow relations between them. 

These theories can be broadly classed within the existing categories of melonic and prismatic, but provide the novel feature of incorporating both within a single model; we also include fermions, which provide additional structure to the fixed point network. The melonic limit is seen to be more fundamental, as it can be described by having only melonic diagrams where all fields are fundamental. The prismatic field instead allows only melonic diagrams where some propagators in the melons are in fact the propagators of non-dynamical auxiliary fields. At the perturbative level, without an auxiliary field, we can access both of these fixed points: the diagrams contributing to each are identical in terms of momentum structure, but differ in tensorial structure. However, if we want to find the prismatic fixed points using the Schwinger-Dyson equations (non-perturbatively), we are required to introduce the auxiliary fields.

While studying the perturbative RG flow between these theories, we identified a candidate line of fixed points perturbatively around $D=3$, and established the non-perturbative existence of the lines of fixed points in general dimension. Using the simple multi-field model of just scalars, the \hprismatic fixed point, we established in \cref{sec:generalCharacteristicsPrismatic,sec:bilinearsCalculationResults} the various elements that enter the analysis of a melonic CFT. Namely: the IR wedge; the infinite branches of IR solutions in general dimension (including collision leading to complex solutions); the apparent breakdown of the conformal analysis at exceptional values of $D$ and $\Delta_\phi$, and the associated breaks in the contours for both the two-point and four-point functions at those exceptional values. We expect these features to be generic for melonic CFTs, and indeed saw them reproduced in the \lammelonic CFT.

Understanding the generic features of the melonic CFTs given above at a deeper level is clearly essential. To that end, further avenues of exploration are suggested by the following:
\begin{enumerate}[noitemsep]
    \item We have focused here on the branch of \lammelonic descending from the free theory in $D=3-\epsilon$. What is described by the other branches of solutions, typically lying outside the IR wedge, of which there are an infinite number for almost every $D$?
    \item Can the fate of the theory in the dimensions that exhibit the spectral divergences mentioned in \cref{sec:zeroScaling} be understood? How does the logarithmic character of the two-point function modify the true IR physics?
    \item We have fixed points that exist only perturbatively around exceptional dimensions, $D=D_0 + \epsilon$, but not for $\epsilon=0$. These exist in the short and long-range models, and there are hints that we should regard these fixed points in $D=D_0$ as the limit of $\epsilon \to 0$, but their nature remains unclear \cite{Benedetti:2019ikb}.
    \item In \cref{sec:stabmats}, we found that in $D=3-\epsilon$, \hmelonic and \hlammelonic appear to sit on a line of fixed points in the direction of $-\hp + 3 h_5 - 3 h_7 + h_8$. Can this line be understood or eliminated without resorting to six-loop computations? 
    \item The 4D Yukawa model is not accessible with rank-$3$ tensors; is there a tractable tensorial $\mathrm{O}(N)^r$ realisation of a melonic 4D Yukawa model? Such a model would enable comparison of the 4D Yukawa model with the $3$D result presented here. The non-perturbative singlet spectrum has been studied already via a disorder realisation in \cite{Prakash:2022gvb}.
    \item We have tacitly assumed that there is no symmetry breaking in the IR, whether of the $\mathrm{O}(N)^3$, $\mathrm{U}(1)$, or conformal symmetry. This is despite the clear indication in \cref{sec:windowsOfStability} of dimensions in which the theory is unstable \cite{Benedetti:2021qyk}. By loosening these assumptions, can we understand the true IR phase? One particular symmetry breaking pattern has been studied in \cite{Benedetti:2019sop}, but this does not apply, for example, to melonic theories realised via disorder.
    \item A number of generalisations of standard melonic theories that exist in the literature could be pursued with this model, due to the presence of fermions in a higher-dimension CFT under good analytic control: for example, consideration of the CFT induced on a defect in this theory, along the lines of \cite{Popov:2022nfq,Gimenez-Grau:2022ebb}; or consideration of the long-range version of this model \cite{Gross:2017vhb,Benedetti:2019rja,Harribey:2021xgh,Shen:2023srk}.
    \item A number of general field-theoretical questions are suggested by this model: The short-range quartic tensor model at its fixed point -- with imaginary tetrahedral coupling -- was found to be asymptotically free in the large-$N$ limit \cite{Berges:2023rqa}, giving access to a strongly coupled theory where analytic progress can be made. The long-range bosonic model was used to verify the F-theorem, even in the absence of unitarity, in \cite{Benedetti:2021wzt}. To what extent can these behaviours be extended to the other theories in this class, such as this one? The presence of fermions complicates the computations, due to the additional $\SO(D)$ representations.
    \item Finally: the bulk dual of these melonic theories remains unknown \cite{Harribey:2022esw}. However, because of the large numbers of a vast number of gauge-invariant operators involving higher powers of the tensor field, the gravity dual is expected to be complicated \cite{Klebanov:2016xxf, Rosenhaus:2018dtp}.
\end{enumerate}

\acknowledgments
LFT is supported by a Dalitz Scholarship from the University of Oxford and Wadham College.

For the purpose of open access, the authors have applied a CC BY public copyright licence to any Author Accepted Manuscript (AAM) version arising from this submission.

\appendix

\section{Potential for the tensor model} \label{app:potential}

The full analytic expression for the potential at optimal $N$ scaling \cite{Benedetti:2023mli} is as follows.
We have normalised each tensor term such that it has overall weight one, i.e. if $N=1$, then $V=\frac{\sum_i\lambda_i}{2} \phi^2 \bar{\psi}\psi +  \frac{\sum_i h_i}{6!}\phi^6$.
For compactness, we present the equations that are not symmetrised (with weight one) under permutation of the bosonic indices; to use these formulae, a full calculation the symmetrised versions must be used\footnote{This symmetrisation is straightforward in each case: on one example $\lambda$ term, we do $r_k^j r_l^i b_k^j b_l^i g_k^j g_l^i \mapsto \frac{1}{2}(r_k^j r_l^i b_k^j b_l^i g_k^j g_l^i + (i \leftrightarrow j))$; on an example $h$ term, $r_j^i r_m^l r_n^k b_l^k b_m^i b_n^j g_l^k g_m^j g_n^i \mapsto \frac{1}{6!}(\text{same} + (6!-1) \text{ perms})$. }. We also use as shorthand $r^i_j = \delta_{i_r j_r}$ (likewise for $g$ and $b$).
\begin{equation}\begin{aligned}\label{eq:explicitPotential}
&\lambda_{(i_r i_g i_b)(j_r j_g j_b)(k_r k_g k_b)(l_r l_g l_b)} =\Big[\\
& +\frac{\lambda _t}{3 N^{3/2}} \left(r_j^i r_l^k b_k^i b_l^j g_k^j g_l^i+r_k^i r_l^j b_j^i b_l^k g_k^j g_l^i+r_k^i r_l^j b_k^j b_l^i g_j^i g_l^k\right)\\
 & +\frac{\lambda _{\text{dD}}}{N^3} \left(r_k^j r_l^i b_k^j b_l^i g_k^j g_l^i\right)\\
 & +\frac{\lambda _{\text{dS}}}{N^3} \left(r_j^i r_l^k b_j^i b_l^k g_j^i g_l^k\right)\\
 & +\frac{\lambda _{\text{pE}}}{3 N^2} \left(r_j^i r_l^k b_j^i b_l^k g_k^j g_l^i+r_j^i r_l^k b_k^j b_l^i g_j^i g_l^k+r_k^j r_l^i b_j^i b_l^k g_j^i g_l^k\right)\\
 & +\frac{\lambda _{\text{pO}}}{3 N^2} \left(r_k^i r_l^j b_k^i b_l^j g_k^j g_l^i+r_k^i r_l^j b_k^j b_l^i g_k^j g_l^i+r_k^j r_l^i b_k^i b_l^j g_k^j g_l^i\right)\\
 & + \frac{\lambda _{\text{pS}}}{3 N^2} \left(r_j^i r_l^k b_k^j b_l^i g_k^j g_l^i+r_k^j r_l^i b_j^i b_l^k g_k^j g_l^i+r_k^j r_l^i b_k^j b_l^i g_j^i g_l^k\right)\Big]_{\text{symmetrised on $i\leftrightarrow j$}}\\
&h_{(j_r j_g j_b)(k_r k_g k_b)(l_r l_g l_b)(m_r m_g m_b)(n_r n_g n_b)} = \Big[\\
 &+\frac{h_1}{N^3} \left(r_j^i r_m^l r_n^k b_l^i b_m^k b_n^j g_l^k g_m^j g_n^i\right)\\
 &+\frac{h_2}{N^3} \left(r_l^j r_m^i r_n^k b_l^i b_m^k b_n^j g_l^k g_m^j g_n^i\right)\\
 &+\frac{h_3}{3 N^{7/2}} \left(r_j^i r_m^l r_n^k b_l^k b_m^i b_n^j g_l^k g_m^j g_n^i+r_k^i r_m^l r_n^j b_l^i b_m^k b_n^j g_l^k g_m^j g_n^i+r_k^j r_m^l r_n^i b_l^i b_m^k b_n^j g_l^k g_m^j g_n^i\right)\\
 &+\frac{h_4}{3 N^4} \left(r_l^i r_m^k r_n^j b_l^i b_m^k b_n^j g_l^k g_m^j g_n^i+r_l^i r_m^k r_n^j b_l^k b_m^j b_n^i g_l^k g_m^j g_n^i+r_l^k r_m^j r_n^i b_l^i b_m^k b_n^j g_l^k g_m^j g_n^i\right)\\
 &+\frac{h_5}{3 N^4} \left(r_l^i r_m^k r_n^j b_l^k b_m^i b_n^j g_l^k g_m^j g_n^i+r_l^k r_m^i r_n^j b_l^i b_m^k b_n^j g_l^k g_m^j g_n^i+r_l^j r_m^k r_n^i b_l^k b_m^i b_n^j g_l^k g_m^j g_n^i\right)\\
 &+\frac{h_6}{N^{9/2}} \left(r_j^i r_l^k r_n^m b_l^k b_m^i b_n^j g_l^k g_m^j g_n^i\right)\\
 &+\frac{h_7}{3 N^5} \left(r_l^k r_m^i r_n^j b_l^k b_m^i b_n^j g_l^k g_m^j g_n^i+r_l^k r_m^i r_n^j b_l^k b_m^j b_n^i g_l^k g_m^j g_n^i+r_l^k r_m^j r_n^i b_l^k b_m^i b_n^j g_l^k g_m^j g_n^i\right)\\
 &+\frac{h_8}{N^6} \left(r_l^k r_m^j r_n^i b_l^k b_m^j b_n^i g_l^k g_m^j g_n^i\right)\Big]_{\text{symmetrised on $S_6$}}
\end{aligned}\end{equation}
The unpleasantness of the notation should make manifest the reason for the visual shorthand. %

\subsection{Potential comparison} \label{app:potentialcomparison}
To ease comparison with the other sextic models of the literature, we directly compare our real bosonic potential sector, $\{h_i\}$, to the \textit{complex} bosonic sextic potential in \cite{Benedetti:2019rja} ($\{\boldsymbol{\lambda}_i (\bar{\phi}\phi)^3\}$), and the real bosonic sextic potential in \cite{Giombi:2018qgp} ($\{\boldsymbol{g}_i \phi^6\}$). 

If we make the fermion sector free ($\lambda_i =0$), and restrict to the couplings that descend from $\mathrm{U}(N)$-invariant couplings (see \ref{app:potential}), we obtain from our beta functions exactly the large-$N$ results of the complex sextic melonic model, up to a rescaling $h= 15 \boldsymbol{\lambda}/s^3$. This is despite their model being complex, and so in theory having double the number of degrees of freedom.%

Since all $O \in \mathrm{O}(N)$ also satisfy $O \in \mathrm{U}(N)$, as $O^\dagger = O^T =O^{-1}$, we note that all possible $\mathrm{U}(N)^3$ interaction terms are manifestly also invariant under $\mathrm{O}(N)^3$. Thus, the $\mathrm{U}(N)^3$ interaction terms are a subset of the $\mathrm{O}(N)^3$ interaction terms, so we can identify:

\begin{center}
\begin{tabular}{|c|c|c|c|}
  \hline
  \textbf{Real sextic} & \textbf{Complex sextic $ (\bar{\phi}\phi)^3$} & \textbf{Real prismatic $\phi^6$} \\
  \hline
  $h_p/N^3$ & Forbidden by $\mathrm{U}(N)$ & $\boldsymbol{g}_1/N^3$ \\
  \hline
  $h_w/N^3$ & $\boldsymbol{\lambda}_1/N^3$ & $\boldsymbol{g}_2/N^5$ \\
  \hline
  $h_3/N^{7/2}$ & Forbidden by $\mathrm{U}(N)$ & $\boldsymbol{g}_5/N^4$ \\
  \hline
  $h_4/N^4$ & $\boldsymbol{\lambda}_3/N^4$ & $\boldsymbol{g}_4/N^5$ \\
  \hline
  $h_5/N^4$ & $\boldsymbol{\lambda}_2/N^4$ & $\boldsymbol{g}_3/N^4$ \\
  \hline
  $h_6/N^{9/2}$ & Forbidden by $\mathrm{U}(N)$ & $\boldsymbol{g}_6/N^5$ \\
  \hline
  $h_7/N^5$ & $\boldsymbol{\lambda}_4/N^5$ & $\boldsymbol{g}_7/N^5$ \\
  \hline
  $h_8/N^6$ & $\boldsymbol{\lambda}_5/N^6$ & $\boldsymbol{h}_8/N^7$ \\
  \hline
\end{tabular}
\end{center}
The prismatic models have picked non-optimal scalings and therefore miss some physics.

\section{Perturbative beta functions}
\subsection{Notes on the calculation} \label{app:calcNotes}
\subsubsection{Gamma matrices in 2+1D} \label{app:gammaMatrices}

Our particles here are 3D Dirac spinors, with 2 complex, and so 4 real, components. The gamma matrices in $(2+1)D$, with signature $(-++)$, can be the Pauli matrices, with up to a factor of $i$ for $\gamma^0$:
\begin{equation}
\gamma^0 = i\sigma_1= \left(
\begin{array}{cc}
 0 & i \\
 i & 0 \\
\end{array}
\right),\gamma^1 = \sigma_2 = \left(
\begin{array}{cc}
 0 & -i \\
 i & 0 \\
\end{array}
\right),\gamma ^2 = \sigma_3 =\left(
\begin{array}{cc}
 1 & 0 \\
 0 & -1 \\
\end{array}
\right)
\end{equation}
In precisely three dimensions, the trace rules are the following:
\begin{equation}\begin{aligned}
\Tr[\gamma^\mu] &=0, \qquad \Tr[\gamma^\mu \gamma^\nu] = 2 \eta^{\mu\nu},\\
\Tr[\gamma^\mu \gamma^\nu \gamma^\rho] &= -2\epsilon^{\mu\nu\rho}, \qquad \epsilon^{\mu\nu\rho} \text{ s.t. } \epsilon^{012} = +1,
\end{aligned}\end{equation}
with $\epsilon^{\mu\nu\rho}$ the alternating tensor with the normalisation indicated. However, it is well known that DReD is algebraically inconsistent; in particular, because different contractions of three or more $\epsilon^{\mu\nu\rho}$ factors yield different results in $D<3$ \cite{Siegel:1980qs}. Our computation of the beta functions in $D=3-\epsilon$ did not suffer from this as we have no more than one momentum at any point in our computation (as we renormalised the couplings at zero momentum), and thus no epsilon factors can ever occur, because we can always reduce any trace using the identity $\not{p}\not{p}=p^2$. Thus, as long as we take traces only at the end, no problems can arise.

\subsubsection{Massive melonic integrals in the epsilon expansion} \label{app:massiveMelonicIntegrals}
We give here the fermionic sunrise integral, in full generality of masses and momenta, along with conventions. The standard Lorentzian sunrise integral, with $(+--)$ conventions, in $D=3-\epsilon$ can be found in the literature \cite{Rajantie:1996np}:
\begin{equation}
\begin{split}
&\mu^{2\epsilon}\int\frac{\mathrm{d}^D k\, \mathrm{d}^D l}{(2\pi)^{2D}} \left(\frac{1}{(k^2-m_1^2+i \eta ) (l^2-m_2^2+i \eta ) (k+l-p)^2-m_3^2+i \eta )}\right),\\
&\frac{1}{32 \pi ^2}\left(\frac{1}{\epsilon }+\log \left(\frac{\bar{\mu }^2}{M^2-p^2}\right)-\frac{2M \tanh ^{-1}\left(\frac{p+0i}{M}\right)}{p}+3+O(\epsilon ^1)\right)
\end{split}
\end{equation}
where we have defined the usual $\overline{MS}$ parameter, $\bar{\mu}^2 = e^{-\gamma_E} 4 \pi \mu^2$, and the mass sum $M\equiv m_1+m_2+m_3$. Next, including fermions:
\begin{equation}
\begin{split}
&\mu^{2\epsilon}\int\frac{\mathrm{d}^{D} k\, \mathrm{d}^{D} l}{(2\pi)^{2D}} \left(\frac{k\cdot l}{(k^2-m_1^2+i \eta ) (l^2-m_2^2+i \eta ) (k+l-p)^2-m_3^2+i \eta )}\right)\\
&=\frac{1}{192\pi^2} \Biggl(\frac{p^2-3 \left(m_1^2+m_2^2-m_3^2\right)}{\epsilon} + \left(p^2 -3 (m_1^2+ m_2^2-m_3^2)\right) \log \left(\frac{\bar{\mu }^2}{M^2-p^2}\right) \\
&\quad +\frac{2 \left(3 m_3 \left(m_1^2+m_2^2-p^2\right)-m_3^3+2 \left(m_1^3+m_2^3\right)\right) \tanh ^{-1}\left(\frac{p+0i}{M}\right)}{p}\\
&\quad -7 m_1^2-7 m_2^2+5 m_3^2-2 m_1 m_2+4 \left(m_1+m_2\right) m_3+3 p^2 +O(\epsilon^1)\Biggr)
\end{split}
\end{equation}
and 
\begin{equation}
\begin{split}
&\mu^{2\epsilon}\int\frac{\mathrm{d}^{D} k\, \mathrm{d}^{D} l}{(2\pi)^{2D}} \left(\frac{k\cdot p}{(k^2-m_1^2+i \eta ) (l^2-m_2^2+i \eta ) (k+l-p)^2-m_3^2+i \eta )}\right),\\
&=\frac{1}{96\pi^2} \Biggl(\frac{p^2}{\epsilon }+ p^2 \log \left(\frac{\bar{\mu }^2}{M^2-p^2}\right)\\
&\quad +\frac{\left(\left(m_2+m_3\right)^3-3 \left(m_2+m_3\right) p^2-2 m_1^3-3 \left(m_2+m_3\right) m_1^2\right) \tanh ^{-1}\left(\frac{p+0i}{M}\right)}{p}\\
&\quad +\left(2 m_1-m_2-m_3\right)M+3 p^2+O(\epsilon^1)\Biggr)
\end{split}
\end{equation}
We have specified the branch of $\mathrm{arctanh}$ that we take for $p = \sqrt{p^2} >M=m_1+m_2+m_3$. All divergent $2$-loop integrals in $3-\epsilon$ dimensions can be obtained from variations on this theme, although integration by parts identities may be required for the more non-trivial ones (say, with a numerator of $(k\cdot p) (l\cdot p)$).

\subsection{Full beta functions} \label{app:betas}

We present the full beta functions for the tensorial $\phi^2 \bar{\psi} \psi$ model at leading order in $N$, with all marginal couplings set to zero. We use $s=1/(8\pi)$ to indicate the order calculated to, since each $\lambda$ comes with an $s$ and each $h$ comes with an $s^2$. Plugging \eqref{eq:explicitPotential} into \eqref{eq:generalVectorResults}, we obtain:
{\scriptsize
\begin{align*}
\beta \left(\lambda_t\right)&=-\epsilon  \lambda_t+\frac{1}{9} s^2 (T+1) \lambda_t^3+s^4 \left(\left(\frac{h_p^2 + 3 h_w^2}{16200}\right) \lambda_t+\frac{1}{486} (-T (11 T+26)-2) \lambda_t^5\right)+O(s^5)\\
\beta \left(\lambda_{d_S}\right)&=-\epsilon  \lambda_{d_S}+\frac{2}{9} s^2 (T+1) \lambda_t^2 \left(\lambda_{p_E}+2 \lambda_{d_S}\right)\\
    & + s^4 \left(\left(h_p^2+3 h_w^2\right)\frac{ 3\lambda_{p_E}+4 \lambda_{d_S}}{4050}-\frac{2}{243} \lambda_t^4 \left(18 \lambda_{p_E}+23 \lambda_{d_S}+5 T^2 \lambda_{d_S}+56 T \lambda_{d_S}+39 \lambda_{p_E} T\right)\right)+O(s^5)\\
\beta \left(\lambda_{d_D}\right)&=-\epsilon  \lambda_{d_D}+\frac{1}{9} s^2 \lambda_t^2 \left((T+7) \lambda_{d_D}+4 \left(\lambda_{p_O}+\lambda_{p_S}\right)\right)\\
    &+ \frac{s^4}{48600} \left(\splitfrac{
(3h_p^2 + 9h_w^2 - 540h_p\lambda_t^2 - 11000\lambda_t^4 - 1100T^2\lambda_t^4 - 11600T\lambda_t^4)\lambda_{d_D}}{-(480h_p\lambda_t^2 + 8400\lambda_t^4 + 6000T\lambda_t^4)(\lambda_{p_O}+\lambda_{p_S})
}\right)+O(s^5)\\
\beta \left(\lambda_{p_E}\right)&=-\lambda_{p_E} \epsilon +\frac{2}{9} \lambda_{p_E} s^2 (T+1) \lambda_t^2+s^4 \left(\frac{h_p^2 \lambda_{p_E}}{4050}+\frac{h_w^2 \lambda_{p_E}}{1350}-\frac{2}{243} \lambda_{p_E} (T (5 T+17)+5) \lambda_t^4\right)+O(s^5)\\
\beta \left(\lambda_{p_O}\right)&=-\epsilon  \lambda_{p_O}+\frac{1}{9} s^2 \lambda_t^2 \left(\lambda_{p_O}(1+T)+2 \lambda_{p_S}\right)\\
    &+s^4 \left(\frac{\left(h_p^2+3 h_w^2\right) \lambda_{p_O}}{16200}-\frac{1}{810} h_p \lambda_{p_O} \lambda_t^2 - \frac{1}{486} \lambda_t^4 \left((T+2) (11 T+4)\lambda_{p_O} + 6 (5 T+3) \lambda_{p_S}\right)\right)+O(s^5)\\
\beta \left(\lambda_{p_S}\right)&=-\epsilon  \lambda_{p_S}+\frac{1}{9} s^2 \lambda_t^2 \left(\lambda_{p_S}(1+T)+ 2 \lambda_{p_O}\right)\\
    &+s^4 \left(\frac{\left(h_p^2+3 h_w^2\right) \lambda_{p_S}}{16200}-\frac{1}{810} h_p \lambda_{p_S} \lambda_t^2 - \frac{1}{486} \lambda_t^4 \left((T+2) (11 T+4) \lambda_{p_S}+ 6 (5 T+3) \lambda_{p_O}\right)\right)+O(s^5)\\
\beta \left(h_p\right)&=-2 \epsilon h_p+s^2 \left(\frac{1}{3} h_p T \lambda_t^2+\frac{h_p^2}{90}\right)+s^4 h_p \left(\frac{ h_p^2 + 3h_w^2}{5400}-\frac{1}{162} T (11 T+10) \lambda_t^4\right)+O(s^5)\\
\beta \left(h_w\right)&=-2\epsilon h_w+\frac{1}{3} h_w s^2 T \lambda_t^2+s^4 h_w \left(\frac{h_p^2 + 3 h_w^2}{5400}-\frac{1}{162} T (11 T+10) \lambda_t^4\right)+O(s^5)\\
\beta \left(h_3\right)&=-2\epsilon h_3 +s^2 \left(\frac{4}{9} h_3 T \lambda_t^2+\frac{h_p h_3}{45}\right)+s^4 h_3 \left(\frac{h_p^2 + 3 h_w^2}{5400}-\frac{1}{162} T (11 T+10) \lambda_t^4\right)+O(s^5)\\
\beta \left(h_4\right)&=-2\epsilon h_4 +s^2 \left(\frac{2}{3} T \left(h_4-10 \lambda_{p_E}^2\right) \lambda_t^2+\frac{h_3^2}{270}\right)+s^4 h_4 \left(\frac{h_p^2 + 3 h_w^2}{5400}-\frac{1}{162} T (11 T+10) \lambda_t^4\right)+O(s^5)\\
\beta \left(h_5\right)&=-2\epsilon h_5 +\frac{1}{270} s^2 \left(30 T \lambda_t^2 (6 h_p + 18 h_w + 5 h_5) +18 h_p^2+81 h_w^2+2 h_3^2-1800 T\lambda_t^2(\lambda_{p_E}^2 + 2 \lambda_t^2)\right)\\
    & +s^4 h_5 \left(\frac{h_p^2 + 3 h_w^2}{5400}-\frac{1}{162} T (11 T+10) \lambda_t^4\right)+O(s^5)\\
\beta \left(h_6\right)&=-2 \epsilon h_6+s^2 \left(\frac{2}{9} \left(h_3+3 h_6\right) T \lambda_t^2+\frac{h_p}{45}  \left(h_3+2 h_6\right)\right)+s^4 h_6 \left(\frac{h_p^2 + 3 h_w^2}{5400}-\frac{1}{162} T (11 T+10) \lambda_t^4\right)+O(s^5)\\
\beta \left(h_7\right)&=-2 \epsilon h_7+\frac{s^2}{270}  \left(\splitfrac{30T \lambda_t^2 (3 h_p + 9 h_w + 6 h_4 + 6h_5 + 8 h_7)}{+9 h_p^2+7 h_3^2+12 h_3 h_6-1800 T \lambda_t^2 (\lambda_t^2+12 \lambda_{p_E} \lambda_{d_S} +11 \lambda_{p_E}^2)}\right)\\
    &+s^4 h_7 \left(\frac{h_p^2 + 3 h_w^2}{5400}-\frac{1}{162} T (11 T+10) \lambda_t^4\right)+O(s^5)\\
\beta \left(h_8\right)&=-2\epsilon h_8 + \frac{s^2}{270}  \left(\splitfrac{30T \lambda_t^2 (h_5 + 4 h_7 + 12 h_8) +9 h_w^2+2 h_3^2+12 h_6^2+12 h_3 h_6}{-1800 T \lambda_t^2 (24 \lambda_{p_E} \lambda_{d_S} + 18 \lambda_{d_S}^2 + 5 \lambda_{p_E}^2)}\right)\\
& +s^4 h_8 \left(\frac{h_p^2 + 3 h_w^2}{5400}-\frac{1}{162} T (11 T+10) \lambda_t^4\right)+O(s^5)\\
\end{align*}
}
The general large-$N$ anomalous dimensions are as given in \cref{eq:anomDimPhiGeneralPert,eq:anomDimPsiGeneralPert}. For completeness, we also give the beta functions for the masses (where necessarily they must now take non-zero values):
{\scriptsize
\begin{subequations}
\begin{align}
    \beta (m^2)&=-2 m^2+\frac{4}{9} s^2 T \left(m^2-3 M^2\right) \lambda_t^2+8 M \sqrt{m^2} s^3 T \lambda_t^2 \left(\lambda_{p_E}+\lambda_{d_S}\right)\\
    & \quad +s^4 \left(\frac{2 (h_p^2 +3 h_w^2) m^2}{2025}+\frac{2}{243} T \lambda_t^4 \left(63 M^2 (T+3)-m^2 (5 T+43)\right)\right)+O(s^5)\notag \\
\beta (M)&=-M+\frac{4}{9} M s^2 \lambda_t^2+\frac{2 M}{243} \lambda_t^2 s^4 \left(648T (\lambda_{d_S}+\lambda_{p_E})^2 -13 T \lambda_t^2-23 \lambda_t^2\right)+O(s^5) \label{eq:perturbativeMbetaFunc}
\end{align}
\end{subequations}
}
Note, that if we consider $\lambda \sim h \sim \epsilon$ of the same magnitude, then these beta functions are only correct to cubic order in the coupling constants. To get the full quintic corrections it would be necessary to renormalise $h$ using the $h^5$ diagrams.

At the \lammelonic{} fixed point, we can use \eqref{eq:perturbativeMbetaFunc} to compute the scaling dimensions of the mass operators;  these match the non-perturbative calculations using \cref{eq:PevenGeneralsSDcondition,eq:PoddGeneralsSDcondition}:
\begin{subequations}
\begin{align}
\Delta_{\phi^2}^{\text{pert}} &=1+\left(3-\frac{4}{T+1}\right) \epsilon + \frac{2 T \left(6 T^2-22 T-41\right)}{3 (T+1)^3} \epsilon^2 +O\left(\epsilon ^3\right) = D+\dv{\beta_{m^2}}{m^2}|_{fp}\\
\Delta_{\bar{\psi}\psi}^{\text{pert}}  &= 2+\left(\frac{4}{T+1}-1\right) \epsilon -\frac{(4 T (T+5)+42) \epsilon ^2}{3 (T+1)^3}+O\left(\epsilon ^3\right) = D+\dv{\beta_{M}}{M}|_{\lambda_*}
\end{align}
\end{subequations}

\subsection{Discrepancy with Jack and Poole} \label{app:JackPooleDiscrepancy}

In $D=3$, we can transform each Dirac fermion into two Majorana fermions via $\psi_a = \frac{1}{\sqrt{2}}(\xi_{a,1} + i \xi_{a,2})$. This keeps the kinetic terms canonical, and  modifies the interaction term from $\half \lambda_{abcd} \phi_a \phi_b \bar\psi_c \psi_d = \frac{1}{4} Y_{abCD} \phi_a \phi_b \bar\xi_C \xi_D $, with $Y_{abCD} = \lambda_{abcd} \delta_{x_c x_d}$, where $C,D$ are superindices combining $c,d$ and $x_{c,d}\in \{1,2\}$. Plugging this into the results of \cite{Jack:2016utw}, ignoring the gauge couplings, we find beta functions for $Y$ and $h$. Though we have an alternate sign for $\lambda$, this does not change the beta functions. These should match ours when we substitute $T=\half T'$ in the equations of \eqref{eq:generalVectorResults}, since in 3D a single Majorana fermion is half of a Dirac fermion:
\begin{equation}\begin{aligned}\label{eq:PooleMatching}
[\beta_{Y}]_{ab(c,x_c)(d,x_d)} = \beta_{\lambda_{abcd}} \delta_{x_c, x_d}|_{T \to \half T'}, \quad [\beta_h]_{abcdef} = [\beta_h]_{abcdef}|_{T \to \half T'}.
\end{aligned}\end{equation}
This is almost the case, except for the coefficient of $h_{B_1 B_2B_3B_4B_5\nu } \lambda _{B_6\rho \tau \sigma } \lambda _{\nu \rho \sigma \tau }$. This matches up with the coefficient $d_{3}^{(2)}$ of the tensor structure $V_3^{(2)}$. To match with our results as per \eqref{eq:PooleMatching}, we need $d_{3}^{(2)}=T' s^2$; the paper \cite{Jack:2016utw} has $d_{3}^{(2)}=2T' s^2$ (taking $T'=2$). %
Since this term contributes to all of the $\beta_{h_i}$s in the melonic limit, the fact that the independent non-perturbative computations of this paper match our perturbative computations suggests that $d_{3}^{(2)}=T' s^2$. %

\section{Vector model analysis} \label{sec:vectorModelAnalysis}

The general analysis can be reduced to the $\mathrm{O}(N_b) \times \mathrm{U}(N_f)$ vector model, with $\phi_{a=1,\ldots,N_b}$, $\psi_{i=1,\ldots,N_f}$ at finite $N_b,N_f$, by keeping only the maximal-trace type couplings: 
\begin{equation}\begin{aligned}
V_{\text{int}}(\phi, \psi) = \frac{\lambda}{2} \phi_a \phi_a \bar{\psi}^i \psi_i + \frac{h}{6!} (\phi_a \phi_a)^2.
\end{aligned}\end{equation} 
In the case $N_b =N_f$, we could also permit the coupling $\half \lambda_{d_O} (\bar\psi_i \phi_i)(\phi_j \psi_j)$, which, if our scalar was complex, would correspond to the Popovi\'c model studied in \cref{sec:Popovic} -- we will not do so here. Substituting into \eqref{eq:generalVectorResults}:
\begin{subequations}
\begin{align}
\beta_\lambda &= -\epsilon \lambda +\frac{8}{3} (T N_f +N_b+3) \lambda^3 s^2  + \frac{16}{3} \frac{(N_b+4)(N_b+2)}{15} \frac{h^2}{15} \lambda \, s^4 + O(\lambda^4, \ldots)\\
\begin{split}
\beta_h &= -2 \epsilon h + 4 (3 N_b + 22)\frac{h^2}{15} s^2 + 32 T N_f h \lambda^2 s^2 - 720 T N_f \lambda^4\\
& \quad \quad -\half \left( \splitfrac{\pi ^2 (((N_b+34) N_b+620) N_b+2720)}{+8 ((53 N_b+858) N_b+3304)}\right) \frac{h^3}{(15)^2} s^4 + O(h^4, \lambda^5,\ldots)
\end{split}\\
\gamma_\phi &=  \frac{(N_b+2) (N_b+4)}{15} \frac{h^2}{15} \frac{s^4}{6} +\frac{1}{3}  TN_f \lambda ^2 s^2 + O(h^3, \ldots) \label{eq:vectorGammaPhi}\\
\gamma_\psi &= \frac{1}{3} N_b  \lambda ^2s^2 + O(\lambda^3,\ldots) \label{eq:vectorGammaPsi}
\end{align}
\end{subequations} %
For the bosonic sector, this agrees with the results of \cite{Pisarski:1982vz}, up to the identification $\frac{h}{6!} = \frac{\pi^2 \lambda}{3}$. 

Thus for all $N_b, N_f, T$, six perturbative fixed points solve $\beta_{\lambda,h}=0$; to leading order in $\epsilon$, which manifestly match up in terms of $\epsilon$ dependence to the prismatic-type fixed points that were identified in \cref{sec:findingPerturbativeBetas}:
\begin{align*}
\text{trivial}: \; &s \lambda= 0, \; s^2 h = 0;\\
\hprismatic:\; &s\lambda = 0, \;  \frac{s^2 h}{15} = \frac{1}{2(22 + 3N_b)}\epsilon;\\
\hlamprismatic:\; &s\lambda = \pm_1 \left(\frac{3}{8(3 + N_b  + TN_f)}\right)^\half \sqrt{\epsilon}, \\
&\qquad \frac{s^2 h}{15} = \frac{ \left(1 -\frac{6 T N_f}{T N_f+N_b+3} \pm_2 \sqrt{\frac{3 T N_f \left(8 T N_f+23 N_b+186\right)}{\left(T N_f+N_b+3\right){}^2}+1}\right)}{4 (3 N_b+22)}\epsilon.
\end{align*}
Thus, these prismatic theories have not only a large-$N$ vector precursor, but also a finite-$N$ precursor. The presence of the $h^2$ and $\lambda^4$ terms in $\beta_h$ means that we are missing any precursors of the melonic fixed points; these are of course suppressed by the melonic limit.

\subsection{Leading order conformal analysis} \label{app:conformalcalc}
\subsubsection{Free propagators}
To set our conventions, we consider the following Lagrangians, in mostly positive signature, where $\bar\psi = \psi^\dagger \beta, \beta = i \gamma^0$:
\begin{equation}\begin{aligned}
\cL = - \bar\psi (\not{\partial} + m) \psi, \quad \cL = -\half \phi (-\partial^2 + m^2) \phi
\end{aligned}\end{equation}
The free Lorentzian position-space propagator is \cite{Zhang:2008jy} (where we set $m=0$): 
\begin{equation} \label{eq:freeBosonPropagator}
    G^\phi_0(x)=\expval{T \phi(x) \phi(0)} = \int \frac{d^D k}{(2\pi)^D} \frac{-i}{k^2 + m^2 -i \epsilon} e^{+i k \cdot x} = \frac{1}{(D-2) \Omega_{D-1}} \left( \frac{1}{x^2 + i \epsilon} \right)^{\frac{D-2}{2}}. %
\end{equation}

We define $\Omega_x$ as the surface area of an $x$-sphere, i.e. $\Omega_x = 2 \pi^{(x+1)/2} /\Gamma((x+1)/2)$.
Once we know this, the fermion propagator is trivial:
\begin{equation}\begin{aligned}%
    G^\psi_0(x) &= \expval{T \psi(x) \bar{\psi}(0)} = \expval{T \psi_\alpha(x) \bar{\psi}_\beta(0)} = \int \frac{d^D k}{(2\pi)^D} \frac{-i(-i \not{k}+m)}{k^2 +m^2 -i \epsilon} e^{+i k \cdot x}\\
    &= \int_k \frac{-i(-\not\partial +m)}{k^2 +m^2 -i \epsilon} e^{+i k \cdot x} \stackrel{m=0}{=} -\not{\partial} G_0^\phi(x) = \frac{\not{x}}{\Omega_{D-1}} \left( \frac{1}{x^2 +i \epsilon} \right)^{\frac{D}{2}}
\end{aligned}\end{equation}
\subsubsection{Anomalous dimensions}
Since we are interested in the fixed points of \eqref{eq:Neq1Lagrangian}, we perform a simple conformal analysis in $D=3-\epsilon$. The equations of motion are
\begin{align}
    &\Box \phi = \partial^\mu \partial_\mu \phi = \lambda \phi \bar{\psi}\psi + \frac{h}{5!} \phi^5 \label{eq:bosonEOM}\\
    &\left(\not{\partial} + \frac{\lambda}{2} \phi^2\right) \psi = 0, \quad \bar\psi\left(-\overleftarrow{\not{\partial}} + \frac{\lambda}{2} \phi^2\right) =0.
\end{align}
Considered as operator equations, these must be normal ordered. However, at the conformal fixed point, we know that the propagator looks like %
\begin{align}
    G(x) = \kappa_{\text{full}} \left(\frac{1}{x^2 +i\epsilon}\right)^{(D-2)/2 + \gamma_\phi} = G_0^\phi(x) + O(\gamma_\phi)
\end{align}
for some constant $\kappa_{\text{full}}$. At precisely $D=3$, we know that the theory is free, and hence near $D=3$ we can assume that $\gamma_\phi$, $\lambda$, and $h$ are small (at least $O(\sqrt\epsilon)$), as with the usual Wilson-Fisher fixed point. Squaring \eqref{eq:bosonEOM}, we can then calculate the following quantity\footnote{We use $\partial_\mu(1/\abs{x}^a) = - a x^\mu / \abs{x}^{a+2}$, which gives $\partial^\mu\partial_\mu(1/\abs{x}^a) = a (a +2-D)/\abs{x}^{a+2}$}:
\begin{align*}
    \expval{:\Box{\phi}(x)::\Box{\phi}(0):} &=\Box^2 G(x)|_{x=0} = \frac{4}{(D-2)\Omega_{D-1}} (D-2) D \gamma_\phi \left(\frac{1}{x^2 + i\epsilon}\right)^{(D+2)/2} + O(\gamma_\phi^2)\\
    &=  \expval{:\lambda \phi \bar{\psi}\psi +\frac{h}{5!} \phi^5:|_{x} :\lambda \phi \bar{\psi}\psi + \frac{h}{5!} \phi^5|_{0}:}
\end{align*}
We can assume the propagators are free to this order, and therefore we obtain
\begin{align}
   \frac{12}{\Omega_{2}} \gamma_\phi \left(\frac{1}{x^2 + i\epsilon}\right)^{5/2} + O(\gamma_\phi^2)= \lambda^2 \expval{:\phi_x\phi_0:}\expval{:\bar{\psi}\psi_x::\bar{\psi}\psi_0:} + \frac{h^2}{(5!)^2}\expval{:\phi_x^5: :: \phi_0^5:}.
\end{align}
 Precisely the same calculation can be performed for the fermionic equation of motion. From the free theory, we know $\kappa_{\text{full}} = (\Omega_2)^{-1}= 2s + O(\epsilon)$, where $s = 1/(8\pi)$ as usual. So performing the free Wick contractions, we obtain that to leading order in $\epsilon$, with $\Tr[\mathbb{I}_s]= T$,
\begin{align}
    \gamma_\phi = \frac{1}{3} T \lambda^2 s^2 + \frac{4}{3} \frac{h^2}{5!}, \quad \gamma_\psi = \frac{\lambda^2}{3} s^2.\label{eq:conformalAnalysisDims} 
\end{align}
This is trivially extensible to the general vector case, which yields precisely \cref{eq:generalVecGammaPhi,eq:generalVecGammaPsi}. In the case of $N_f$ fermions and $N_b$ scalars, satisfying an $\mathrm{O}(N_b)\times \mathrm{U}(N_f)$ symmetry, adding the appropriate $\delta_{ij}$ and $\delta^m_n$s to the Lagrangian we find exactly \cref{eq:vectorGammaPhi,eq:vectorGammaPsi}.
\subsubsection{Beta functions}
As demonstrated by Cardy \cite{Cardy:1996:SRinSP,Gaberdiel:2008fn}, we can use conformal perturbation theory \cite{Komargodski:2016auf} (see also section 4 of \cite{Fei:2015oha}) to obtain the quadratic terms in the $\beta_h$, in a particular scheme, using the OPE coefficients $C^I_{JK}$ of the CFT about which we are perturbing. Specifically, if perturbing a CFT with the set of dimensionless couplings $S_{\text{int}} = \sum_I \lambda^I \mu^{D-\Delta_I} \int \cO_I$, their beta functions should be
\begin{equation}\begin{aligned}\label{eq:CardyBetas}
\beta^I = (\Delta_I - D) \lambda^I + \half \Omega_{D-1} C^I_{JK} \lambda^J \lambda^K + O(\lambda^3),
\end{aligned}\end{equation}
where $C^{I}_{JK}$ is the OPE coefficient in the conformal theory: $\cO_J\times \cO_K \sim \sum_I C^I_{JK} \cO_I$. The particular scheme is similar to but not precisely the same as MS scheme; however, to this order there is no difference in the beta function, because we are only considering the marginal couplings and MS scheme is massless. For our case, take $\lambda^I = h$, with $\cO_I =\phi^6/6!$ as the operator in the Gaussian CFT. From the Gaussian OPE, it is easy to calculate that the OPE coefficient $C^h_{hh}$ is ${6 \choose 3} \times 6 \times 5 \times 4 = 2400$ multiplied by terms coming from the normalisation of $h$ and $\phi$:
\begin{equation}\begin{aligned}
\frac{\phi^6(x)}{6!} \times \frac{\phi^6(y)}{6!} \supset 2400 \times \frac{1}{6!} G_0(x-y)^3 \frac{\phi^6(y)}{6!}+ \cdots,
\end{aligned}\end{equation}
where $G_0^\phi(x)$ is the usual free propagator, which in mostly positive or positive signature is \eqref{eq:freeBosonPropagator}. Hence, we find that the coefficient of the $h^2$ term is indeed as given in \eqref{eq:generalVectorResults}
\begin{equation}\begin{aligned}
\half \Omega_{D-1} C^h_{hh} = \left[\half \Omega_{D-1} \times\frac{2400}{6!} \times \left(\frac{1}{(D-2)\Omega_{D-1}}\right)^3\right]_{D=3} = \frac{20}{3} s^2.
\end{aligned}\end{equation}

\section{Analysis of the Popović model} \label{sec:Popovic}

Here we resolve some unclear points with the original model of \cite{Popovic:1977cq}, which is the large-$N$ theory of two $N$-vector fields, a Dirac fermion and a complex scalar. The corrected\footnote{The paper  gives Lagrangian containing ``$\half \phi^\dagger \phi$'', but the $\phi$ field is clearly intended to be complex (for example, using for the one-boson-loop correction to the $\phi$ propagator a symmetry factor of 1). They use $T=\Tr[\mathbb{I}_s] = 2$, due to working in exactly $D=3$ -- we will leave $T$ general here for clarity.} $\mathrm{U}(N)$-symmetric Lagrangian there corresponds to our model but with a complex $\phi$ and the only non-zero coupling being $\lambda_{d_O}$:
\begin{equation}\begin{aligned}
\cL &= \phi_i^\dagger (-\partial^2 + m_0^2)\phi_i + \bar\psi_i(i \gamma_\mu \partial^\mu - M_0) \psi_i + g_0 \sum_{i,j=1}^N (\bar\psi_i \phi_i^\dagger) (\phi_j \psi_j).
\end{aligned}\end{equation}
We stress that the metric signature in \cite{Popovic:1977cq} is $(+--)$, and take
\begin{equation}\begin{aligned}
\Tr[\mathbb{I}_s] &= T, \quad \phi_{i=1,\ldots,N}, \quad \psi_{i=1,\ldots,N}.
\end{aligned}\end{equation}
They analyse this theory in the large-$N$ limit, finding that the beta function is zero in exactly three spacetime dimensions. Beginning with the four-point vertex function
\begin{equation}\begin{aligned}
i g_0 \Gamma_4(p) = \frac{i g_0}{1-ig_0\tilde\Sigma(p)} = \frac{1}{1-N g_0 G \not{p} (-p^2)^{(D-4)/2}},\quad G\equiv \frac{\Gamma(\frac{4-D}{2})B(\frac{D}{2}, \frac{D-2}{2})}{(4\pi)^{D/2}}
\end{aligned}\end{equation}
We define the renormalised and rescaled coupling $g$ by
\begin{equation}\begin{aligned}
g_0 N &\equiv Z_g Z_\phi Z_\psi g \mu^{3-D} =  Z_g g \mu^{3-D} + O(1/N),\\
i g &\equiv i g_0 [\Gamma_4(p^2 =\mu^2)]_{\mathbb{I}},
\end{aligned}\end{equation}
where $[\cdot]_{\mathbb{I}}$ indicates that we drop the part proportional to $\not{p}$, and we have used the fact that the fields are not renormalised to leading order. This gives that $Z_g = 1 + g_0^2 N^2 G^2 \mu^{2(D-3)} + O(1/N) = 1+ Z_g^2 g^2 G^2 +O(1/N)$, and so almost every value of the renormalised coupling in the allowed range $g \in [-1/(2G), +1/(2G)]$ can be reached by two values of the bare coupling (which can take any real value):
\begin{equation}\begin{aligned}
Z_g &= \frac{1\pm \sqrt{1-4 g^2 G^2}}{2 g^2 G^2} +O(1/N)\\
\implies \beta_g &= \mu \od{g}{\mu} = \pm (3-D)\sqrt{1- 4g^2 G^2} +O(1/N)
\end{aligned}\end{equation}
Note that in the limit $D=4-\epsilon$, $G \propto 1/\epsilon$, so this theory has a Wilson-Fisher-like fixed point that is perturbative around $D=4$. They then move on to calculate the anomalous dimension of $\phi$. However, in their equation (3.4) the $\frac{1}{\sqrt{-k^2}}$ should be a $\frac{1}{(-k^2)^{(D-2)/2}}$ -- they have taken the limit too early. However, in trying to resolve this, we then have to evaluate the integral %
\begin{equation}\begin{aligned}
i\tilde\Pi(p^2)&= (-1) \Tr \int \frac{d^D k}{(2\pi)^D} \frac{i}{\cancel{(k-p)}} (ig_0) i\Gamma_4(k) = i g_0\int_k \frac{1}{(k-p)^2} \Tr\left[\cancel{(k-p)} \Gamma_4(k)\right]\\
&= -i Tg_0 \int_k   \frac{(k-p)\cdot k}{[-(k-p)^2]^1} \frac{g_0 N G (-k^2 )^{\frac{D-4}{2}}}{1 +  g_0^2 N^2 G^2 (-k^2)^{D-3}}\\ 
&\propto \int dq q^{D-1}  \frac{q^2}{q^2}\frac{q^{D-4}}{q^{2D-6}} \sim \int_0^\infty dq q^{D-1} \frac{1}{q^{D-2}} \to \infty,
\end{aligned}\end{equation}
which is manifestly divergent for all values of $D$. A neat way to proceed in precisely $D=3$ is to regulate by consistently shifting the conformal scaling dimension of the Hubbard-Stratonovich field \cite{Goykhman:2019kcj}; that is, we switch $(-k^2)^{\frac{D-4}{2}} |_{D=3} \to (-k^2)^{-\half -\eta} \mu^{2\eta}$, for some infinitesimal $\eta$, which becomes the regulator:
\begin{equation}\begin{aligned}
\tilde\Pi(p^2)|_{D=3} &= -T \frac{g_0^2 N G }{1 +  g_0^2 N^2 G^2}\int_k   \frac{(k-p)\cdot k}{[-(k-p)^2]^1(-k^2 )^{\half + \eta}} \\ 
 &= \frac{-i(-p^2)^{1-\eta}}{\eta} T \frac{\tilde{A}}{N} \times \mu^{2\eta} + O(\eta^0), \\\tilde{A}&\equiv \frac{1}{12 \pi^2} \frac{(Ng_0)^2 G}{1+(Ng_0 G)^2} =\frac{4}{3 \pi^2} \frac{Z_g -1}{Z_g} 
\end{aligned}\end{equation}
where $T\tilde{A} =A_{\text{Popovi\'c}}$. Then we can define the field renormalisation $Z_\phi \phi_R \equiv \phi$ with the renormalisation condition $\Delta_R(p^2) =\frac{i}{p^2}$ at $p^2=-\mu^2$. This gives $Z_\phi = (1 + \frac{T\tilde{A}}{N} \frac{1}{\eta})^{-1}$, and so the anomalous dimension is simply $\gamma_\phi = \half \dv{\log Z_\phi^2}{\log \mu} = T\frac{\tilde{A}}{N} + O(1/N^2)$.
Playing the same game for the fermion, we find an anomalous dimension $\gamma_\phi =N_b \frac{\tilde{A}}{N} + O(1/N^2)$, where we define $N_b=2$ as the number of real degrees of freedom of each entry of the scalar field. These results agree with those of \cite{Popovic:1977cq}. In the strong coupling limit, $g_0 N\to\infty$, $Z_g \to \infty$, $g\to 0$ these anomalous dimensions also precisely match the leading order results of a $\phi_i \bar\psi_i \lambda$-type melonic theory in $D=3$ \cite{Fraser-Taliente:2024prep}, where $\lambda$ is a fermionic Hubbard-Stratonovich field. We note that with canonical kinetic terms, the IR scaling for this melonic theory ($\gamma_\phi, \gamma_\psi>0, \gamma_\lambda >0$) is in fact only consistent for $2<D<3$; in particular, the requirement $\gamma_\lambda \ge 0$ invalidates these solutions for $D \ge 3$.

\section{SDE Euclidean integral calculations in general D} \label{app:loopIntegrals}

For ease of use, we put here some standard Euclidean integrals, which are calculated in a straightforward fashion described in \href{https://arxiv.org/pdf/1706.05362.pdf#appendix.B}{\color{black}{Appendix B}} of \cite{Murugan:2017eto}. %

The first is symmetric in $\alpha \longleftrightarrow \beta$, which we emphasise by defining $X_2 \equiv D-\alpha-\beta$.
\begin{equation}\begin{aligned}
    I_{\alpha,\beta}(p) &= \int \frac{\mathrm{d}^D k} {(2\pi)^D} \frac{1}{(k^2)^\alpha ((k+p)^2)^\beta} = \frac{1}{(4\pi)^{D/2}} \frac{\Gamma \left(\frac{D}{2}-\alpha \right) \Gamma \left(\frac{D}{2}-\beta \right) \Gamma \left(\alpha +\beta -\frac{D}{2}\right)}{\Gamma (\alpha ) \Gamma (\beta ) \Gamma (D -\alpha -\beta)(p^2)^{\alpha+\beta-\frac{D}{2}}}\\
    &= \frac{1}{(4\pi)^{D/2}} \frac{1}{(p^2)^{\frac{D}{2}-X_2}} \frac{\Gamma \left(\frac{D}{2}-\alpha \right) \Gamma \left(\frac{D}{2}-\beta \right) \Gamma \left(\frac{D}{2}-X_2\right)}{\Gamma (\alpha ) \Gamma (\beta ) \Gamma (X_2)}. \label{eq:singlemelonstep} %
\end{aligned}\end{equation}
Another basic integral is
\begin{align}
    \int_{k} \frac{k \cdot p}{k^{2\alpha} (k+p)^{2\beta}}  &= \frac{1}{2} \int_k \frac{(k+p)^2 - k^2 -p^2}{k^{2\alpha} (k+p)^{2\beta}} = \frac{1}{2}\left(I_{\alpha,\beta-1}(p) - I_{\alpha-1, \beta}(p) - p^2 I_{\alpha,\beta}(p) \right).
\end{align}
Next, the quartic bosonic melon, symmetric in $\alpha,\beta,\gamma$, with $X_3\equiv \frac{3D}{2} -\alpha -\beta -\gamma$:
\begin{subequations}
\begin{align}
I_{\alpha\beta\gamma} = \int_{k,l} \frac{1}{(k+l+p)^{2\alpha} k^{2\beta} l^{2\gamma}} = \frac{1}{(4\pi)^D} \frac{\Gamma(\frac{D}{2} - \alpha)\Gamma(\frac{D}{2} - \beta)\Gamma(\frac{D}{2}- \gamma) \Gamma(\frac{D}{2} - X_3)}{\Gamma(\alpha)\Gamma(\beta)\Gamma(\gamma) \Gamma(X_3) (p^2)^{\frac{D}{2} - X-3}},
\end{align}
which has similar variations
\begin{align}
J_{\alpha\beta\gamma}(p) = \int_{k,l} \frac{k\cdot p}{(k+l+p)^{2\alpha} k^{2\beta} l^{2\gamma}} = \frac{-(D-2\beta) p^2}{3D - 2 (\alpha+ \beta+\gamma)}I_{\alpha\beta\gamma}(p),
\end{align}
\begin{align}
K_{\alpha\beta\gamma}(p) = \int_{k,l} \frac{k\cdot l}{(k+l+p)^{2\alpha} k^{2\beta} l^{2\gamma}} = \frac{-(D-2\beta)(D-2\gamma) p^2 \, I_{\alpha\beta\gamma}(p)}{2(\alpha+\beta + \gamma - D- 1)(3D - 2 (\alpha+ \beta+\gamma))}.
\end{align}
\end{subequations}
Hence, using that integrals containing $\not{k}$ or $\not{l}$ must be proportional to $\not{p}$, we find:
\begin{subequations}
\begin{align}
    \Sigma_F^{FBB}(p) &= \frac{\lambda_{t}^2}{2} fb^2 \left(I_{\alpha\beta\beta} + \frac{1}{p^2} (J_{\alpha \beta \beta} + J_{\alpha\beta\beta})\right),\\
    \Sigma_B^{BBB}(p) &= \frac{g_{t}^2}{3!} b^3 I_{\beta\beta\beta},\\
    \Sigma_B^{BFF}(p) &= -\frac{\lambda_{t}^2}{1} b f^2 K_{\beta\alpha\alpha} \Tr[\mathbb{I}_s].
\end{align}
\end{subequations}
We also require the sextic bosonic melon, which in Euclidean signature is \cite{Benedetti:2019rja}:
\begin{equation}
    \Sigma_B^{B^5}(p) = \frac{h_t^2 b^5}{5!} \left(\frac{p^{4D-10}}{(4\pi)^{2D}} \frac{\Gamma(\frac{D}{2}- b)^5 \Gamma(5b- 2D)}{\Gamma(b)^5 \Gamma(5\frac{D}{2} - 5 b )} \right).
\end{equation}

 \subsection{Euclidean integrals for q-model} \label{sec:qModelIntegrals}

The general Euclidean scalar melon with $q-1$ different arbitrary conformal propagators $\sim p^{-2 \alpha_i}$ is found by induction to be:
\begin{subequations}
\begin{equation}\begin{aligned}
&I_{\alpha_1, \alpha_2, \ldots, \alpha_{q-1}} (p) = \int \prod_{i=1}^{q-2} \left(\frac{d^d k_i}{(2\pi)^d} \frac{1}{k_i^{2\alpha_i}}\right) \frac{1}{(\sum_i^{q-2} k_i + p)^{2\alpha_{q-1}}} \\
&=\frac{1}{(4\pi)^{D(q-2)/2}} \left( \prod_i^{q-1} \frac{\Gamma(\frac{D}{2} - \alpha_i)}{\Gamma(\alpha_i)}\right) \frac{\Gamma(\frac{D}{2} - \sum_i^{q-1} (\frac{D}{2} -\alpha_i))}{\Gamma( \sum_i^{q-1} (\frac{D}{2} -\alpha_i))} \frac{1}{p^{2(\frac{D}{2} - \sum_i^{q-1} (\frac{D}{2} -\alpha_i))}}
 \ee
The induction proceeds by noting that $I_{\alpha_1, \alpha_2, \ldots, \alpha_{q-1}} (p) = \int_k k^{-2 \alpha_{q-1}} I_{\alpha_1, \alpha_2, \ldots, \alpha_{q-2}} (p+k)$; the base case and inductive step are both \eqref{eq:singlemelonstep}. The fermion integrals, with numerator $k \cdot l$, can also be found by induction from the base case, with the result:
\begin{equation}\begin{aligned}
&K_{mel}(p, \{\alpha_1, \alpha_2\}, \{\alpha_3, \ldots, \alpha_{q-1}\})\\
&= \int \frac{d^d k_1}{(2\pi)^d} \frac{d^d k_2}{(2\pi)^d} \frac{k_1 \cdot k_2}{k_1^{2\alpha_1 +1} k_2^{2\alpha_2+ 1}} \, \prod_{i=3}^{q-2} \left(\frac{d^d k_i}{(2\pi)^d} \frac{1}{k_i^{2\alpha_i}}\right) \frac{1}{(\sum_{i=1}^{q-2} k_i + p)^{2\alpha_{q-1}}} \\
&=\frac{-1}{(4\pi)^{D(q-2)/2}} \left( \prod_{i=1}^{2} \frac{\Gamma(D/2 - \alpha_i + 1/2)}{\Gamma(\alpha_i + 1/2)}\right) \left( \prod_{i=3}^{q-1} \frac{\Gamma(D/2 - \alpha_i)}{\Gamma(\alpha_i)}\right) \\ & \quad \times \frac{\Gamma(D/2 - \sum_i^{q-1} (D/2 -\alpha_i))}{\Gamma( \sum_i^{q-1} (D/2 -\alpha_i))} \frac{1}{p^{2(D/2 - \sum_i^{q-1} (D/2 -\alpha_i))}} %
\end{aligned}\end{equation}
We can work out the propagators with $k \cdot p$ in the same way:
\begin{equation}\begin{aligned}
&J_{mel}(p, \{\alpha_1\}, \{\alpha_2, \ldots, \alpha_{q-1}\})\\
&= \int \frac{d^d k_1}{(2\pi)^d} \frac{k_1 \cdot p}{k_1^{2\alpha_1 +1}} \, \prod_{i=2}^{q-2} \left(\frac{d^d k_i}{(2\pi)^d} \frac{1}{k_i^{2\alpha_i}}\right) \frac{1}{(\sum_{i=1}^{q-2} k_i + p)^{2\alpha_{q-1}}} \\
&=\frac{-1}{(4\pi)^{D(q-2)/2}} \left(\frac{\Gamma(D/2 - \alpha_1 + 1/2)}{\Gamma(\alpha_1 + 1/2)}\right) \left( \prod_{i=2}^{q-1} \frac{\Gamma(D/2 - \alpha_i)}{\Gamma(\alpha_i)}\right) \\ & \quad \times \frac{\Gamma(D/2 +1/2 - \sum_i^{q-1} (D/2 -\alpha_i))}{\Gamma(1/2 + \sum_i^{q-1} (D/2 -\alpha_i))} \frac{1}{p^{2(D/2 -1/2 - \sum_i^{q-1} (D/2 -\alpha_i))}} %
\end{aligned}\end{equation}
\end{subequations}

\include{99-GeneralAnalysis}

\bibliographystyle{JHEP}%
{\raggedright  %
\bibliography{references}    %
}              %

\end{document}